\documentclass[aps,pre,onecolumn,11pt,tightenlines,superscriptaddress,nofootinbib,longbibliography]{revtex4-1}

\usepackage{graphicx}%
\usepackage{mathrsfs}
\usepackage{euscript}
\usepackage{bm}%
\usepackage{lineno}
\usepackage{amsmath}
\usepackage{amssymb}
\usepackage{appendix}

\usepackage{xcolor}
\usepackage{float}

\bibpunct{[}{]}{,}{n}{}{;}

\makeatletter
\renewcommand{\p@subsection}{\thesection~}
\renewcommand{\p@subsubsection}{\thesection~\thesubsection~}
\def\@hangfrom@appendix#1#2#3{\large#1\@if@empty{#2}{#3}{#2\@if@empty{#3}{}{:\ #3}}\par}%
\makeatother
\newcommand{\ocd}[1]{\mkern1mu\overline{\mkern1.5mu;\mkern1.5mu#1\mkern0.5mu}}
\begin{document}

	\title{Boundary Layers, Displacement Flow, and the High-Frequency Response of Porous Media: Beyond the Viscous and Thermal Characteristic Lengths, Order by Order}

	\author{D. Lafarge}
	\email[Email: ]{denis.lafarge@univ-lemans.fr}
	\affiliation{Acoustics Laboratory of the University of Le Mans (LAUM), UMR 6613, Institut d'Acoustique Graduate School (IA-GS), CNRS, Le Mans University, Le Mans, France}
	\author{A. Bouchendouka}
	\author{Z. E. A. Fellah}
	\affiliation{Aix Marseille Univ, CNRS, Centrale Med, LMA, Marseille, France}

	\date{November 3, 2026}   %

	\begin{abstract}
		The Johnson--Koplik--Dashen viscous characteristic length $\Lambda$ and its thermal analogue $\Lambda'$ introduced by Champoux and Allard play a central role in describing boundary-layer dissipation and high-frequency dispersion in fluid-saturated porous media, both in Biot's theory with frame motion and in the equivalent-fluid (rigid-frame) limit. While these lengths govern the leading $O(\omega^{-1/2})$ boundary-layer behaviour, they omit higher microgeometric effects associated with interface curvature and field gradients. As shown in recent investigations of the breakdown of the classical Johnson--Allard description in highly absorbing media, the influence of higher-order high-frequency terms is not merely an incremental correction, but can become essential to capturing the true dispersive and dissipative response. In this work, we present an exact analytical framework developing the asymptotic high-frequency expansions of dynamic tortuosity and dynamic compressibility beyond the leading boundary-layer terms. By systematically resolving the oscillatory Stokes and heat conduction problems in the thin boundary-layer limit, we establish a recursive hierarchy, which can be continued to any order, defining successive microgeometric parameters. These higher-order characteristic parameters of dimension surface, volume, etc., account for boundary-layer curvature, tangential gradients, and the local geometry of pore boundaries. Each order requires transport equations of WKB type across the boundary layer, closed by a non-secularity condition, together with one Neumann problem for the potential displacement flow that the boundary layer induces. The parameters are given by wall integrals; the first ones are obtained explicitly. Applied to the values recently fitted on highly absorbing foams, these expressions show that the negative surface parameter is consistent with strong in-parallel disorder of the pore sizes. The third-order term of the expansion, large and positive where any in-parallel arrangement makes it negative, points instead to strong constrictions or enlargements along part of the flow paths, at which the boundary layer acts back on the core flow; the fitted volume parameter is an effective value at the measurement frequencies. The thermal parameters reduce to purely geometrical quantities: the integral mean curvature of the pore walls, a Willmore-type measure of their departure from umbilicity and, at the next order, this measure weighted by the mean curvature, in agreement with the heat-content asymptotics of van den Berg and Gilkey. Using the Avellaneda--Torquato relaxation-mode representation, we further show that all these parameters are spectral data of the distribution of relaxation times. The odd-order coefficients fix the shape of its short-relaxation-time end---beyond $\Lambda$, through a single combination of $\Lambda$ and of the surface and volume parameters---whereas the even-order ones are regularized moments of the whole distribution, global sum rules complementing the low-frequency parameters. The same analysis yields the short-time impulse response, an extended Curie--von Schweidler law, and the structure of the pressure field. The analysis assumes smooth pore walls; for isolated sharp edges across which the fluid flows, we show that the singular terms of the formal expansion cancel, so that the description persists with apparent, logarithmically varying values of the parameters. The resulting analytical expansions provide a rigorous theoretical closure for wideband asymptotic matching and establish the exact geometrical foundation required to describe high-frequency visco-inertial and thermal transport beyond the classical Johnson--Allard paradigm.

	\end{abstract}

	\maketitle %

	\section{Introduction}\label{sec1}

	Air-saturated porous materials are among the main sound absorbers used in airborne noise control: fibrous mats, mineral wools and open-cell and reticulated polyurethane foams, with low, intermediate or high airflow resistivities, are familiar examples. As shown principally by Allard and co-workers~\cite{ALLAT}, their acoustic response results both from the mechanical behaviour of the solid frame and from the visco-inertial coupling and thermal exchanges between frame and pore fluid. For many of these materials it is satisfactorily described by Biot's theory~\cite{BIOT}, supplemented by Johnson--Allard-type models~\cite{JKD, A6} of the frequency-dependent fluid response; when the frame motion can be neglected, this reduces to the rigid-frame, equivalent-fluid description. The Johnson--Champoux--Allard--Lafarge (JCAL)~\cite{LLAT} and Johnson--Champoux--Allard--Pride--Lafarge (JCAPL)~\cite{LAF_PhD}, \cite[Appendix]{LAF2} models are widely used refinements of this framework (see e.g.\ Ref.~\cite{LLN}). They add low-frequency information, but keep the classical high-frequency description unchanged.

	This practical success must not hide what the underlying theory is, and what it is not. It is a local theory: its response functions depend on frequency, but not on the spatial variations of the macroscopic excitation. It is local because it rests on a simplification of the pore-scale physics, which we shall call the ``divergence-free'' simplification. At the pore scale, the motion of the fluid relative to the frame is taken as divergence-free, and the heat source due to compression is set by the macroscopic pressure alone; the incompressible microscopic Stokes problem and a Fourier problem with a uniform source follow. The thermal part is thus neither an independent approximation nor a condition on the temperature field: it comes from the same rationale that makes the velocity field divergence-free (Sec.~\ref{sec2}). In straight cylindrical tubes, this is essentially the passage from Kirchhoff's full theory to Zwikker and Kosten's treatment~\cite{ZK, KIRCH}. The simplification is well justified for conventional sound-absorbing materials without locally resonant structures; but it is a simplification of the real problem, which restricts the microscopic dynamics retained, and not a general theory valid for arbitrarily complex pore geometries.

	The general description is nonlocal. It keeps the dependence of the response on the spatial variations of the excitation, and no longer imposes microscopic incompressibility~\cite{LN,NL,NLL,NKL,LAF1,LAF2}. This is not a matter of adding further frequency-dependent coefficients to the classical models: the microscopic problem itself is different, and so are the nature and meaning of its asymptotic limits. In particular, as discussed in Ref.~\cite{LAF_HF}, the high-frequency description must then be generalized, together with the characteristic lengths and the high-frequency limits $\alpha_\infty$ and $\beta_\infty$ of the tortuosity and of the compliability defined below. This broader setting is not our subject here. We deliberately place ourselves within the classical local theory, and ask a precise question about it: what microgeometric information does its high-frequency response contain? As in the classical literature~\cite{JKD,AT,LLAT}, we work in the isotropic setting, in which the visco-inertial response reduces to a scalar dynamic tortuosity, and leave the anisotropic formulation for later developments.

	Within the local theory, two levels must be kept apart. The first is the microscopic response itself, which is well defined once the physical approximations of the theory are granted, together with its exact macroscopic expression. The second is the level of the models used to represent it. In principle, the full-frequency visco-inertial and thermal response functions are fixed by the time-harmonic Stokes and Fourier problems in the pore space, solved within the ``divergence-free'' simplification and averaged over the pore volume; the Stokes problem involves a divergence-free velocity field explicitly, the scalar Fourier problem the corresponding thermal simplifications~\cite{LAF0}. The dynamic tortuosity and the dynamic normalized compressibility are therefore determined by the microgeometry, independently of any analytical interpolation.

	We call the dynamic normalized compressibility the dynamic compliability, a term coined in Ref.~\cite{LAF_HF} that echoes the acoustic compliance, the partner of the inertance in lumped-element acoustics. The compliability is not the thermal counterpart of the tortuosity, but its counterpart on the side of compressibility; the thermal counterpart is the thermal tortuosity, introduced as such in Ref.~\cite{LAF_PhD} and recalled in Sec.~\ref{sec2}. The term is meant to place the compliability on the same footing as the tortuosity: the tortuosity turns the density of the pore fluid into an apparent, frequency-dependent density, and the compliability turns its compressibility into an apparent, frequency-dependent compressibility. In the local theory, this equal footing is partly hidden: the compliability is then fixed by the thermal tortuosity [Eq.~\eqref{eq:dynamic-compliability}], and its high-frequency limit $\beta_\infty$ is 1, whereas the high-frequency limit $\alpha_\infty$ of the tortuosity depends on the geometry. In the nonlocal theory, $\beta_\infty$ is no longer 1 in general, and the compliability becomes as nontrivial as the tortuosity~\cite{LAF_HF}.

	The second level is that of the Johnson--Allard-type models: compact representations of the dynamic tortuosity and of the dynamic compliability, built from a few geometrical and transport parameters. They do not merely interpolate between the low- and high-frequency limits: the divergence-free simplification itself constrains the form of the responses. Within it, the pore fluid in the rigid frame stores kinetic energy only, and the excess temperature obeys a pure diffusion equation, so that a free motion can only decay without oscillating; all the singularities of the response functions then lie on the negative imaginary frequency axis~\cite{JKD,LAF_PhD}. The responses are purely relaxational, a property stronger than causality (Sec.~\ref{sec5-time}), and the model functions are constructed to respect it~\cite{JKD,LAF_PhD,LLAT}.

	At high frequencies, viscous and thermal diffusion are confined to thin layers along the pore walls, and the leading departures from inviscid and adiabatic behaviour are governed by two lengths. The viscous characteristic length $\Lambda$ of Johnson, Koplik and Dashen involves the distribution on the walls of the microscopic potential-flow field; the thermal characteristic length $\Lambda'$ of Champoux and Allard is fixed by the ratio of pore volume to pore surface. They are two different geometrical weightings of the interface, and account for the leading boundary-layer contributions to the macroscopic response. For many conventional materials this is enough for a useful description of high-frequency dispersion and dissipation, and the next asymptotic coefficients need not be known.

	The usual derivation of these characteristic lengths assumes a smooth pore boundary, but otherwise leaves the pore geometry arbitrary. We keep this hypothesis throughout: the formulas derived below apply to pore boundaries that are smooth at the scale of the viscous and thermal skin depths. Sharp geometrical features, such as edges across which the fluid flows, make the ideal field singular and call for a separate asymptotic treatment~\cite{FGL,AA,CSGL,LANG}. We show in Appendix~\ref{app:wedge} that, in practice, they do not invalidate the present description. For wedge-shaped edges of apex angle $\gamma$, the formal high-frequency expansion of the response, in powers of the complex viscous skin depth $\varepsilon\propto\omega^{-1/2}$ [Eq.~\eqref{eq:complex-skin-depths}], contains, besides the Johnson term in $\varepsilon/\Lambda$, a nonanalytic term of order $\varepsilon^{w}$, with $w=2\pi/(2\pi-\gamma)$ between 1 and 2. As $\gamma\to0$, both terms diverge, their coefficients behaving as $\pm1/(w-1)$ while $\varepsilon^w\to\varepsilon$, but their divergent parts cancel, and their sum tends to a finite knife-edge law in $\varepsilon\ln\varepsilon$. This is consistent with the numerical study of Langlois~\cite{LANG}. Revisiting by finite-element computations the corrugated channel with wedge-shaped constrictions of Ref.~\cite{CSGL}, Langlois confirmed the form proposed there for the real part of the dynamic permeability $k(\omega)$, $\Re k\propto\omega^{-3/2}\big[1+C_1(\omega/\omega_c)^{(1-w)/2}\big]$, in which $\omega_c$ is a characteristic frequency and $C_1$ a dimensionless coefficient (Appendix~\ref{app:wedge-ReIm}), and found that $C_1$ tends to $-1$ for small wedge angles. The second term is then not a small correction to the Johnson term: the two nearly cancel, and the bracket becomes proportional to $\gamma$, while the coefficient of the Johnson term, proportional to $1/\Lambda$, grows as $1/\gamma$. Their product is the finite law $\Re k\propto\omega^{-3/2}(\ln\omega+\mathrm{const})$, which Langlois obtained by expanding the same expression for small angles. He also showed that the classical high-frequency limit is recovered with good accuracy, for both the real and imaginary parts of $k$, if the electrical (ideal-flow) problem is solved in a fluid domain truncated by a layer of thickness comparable to the viscous skin depth, the wedge tips being rounded at that scale. The two terms separate only when the skin depth becomes much smaller than $\ell\,e^{-2\pi/\gamma}$, with $\ell$ a length of the order of the pore size defined in Appendix~\ref{app:wedge}. For small angles this happens at no physical frequency: the skin depths required lie far below the mean free path of the gas molecules, where the continuum description itself fails. A vanishing $\Lambda$ therefore does not signal a breakdown of the high-frequency response. What measurements yield are apparent values of $\Lambda$, and likewise of the higher-order parameters, which depend only logarithmically on frequency and keep a reasonable magnitude: they are those of the geometry with its sharp features rounded at the scale of the skin depth, to which the smooth-boundary expressions obtained here apply.

	For some materials, however, $\Lambda$ and $\Lambda'$ are not enough: recent observations show that, even within the local theory, the next terms of the high-frequency expansions, with their own microgeometric parameters, are needed. From ultrasonic reflected waves, Roncen \emph{et al.}~\cite{RFP} identified the viscous surface parameter $\Sigma$ appearing in Kergomard \emph{et al.}~\cite{KLG}, and Bouchendouka \emph{et al.}~\cite{BFN} measured both $\Sigma$ and the volume parameter $V$ in highly attenuating materials, with positive values of $\Sigma$. In our recent work on the breakdown of classical descriptions~\cite{LBF}, anomalous absorption was observed in similar polyurethane foams fabricated with graphite particles, which exhibit numerous internal surfaces and membranes. Their ultrasonic transmission could be reproduced only when both higher-order terms were included, the second, volume-related correction playing a central role, rather than the first, surface-related one alone, and with form factors far outside the usual range, which no classical model, JCAPL included, could accommodate. The form factors $M$ and $P$ are dimensionless numbers that combine $\Lambda$ with the static permeability $k_0$ and the static viscous tortuosity $\alpha_0$; with $\Lambda$, they fix $\Sigma$ and $V$ in the JCAPL model. A negative $\Sigma$, together with a small positive $V$, gave a slightly better agreement than a positive $\Sigma$ and more plausible values of $\Lambda$ and $V$, in line with the sign of $\Sigma$ that the JCAPL model gives for such form factors. This raised the question whether still higher-order terms are needed.

	A point of principle must be made here. That a term is of higher order in an asymptotic expansion means that it is smaller in the strict high-frequency limit; it does not mean that it is physically negligible over the frequency range that matters for sound absorption. Its coefficient, and the way it enters the full-frequency response, can make the corresponding microgeometric information essential. The sharp edges discussed above give an extreme example. The edge term in $\varepsilon^w$ is of higher order than the Johnson term, yet, for small angles, $w$ is so close to 1 that the two terms keep the same magnitude at every physical frequency, with large and nearly opposite coefficients; the formal ordering would become effective only where the continuum description itself fails. Discarding the edge term because of its order would leave a Johnson term that diverges as the edge sharpens. For the materials that motivate this work, the problem is therefore not to refine an adequate description by small corrections. It is to identify information that the classical models leave out, and that the observed response requires.

	The series--parallel pore-network model of our accompanying work, ``Mechanisms Behind Higher-Order Visco-Thermal Parameters in Porous Media: A Series--Parallel Pore Network Model,''~\cite{BLF} provides a tractable setting in which the emergence and influence of these higher-order parameters can be examined, shows why the classical lengths cannot exhaust the relevant information, and traces the large form factors and the negative $\Sigma$ to strong in-parallel disorder. With the log-normal distributions considered there, however, it does not reproduce the positive values of $V$ found together with negative values of $\Sigma$ at the measured form factors, a discrepancy attributed there to transverse flow and three-dimensional interconnectivity. With other distributions of radii, the same model can match the fitted values, but only by including passages narrower than the viscous skin depth at the measurement frequencies, with radius contrasts of order 100, where neither its neglect of the transition zones at the junctions nor the high-frequency expansion itself is justified: such networks are configurations possible within the model, not a representation of the actual foams (Sec.~\ref{sec7}). More fundamentally, being one-dimensional, a model of this kind cannot tell what the parameters are for an arbitrary smooth pore geometry. The essential question thus remains open: which microscopic fields and which geometrical quantities determine the successive coefficients of the high-frequency visco-inertial and thermal responses?

	The present paper answers this question analytically. We develop the high-frequency expansions beyond the leading Johnson--Allard boundary-layer terms and identify the successive microgeometric parameters they contain. The thermal identification follows rather directly from the scalar heat-conduction problem. The visco-inertial one is more delicate, because velocity and pressure are coupled: the boundary layer acts back on the pore space through a potential displacement flow, which must be retained at every order. A two-scale asymptotic description, which separates the rapid variations across the boundary layer from the variations on the scale of the pore geometry, makes this structure explicit and allows the successive contributions to be identified. The resulting definitions hold for an arbitrary smooth geometry, three-dimensional flow included, and can serve as a basis for three-dimensional numerical evaluations.

	A second ingredient is the relaxation-mode representation of Avellaneda and Torquato~\cite{AT} (see also Ref.~\cite{L-AT}). It was already used, with the notations adopted here, in an earlier study of the high-frequency velocity pattern~\cite{CSGL}, where the potential flow induced in the pore volume by the boundary layer was identified and shown to be necessary for a correct derivation of $\Lambda$; the present work extends that analysis to all orders. The representation writes the response functions exactly as superpositions of exponentially relaxing viscous modes. Since the representation is exact, it does not alter the meaning of the parameters, which are fixed by the microscopic problems; but it shows that all of them, low- and high-frequency alike, are properties of a single spectral object, the distribution of relaxation times, and thereby tells what kind of information each one carries. We introduce it early, for two reasons. First, it shows how to normalize the viscous part of the pore-scale pressure so that it is, mode by mode, the partner of the velocity; second, it makes this spectral reading available from the outset. The odd-order coefficients of the high-frequency expansion, which involve $\Lambda$ and then a single combination of $\Lambda$ and of the surface and volume parameters $\Sigma$ and $V$, fix the shape of the short-relaxation-time end of the spectrum, generated by the modes confined near the walls. The even-order ones are regularized moments of the whole distribution: global sum rules, of the same family as $\alpha_\infty$ and the static parameters $k_0$ and $\alpha_0$. The representation also gives the short-time impulse response and the structure of the pressure field, and clarifies some subtle convergence properties of the modal sums. The thermal problem admits the same representation, with scalar modes and no pressure partner, and all of the above carries over to it in simpler form: no displacement flow arises, the thermal parameters reduce to purely geometrical properties of the walls, and the modal sums converge without the subtleties of the visco-inertial case.

	The parameters obtained in this way are defined by the microscopic problems and their geometrical structure, independently of any model. We do not propose or validate a new full-frequency model here; our aim is to establish what such a model must contain. The analysis is thus a first step toward extended relaxational descriptions, able to incorporate information that can be essential, and not merely incremental, in the new absorbing materials.

	The paper is organized as follows. Section~\ref{sec2} recalls the local formulation and the resulting response functions. Section~\ref{sec3} introduces Gauss coordinates adapted to the pore walls. Section~\ref{sec4} sets up the canonical microscopic problems and exact boundary representations of the inverse tortuosities. Section~\ref{sec5} presents the relaxation-mode representation and its low-frequency content. Sections~\ref{sec6} and~\ref{sec7} develop the visco-inertial boundary-layer analysis and the microgeometric meaning of $\Lambda$, $\Sigma$ and $V$; Secs.~\ref{sec8} and~\ref{sec9} do the same for the thermal problem. Section~\ref{sec10} establishes the spectral and temporal significance of the high-frequency parameters, and Sec.~\ref{sec11} concludes. Appendix~\ref{app:wedge} treats sharp wedges and shows how apparent parameters arise, and Appendix~\ref{app:greek} expresses the high-frequency coefficients and parameters in terms of the unit normal field.

	\section{The ``Divergence-Free'' (Zwikker--Kosten) Simplification and the Resulting Local Biot Theory}
	\label{sec2}

	We consider a porous material consisting of two connected, interpenetrating phases: a solid frame and a fluid occupying the pore space. No particular global topology is prescribed. The pore space may contain interconnected channels, branches, and loops, and the solid frame need not admit a simple geometrical representation. The fluid--solid interface is assumed to be smooth, in accordance with the hypothesis retained in the Introduction.

	The objective is a macroscopic description, in which the detailed microscopic fields are replaced by suitably averaged quantities. The averaging procedure is that introduced by Lorentz for the derivation of the macroscopic equations of electromagnetism, in which the microscopic fields are averaged over small spheres~\cite[\S\S\,112--113, pp.~133--134]{LOR}, best performed, following Russakoff~\cite{RUS}, as a convolution with a smooth normalized test function. Reference~\cite[Appendix~A]{LAF0} discusses the meaning of these averages, conceived either in Lorentz's manner or, when ergodicity applies, as Gibbs ensemble averages, the choice of normalization, and the equivalence, in the local theory, of the direct and Umov--Heaviside--Poynting definitions of the macroscopic pressure. We introduce a representative elementary volume (REV), denoted by $V$, with fluid and solid portions $V_p$ and $V_s$, respectively. Its fluid--solid interface is denoted by $S_p$, and its porosity is
	\begin{equation}
		\phi=\frac{|V_p|}{|V|}.
	\end{equation}
	For a field $a$ defined in the fluid, we distinguish its intrinsic fluid average,
	\begin{equation}
		\langle a\rangle_p
		=\frac{1}{|V_p|}\int_{V_p}a\,dV,
	\end{equation}
	from its average per unit total volume,
	\begin{equation}
		\langle a\rangle
		=\frac{1}{|V|}\int_{V_p}a\,dV
		=\phi\langle a\rangle_p.
	\end{equation}
	These definitions apply componentwise to vector and tensor fields. The distinction matters, in particular, when relating an intrinsic fluid velocity to a volume flux per unit total cross-sectional area.

	For a periodic material, the averaging volume may be chosen as a periodic cell, and the microscopic problems can be completed by appropriate periodic conditions. Periodicity is a convenient means of specifying the geometry and the averaging procedure, not a physical requirement of the description. In a statistically homogeneous material, representative volume averages, or ensemble averages where appropriate, take its place.

	It should be noted that once the response depends on the spatial variation of the excitation, the microscopic problem cannot in general be reduced to a cell response depending on frequency alone.
	A complete description should retain this spatial dependence and hence be nonlocal. Here, however, we deliberately restrict attention to the usual local setting. We assume that the geometry contains no locally resonant elements, such as Helmholtz-type resonators in the pore space or solid mass--spring structures, and that the acoustic wavelength is large compared with the size of the REV. In the regime considered, the solid motion within the REV can be treated as quasi en bloc, and the microscopic fluid motion relative to the solid as quasi divergence-free. Each property requires, besides the separation of scales, the absence of the corresponding locally resonant elements: solid mass--spring structures for the first, Helmholtz-type resonators in the pore space for the second. These are properties of the particular regime and geometry under consideration, rather than general consequences of long wavelengths in porous materials, although the long wavelength alone is commonly invoked to justify them (see, e.g., Ref.~\cite{JKD}): long wavelengths exclude neither inner resonances in the pore space~\cite{B1,BB} nor pore-scale deformations of the solid, such as those of thin films attached to the frame~\cite{VB1}.

	Under these assumptions, the macroscopic description takes the usual Biot form. The visco-inertial and thermal responses of the pore fluid are represented by local constitutive operators: they exhibit temporal dispersion, but spatially nonlocal effects are disregarded. The dynamic tortuosity and dynamic compliability specify, with the chosen averaging conventions, the effective fluid density and effective fluid compressibility entering this description. The relation between this local theory and the more general description, already nonlocal, that relaxes the divergence-free condition on the fluid motion while keeping the en-bloc solid motion, is discussed in Ref.~\cite{LAF_HF}, particularly in its Appendix.

	Although these fluid-response operators enter a theory that may retain solid-frame motion, their determination does not require solving the coupled macroscopic poroelastic problem as long as we maintain the hypothesis of solid en bloc motion. With this hypothesis, the same operators can be studied with the frame at rest, using the fluid motion relative to the frame. We therefore consider a stationary solid geometry in what follows. This choice is a means of determining the fluid-response functions; it does not restrict their subsequent use to rigid-frame materials, and they enter Biot's theory with frame motion as well.

	Before introducing the divergence-free simplification, it is useful to state the coupled microscopic equations from which it proceeds. Let $\mathbf{v}$ be the fluid velocity, $p$ the excess pressure, and $\tau$ the excess temperature. We introduce the condensation $b$ as the negative divergence of the fluid displacement $\mathbf{u}$:
	\begin{equation}
		b=-\nabla\cdot\mathbf{u},
		\qquad
		\mathbf{v}=\frac{\partial\mathbf{u}}{\partial t}.
	\end{equation}
	By linearized mass conservation, the condensation also equals the relative density variation,
	\begin{equation}
		b=\frac{\rho-\rho_0}{\rho_0},
		\label{eq:condensation-density}
	\end{equation}
	with perturbations measured from the equilibrium state.

	The equilibrium density and temperature are $\rho_0$ and $T_0$. We denote by $\eta$ and $\zeta$ the shear and bulk viscosities, by $\kappa$ the thermal conductivity, by $c_P$ the specific heat at constant pressure, by $\beta_0$ the volumetric thermal expansion coefficient, and by $\chi_0$ the adiabatic compressibility, with $\gamma=c_P/c_V$. All thermodynamic coefficients are evaluated at equilibrium, and we have the general thermodynamic identity \cite{LL_FM}
	\begin{equation}
		\gamma-1=\frac{T_0\beta_0^2}{\rho_0c_P\chi_0}.
		\label{eq:thermodynamic-identity}
	\end{equation}

	We retain a force density $\mathbf{F}$ in the momentum equation and a heat-supply rate per unit fluid volume, written as $\dot Q$, in the energy equation. The linearized microscopic equations are given below. Equation~\eqref{eq:full-momentum} expresses momentum balance with the Stokes constitutive laws for the viscous stress, while Eq.~\eqref{eq:full-energy} expresses energy balance with Fourier's constitutive law for heat conduction.
	\begin{align}
		\rho_0\frac{\partial\mathbf{v}}{\partial t}
		&=-\nabla p+\eta\nabla^2\mathbf{v}
		-\left(\zeta+\frac{\eta}{3}\right)
		\nabla\frac{\partial b}{\partial t}
		+\mathbf{F}
		\notag\\
		&=-\nabla p
		-\eta\nabla\times(\nabla\times\mathbf{v})
		-\left(\zeta+\frac{4\eta}{3}\right)
		\nabla\frac{\partial b}{\partial t}
		+\mathbf{F},
		&&\text{in }V_p,
		\label{eq:full-momentum}\\
		\frac{\partial b}{\partial t}
		&+\nabla\cdot\mathbf{v}=0,
		&&\text{in }V_p,
		\label{eq:full-continuity}\\
		\gamma\chi_0 p
		&=b+\beta_0\tau,
		&&\text{in }V_p,
		\label{eq:full-state}\\
		\rho_0c_P\frac{\partial\tau}{\partial t}
		&=\beta_0T_0\frac{\partial p}{\partial t}
		+\kappa\nabla^2\tau+\dot Q,
		&&\text{in }V_p.
		\label{eq:full-energy}
	\end{align}

	With condensation defined kinematically, Eq.~\eqref{eq:full-continuity} follows directly from the displacement--velocity relation. It also expresses linearized mass conservation when $b$ is identified with the relative excess density. The condensation terms in Eq.~\eqref{eq:full-momentum} follow by substituting
	$\nabla\cdot\mathbf{v}=-\partial b/\partial t$
	into the viscous stress contribution. The equivalence of the two displayed forms of the momentum equation then follows from the vector identity relating the Laplacian of a vector, the gradient of the divergence, and the curl of the curl.

	Equation~\eqref{eq:full-state} is the linearized thermodynamic equation of state, written with the condensation $b$ in place of the relative density variation, the two being equal by mass conservation [Eq.~\eqref{eq:condensation-density}].

	At the stationary fluid--solid interface, we impose no slip and vanishing excess temperature:
	\begin{align}
		\mathbf{v}&=\mathbf{0},
		&&\text{on }S_p,
		\label{eq:no-slip}\\
		\tau&=0,
		&&\text{on }S_p.
		\label{eq:isothermal-wall}
	\end{align}
	The latter condition is the isothermal-wall assumption adopted here. Conditions completing the microscopic problem at the outer boundary of the averaging volume depend on the geometrical and averaging formulation; the interface conditions alone do not specify them.

	Equations~\eqref{eq:full-momentum}--\eqref{eq:full-energy} retain microscopic compressibility and the coupling between pressure, density, temperature, and velocity. They are therefore the starting equations that would have to be retained in a nonlocal treatment, together with the appropriate excitation and averaging procedure. In particular, the pressure field then participates in the coupled compressible dynamics and cannot generally be identified with the pressure enforcing an incompressibility constraint.

	The local approximation/simplification selects a simpler regime of these equations. The relative fluid motion driven within the REV is quasi-divergence-free, while the macroscopic compression is retained in the averaged description. This does not mean setting every density variation to zero in the full coupled problem. Rather, the approximation/simplification separates the microscopic transport problem from the macroscopic compression that drives the thermal response. The resulting ``divergence-free'' simplification yields two separate microscopic problems: an incompressible visco-inertial problem and a scalar thermal diffusion problem. As emphasized in the Introduction, the thermal simplification follows from the same rationale as the incompressible-flow simplification, although it is not a divergence condition imposed on the temperature field. It has not been sufficiently recognized in the literature that the divergence-free simplification, although not explicitly formulated as such in their work, is the essence of the Zwikker and Kosten simplification of the full Kirchhoff theory, and that, for materials with complex microgeometry, it is likewise the essence of the corresponding local theory, which neglects spatial dispersion. The two response functions that follow from this simplification, the dynamic tortuosity and the thermal tortuosity, were defined precisely on this basis in earlier work~\cite{LAF_PhD,LAF0} and in the Appendix of Ref.~\cite{LAF2}, the rest of which deals with the nonlocal theory; they are recalled below in the notation of the present paper.

	Before writing these two problems in their final form, the interpretation of their forcing must be specified. In what follows, the only driving is the macroscopic pressure, as in acoustic propagation with the frame at rest: $\mathbf F$ and $\dot Q$ do not appear in the microscopic problems written below. The next two paragraphs say how this driving can be represented, and how other excitations, a frame acceleration or a heat supply, would enter and be interpreted. In the momentum equation, the fluid responds to the combination
	\begin{equation}
		-\nabla p+\mathbf{F}.
	\end{equation}
	A prescribed macroscopic pressure gradient can be represented explicitly as part of the pressure field, or transferred to an equivalent driving force in a microscopic cell problem. In the latter representation, the remaining microscopic pressure enforces the divergence-free constraint and accommodates the pore geometry. It is not, by itself, the complete thermodynamic pressure appearing in the compressible formulation. Likewise, when the excitation represents the acceleration of a moving frame, the corresponding inertial driving can be included in the force term of the relative-motion problem. These interpretations require a consistent distinction between the prescribed macroscopic driving and the microscopic pressure correction.

	A related distinction is needed in the thermal problem. In the coupled energy equation, the source is the combination
	\begin{equation}
		\beta_0T_0\frac{\partial p}{\partial t}+\dot Q.
	\end{equation}
	The first term represents the thermal driving associated with pressure variations, whereas the second permits an independently prescribed heat supply. Within the divergence-free simplification, the macroscopic pressure driving can be used to define the thermal response through an equivalent prescribed source. The thermal field obtained in this way is a response to that source; its interpretation in terms of dynamic compressibility requires restoring the relation between the source and the macroscopic pressure. An independently imposed heat supply and compression-induced thermal driving can therefore lead to the same diffusion problem without having the same physical interpretation.

	The subsequent formulation must preserve these distinctions while separating the amplitude of the prescribed driving from the geometrical response. Only after this separation will we introduce the canonical microscopic problems and their normalized field quantities.

	In the local description, the quasi-divergence-free microscopic motion within the REV determines a frequency-dependent Newton-type relation between the macroscopic fluid acceleration and the force density producing it. For acoustic propagation with the frame at rest, this driving force density is the negative macroscopic pressure gradient. In its most general form, this relation involves a dynamic tortuosity tensor. At this first stage, however, we adopt from the outset the isotropic setting of Biot, of Johnson, Koplik and Dashen, of Avellaneda and Torquato, and of our earlier work: the material is taken to be macroscopically isotropic, so that the tensorial response reduces to a scalar multiple of the identity. This is a convenient simplification, to be reconsidered later. The general anisotropic formulation is deferred to a subsequent treatment of the tensorial aspects of oblique propagation and, later still, of the specific tensorial aspects arising in a nonlocal extension. The direction of the macroscopic pressure gradient is denoted by a unit vector $\mathbf{e}$, which may be taken arbitrarily. For time-harmonic fields with time dependence $e^{-i\omega t}$, the relation reads
	\begin{equation}
		-i\omega\rho_0\,
		\alpha(\omega)\,
		\langle\mathbf{v}\rangle_p
		=-\nabla\langle p\rangle_p .
		\label{eq:local-dynamic-tortuosity}
	\end{equation}
	Here $\alpha(\omega)$ is the scalar dynamic tortuosity, and $\langle\mathbf{v}\rangle_p$ is the intrinsic fluid-average velocity, which by isotropy is collinear with the driving, hence directed along $\mathbf{e}$. The left-hand side represents the macroscopic acceleration multiplied by the frequency-dependent effective inertia per unit fluid volume. The frequency dependence of $\alpha(\omega)$ accounts for the visco-inertial coupling between the pore fluid and the solid frame. In the time domain, this multiplication corresponds to a temporally dispersive operator rather than an instantaneous Newton law.

	To determine this response, the macroscopic pressure gradient is treated as a prescribed driving that is uniform on the scale of the REV. The pressure entering the microscopic incompressible-flow problem is then separated into this prescribed macroscopic contribution and a fluctuating microscopic correction, denoted by $\widetilde p$. The latter enforces the divergence-free constraint and accommodates the pore geometry.

	The microscopic visco-inertial problem is consequently
	\begin{align}
		-i\omega\rho_0\mathbf{v}
		&=-\nabla\widetilde p
		+\eta\nabla^2\mathbf{v}
		-\nabla\langle p\rangle_p,
		&&\text{in }V_p,
		\label{eq:local-viscous-momentum}\\
		\nabla\cdot\mathbf{v}
		&=0,
		&&\text{in }V_p,
		\label{eq:local-viscous-incompressibility}\\
		\mathbf{v}
		&=\mathbf{0},
		&&\text{on }S_p.
		\label{eq:local-viscous-no-slip}
	\end{align}
	The disappearance of the condensation-gradient terms from the momentum equation follows from the divergence-free constraint. The problem must also be completed by conditions consistent with the geometrical and averaging formulation. In a periodic cell, $\mathbf{v}$ and $\widetilde p$ are periodic; in a stationary random geometry, the corresponding statistical formulation is used, with the macroscopic pressure variation entirely assigned to $\langle p\rangle_p$. The fluctuating pressure $\widetilde p$ therefore contains no macroscopic drift and, by the definition of the decomposition $p=\langle p\rangle_p+\widetilde p$, satisfies $\langle\widetilde p\rangle_p=0$. Since only its gradient enters the momentum equation, this condition is what fixes the additive constant in $\widetilde p$, and with it the meaning of the macroscopic pressure, defined here as the volume average $\langle p\rangle_p$. The physically meaningful definition in the nonlocal theory \cite{LAF1,LAF2,LAF_HF} is that of Umov--Heaviside--Poynting: the macroscopic pressure $P$ is such that $\langle p\mathbf v\rangle_p=P\langle\mathbf v\rangle_p$, so that the macroscopic energy flux transported in acoustic form reproduces the averaged microscopic one. In the present setting the two definitions coincide, $P=\langle p\rangle_p$.

	Solving this problem for a prescribed direction $\mathbf{e}$ of the macroscopic pressure gradient and averaging the resulting velocity determines the dynamic tortuosity through Eq.~\eqref{eq:local-dynamic-tortuosity}; by isotropy, the result does not depend on $\mathbf{e}$. Thus the macroscopic driving fixes the excitation amplitude, whereas the microscopic solution determines the geometrical and frequency-dependent response.

	There is an entirely analogous thermal formulation, in which a scalar ``thermal tortuosity'' $\alpha'(\omega)$ relates the intrinsic fluid-average excess temperature to the thermal driving associated with the macroscopic pressure:
	\begin{equation}
		-i\omega\rho_0c_P\,
		\alpha'(\omega)\langle\tau\rangle_p
		=-i\omega\beta_0T_0\langle p\rangle_p.
		\label{eq:local-thermal-tortuosity}
	\end{equation}
	This quantity was first introduced as such in Ref.~\cite{LAF_PhD}; an equivalent one was defined earlier by L\'evy and Sanchez-Palencia~\cite{LSP}, whose complex constant $k'$, the ratio of the cell-averaged excess temperature to the pressure [their Eq.~(2.14)], is related to $\alpha'$, for a perfect gas, by $k'=\phi(\gamma-1)/[\gamma\,\alpha'(\omega)]$. The thermal tortuosity accounts for the frequency-dependent effect of heat exchange between the pore fluid and the solid frame. It is determined by the microscopic thermal problem
	\begin{align}
		-i\omega\rho_0c_P\tau
		&=\kappa\nabla^2\tau
		-i\omega\beta_0T_0\langle p\rangle_p,
		&&\text{in }V_p,
		\label{eq:local-thermal-energy}\\
		\tau&=0,
		&&\text{on }S_p.
		\label{eq:local-thermal-wall}
	\end{align}
	As in the visco-inertial problem, these equations must be completed by conditions consistent with the geometrical and averaging formulation.

	The prescribed thermal source is thus
	\begin{equation}
		-i\omega\beta_0T_0\langle p\rangle_p.
		\label{eq:local-thermal-source}
	\end{equation}
	In Eq.~\eqref{eq:local-thermal-energy}, the macroscopic pressure alone drives the heat source; the pressure correction $\widetilde p$ does not enter it. This is the thermal part of the divergence-free simplification, not an independent assumption, and it rests on the same separation of scales. The pressure correction is small compared with the macroscopic pressure, but varies over pore-scale distances, so that its gradient can be comparable to the macroscopic pressure gradient, which varies over the wavelength: it cannot be neglected in the visco-inertial momentum balance. In the thermal equation, however, the pressure enters the source through its time derivative, without a spatial gradient, so that $\widetilde p$ produces only a small correction to the source $-i\omega\beta_0T_0\langle p\rangle_p$, neglected at the order retained in the local description.

	The macroscopic pressure itself is locally affine on the scale of the REV. In the macroscopically homogeneous medium considered here, translational invariance and reciprocity of the thermal diffusion problem imply that this linear variation contributes no additional macroscopic mean-temperature response, so that the thermal driving may be represented by the uniform source associated with the local macroscopic pressure.

	Solving this diffusion problem and averaging the resulting excess temperature determines $\alpha'(\omega)$ through Eq.~\eqref{eq:local-thermal-tortuosity}.
	The thermal tortuosity also determines the dynamic compliability. Indeed, its definition gives
	\begin{equation}
		\langle\tau\rangle_p
		=
		\frac{\beta_0T_0}{\rho_0c_P}\,
		\frac{\langle p\rangle_p}{\alpha'(\omega)}.
	\end{equation}
	Averaging the linearized equation of state,
	$\gamma\chi_0p=b+\beta_0\tau$, and using the thermodynamic identity
	given in Eq.~\eqref{eq:thermodynamic-identity}, we obtain
	\begin{equation}
		\langle b\rangle_p
		=
		\chi_0
		\left[
		\gamma-\frac{\gamma-1}{\alpha'(\omega)}
		\right]
		\langle p\rangle_p.
	\end{equation}
	Thus, defining the dimensionless dynamic compliability
	$\beta(\omega)$ by
	\begin{equation}
		\langle b\rangle_p
		=\chi_0\beta(\omega)\langle p\rangle_p,
	\end{equation}
	we have
	\begin{equation}
		\beta(\omega)
		=\gamma-\frac{\gamma-1}{\alpha'(\omega)}.
		\label{eq:dynamic-compliability}
	\end{equation}
	The effective compressibility of the pore fluid is therefore
	$\chi_0\beta(\omega)$.

	To exploit the boundary-layer structure of these two problems, we shall use Gauss coordinates adapted to the pore boundary. Section~\ref{sec3} introduces them and writes the microscopic equations in them, under the two possible conventions for counting volume. We then separate the prescribed excitation amplitudes from the microscopic response fields and put the two problems into canonical scaled forms in Section~\ref{sec4}.

	\section{Gauss Coordinates}\label{sec3}
	The pore domain $V_p$ has an irregular, generally non-simply-connected
	geometry. Cartesian coordinates are therefore poorly adapted to both
	the boundary conditions and the boundary-layer structure that governs
	the high-frequency regime. We introduce Gauss (curvilinear)
	coordinates as the natural setting for the analysis that follows.
	Our notation is chosen to recall that used by Weinberg
	\cite[Chap.~4]{WEIN}, whose exposition provides a clear account
	of the coordinate and tensor calculus needed here.
	We first recall the general principle of these coordinates and the associated tensor calculus, then the specific coordinates used in the vicinity of the pore walls, and finally we write the microscopic equations of Section~\ref{sec2} in Gauss coordinates, in their most general form and under the two possible conceptions of the counting of volume.

	\subsection*{Cartesian coordinates, Gauss coordinates and atlas of patches}

	A point of the fluid is located either by its Cartesian coordinates $\xi^{\alpha}$ $(\alpha=1,2,3)$, in which
	$ds^{2}=\delta_{\alpha\beta}\,d\xi^{\alpha}d\xi^{\beta}$,
	or by Gauss coordinates $x^{\mu}$ $(\mu=1,2,3)$, which are arbitrary smooth labels of the same point. The letter $\xi$ is used for the Cartesian coordinates, as for the inertial coordinates of Weinberg, to distinguish them from the $x^{\mu}$. Greek indices run over $\{1,2,3\}$. Because the underlying space is a single Euclidean space, the generality of the Gauss coordinates reflects only arbitrary choices of labelling, not any curvature of space itself.

	A single coordinate system cannot in general cover all of
	$V_p$ smoothly and invertibly. We therefore cover
	$V_p$ with a finite collection of \emph{coordinate
		patches} $x$, $x'$, ... On each patch, a smooth, invertible map relates the
	Gauss coordinates $x^{\mu}$ to the ambient
	Cartesian coordinates $\xi^{\alpha}$:
	\begin{equation}
		\begin{aligned}
		&\xi^{\alpha} = \xi^{\alpha}(x^{1},x^{2},x^{3}),&\qquad& x^{\mu} = x^{\mu}(\xi^{1},\xi^{2},\xi^{3}),\\
		&\xi^{\alpha} = \xi^{\alpha}(x^{1'},x^{2'},x^{3'}),&\qquad& x^{\mu'} = x^{\mu'}(\xi^{1},\xi^{2},\xi^{3}),
		\end{aligned}
		\label{eq:coord_change}
	\end{equation}
	with non-vanishing Jacobian $J=\det(\partial x^{\mu'}/\partial x^{\mu})\neq 0$
	on every patch overlap (we use Schouten's convenient kernel-index notation~\cite{SCH}, in which the primes distinguishing the coordinate systems are placed on the indices rather than on the kernel letters); we consider orientation-preserving changes, $J>0$. The patches overlap sufficiently to leave no part of $V_p$ undescribed, and integrals over $V_p$ are understood as sums of integrals over pieces covered by different patches. For simplicity of notation, the patch label is left implicit, and we write as if a single system $x^\mu$ covered the whole fluid. On each patch, the fields are represented by components, related on overlaps by tensor transformation laws. The kinematic fields, such as the velocity, and the intensive thermodynamic variables, such as the temperature, transform as true tensors in any case; the quantities concerned by the counting of volume, such as the pressure and the mass density, transform as true tensors or as tensor densities according to the way the volume is counted, as explained below (Table~\ref{tab:weights}).

	\subsection*{Metric, connection and covariant derivatives}

	The metric is the true twice-covariant tensor
	\begin{equation}
		g_{\mu\nu}
		=\delta_{\alpha\beta}\,
		\frac{\partial\xi^{\alpha}}{\partial x^{\mu}}
		\frac{\partial\xi^{\beta}}{\partial x^{\nu}},
		\qquad
		g_{\mu'\nu'}
		=\frac{\partial x^{\mu}}{\partial x^{\mu'}}
		\frac{\partial x^{\nu}}{\partial x^{\nu'}}\,g_{\mu\nu},
		\label{eq:metric_tensor}
	\end{equation}
	as follows from $ds^{2}=g_{\mu\nu}dx^{\mu}dx^{\nu}$. Its inverse $g^{\lambda\kappa}$ is defined by
	$g^{\lambda\kappa}g_{\kappa\nu}=\delta^{\lambda}_{\nu}$, where the Kronecker symbol is an invariant tensor; $g^{\lambda\kappa}$ is therefore a true twice-contravariant tensor. Indices are raised and lowered with $g^{\lambda\kappa}$ and $g_{\kappa\nu}$. We write $g=\det(g_{\mu\nu})$, so that the physical volume element is $dV=\sqrt{g}\,d^{3}x$.

	The non-tensorial affine connection coefficients are
	\begin{equation}
		\Gamma^{\lambda}_{\mu\nu}
		=
		\frac{\partial x^{\lambda}}{\partial \xi^{\alpha}}\,
		\frac{\partial^{2}\xi^{\alpha}}
		{\partial x^{\mu}\partial x^{\nu}}
		=
		\frac{1}{2}\,g^{\lambda\kappa}
		\left[
		\frac{\partial g_{\kappa\nu}}{\partial x^{\mu}}
		+\frac{\partial g_{\kappa\mu}}{\partial x^{\nu}}
		-\frac{\partial g_{\mu\nu}}{\partial x^{\kappa}}
		\right],
		\label{eq:christoffel_def}
	\end{equation}
	and satisfy $\Gamma^{\kappa}_{\kappa\mu}=\partial_{\mu}\ln\sqrt{g}$. The covariant derivatives are denoted with a semicolon. For a true scalar and a true vector, they are
	\begin{equation}
		\varphi_{;\lambda}=\partial_{\lambda}\varphi,
		\qquad
		a^{\mu}{}_{;\nu}=\partial_{\nu}a^{\mu}+\Gamma^{\mu}_{\nu\kappa}a^{\kappa},
		\qquad
		a_{\mu;\nu}=\partial_{\nu}a_{\mu}-\Gamma^{\lambda}_{\mu\nu}a_{\lambda},
		\label{eq:covariant_derivatives}
	\end{equation}
	and, for a tensor with several indices, one adds one signed connection term per index, positive for a contravariant index and negative for a covariant one. The metric is covariantly constant, $g_{\mu\nu;\lambda}=0$, and the Leibniz rule holds. Since the ambient space is Euclidean, covariant differentiation is commutative. In Cartesian coordinates, $\Gamma^{\lambda}_{\mu\nu}=0$ and covariant derivatives reduce to partial derivatives. In particular, a vector that is constant in Cartesian coordinates, such as the unit vector $\mathbf{e}$ used below, generally has position-dependent components $e^{\mu}$ in a curvilinear patch, but its covariant derivative vanishes, $e^{\mu}{}_{;\nu}=0$.

	\subsection*{True tensors and tensor densities}

	A tensor density of weight $W$ transforms like the corresponding true tensor, multiplied in addition by $J^{W}$. For instance, for a vector density,
	\begin{equation}
		T^{\mu'}
		=J^{W}\,\frac{\partial x^{\mu'}}{\partial x^{\nu}}\,T^{\nu}.
	\end{equation}
	Since $g'=J^{-2}g$, the determinant $g$ is a scalar density of weight $-2$, and $\sqrt{g}$ of weight $-1$. The product $g^{W/2}T$ of a density of weight $W$ by $g^{W/2}$ is then a true tensor: the Jacobian factors cancel. Covariant differentiation of a density is obtained by applying the rules above to this true tensor; one finds, for a scalar density $a$ of weight $W$,
	\begin{equation}
		a_{;\lambda}
		=\partial_{\lambda}a+\frac{W}{2}\,g^{-1}\partial_{\lambda}g\;a
		=\partial_{\lambda}a+W\,\Gamma^{\kappa}_{\kappa\lambda}\,a,
		\label{eq:density_covariant}
	\end{equation}
	and, for a tensor with several indices, the usual connection terms of the true-tensor rule are supplemented by the single term $W\Gamma^{\kappa}_{\kappa\lambda}$ times the tensor itself.

	The invariant volume element is $dV=\sqrt{g}\,d^{3}x$, and the pore volume is $|V_p|=\int_{V_p}dV$. The direct integral $\int_{V_p}a\,d^{3}x$ has a physical meaning only if $a$ is a scalar density of weight $-1$; it is then the physical integral $\int_{V_p}(g^{-1/2}a)\,dV$ of the corresponding true scalar $g^{-1/2}a$.

	\subsection*{Differential operators}

	For a true scalar field $\varphi$ and a true contravariant vector
	field $v^{\alpha}$, the \emph{gradient}, \emph{divergence}, and
	\emph{scalar Laplacian} take the standard covariant forms
	\begin{equation}
		(\nabla\varphi)^{\alpha}
		= g^{\alpha\beta}\,\partial_{\beta}\varphi,
		\label{eq:gradient}
	\end{equation}
	\begin{equation}
		\nabla\cdot\mathbf{v}
		= v^{\alpha}{}_{;\alpha}
		= \frac{1}{\sqrt{g}}\,\partial_{\alpha}\!\bigl(\sqrt{g}\,v^{\alpha}\bigr),
		\label{eq:divergence}
	\end{equation}
	\begin{equation}
		\Delta\varphi
		= g^{\mu\nu}\varphi_{;\mu;\nu}
		= \frac{1}{\sqrt{g}}\,
		\partial_{\alpha}\!\bigl(\sqrt{g}\,g^{\alpha\beta}\,\partial_{\beta}\varphi\bigr).
		\label{eq:scalar_laplacian}
	\end{equation}
	The \emph{vector Laplacian} (connection Laplacian) acting on
	a true vector field is
	\begin{equation}
		(\Delta\mathbf{v})^{\alpha}
		= g^{\mu\nu}\,v^{\alpha}{}_{;\mu;\nu}.
		\label{eq:vector_laplacian}
	\end{equation}
	In Gauss coordinates the curl is expressed through the alternating symbol
	$[\mu\nu\rho]$ ($+1$ for even permutations of $(1,2,3)$, $-1$ for odd ones,
	$0$ otherwise). Following the conception of Weinberg~\cite{WEIN}, we take as
	Levi-Civita density the contravariant object equal to the alternating symbol,
	the coordinates being right-handed, and obtain the covariant one by lowering indices:
	\begin{equation}
		\epsilon^{\mu\nu\rho}=[\mu\nu\rho],
		\qquad
		\epsilon_{\mu\nu\rho}=g_{\mu\alpha}g_{\nu\beta}g_{\rho\gamma}\,\epsilon^{\alpha\beta\gamma}=g\,[\mu\nu\rho],
	\end{equation}
	with $g=\det(g_{\alpha\beta})>0$. The symbol $\epsilon$ is not to be confused with the complex skin depth $\varepsilon$ introduced below.
	The combinations $\epsilon^{\mu\nu\rho}/\sqrt g$ and $\epsilon_{\mu\nu\rho}/\sqrt g$
	are true tensors, covariantly constant. The curl of a vector field $\mathbf v$
	is then the vector whose contravariant components are
	\begin{equation}
		(\nabla\times\mathbf v)^{\rho}
		=\frac{1}{\sqrt g}\,\epsilon^{\rho\sigma\lambda}\,v_{\lambda;\sigma}
		=\frac{1}{\sqrt g}\,\epsilon^{\rho\sigma\lambda}\,\partial_\sigma v_{\lambda},
	\end{equation}
	where the Christoffel terms of the covariant derivative
	cancel because $\epsilon^{\rho\sigma\lambda}$ is antisymmetric in $\sigma\lambda$
	while $\Gamma^{\kappa}_{\sigma\lambda}$ is symmetric. The covariant components are
	$(\nabla\times\mathbf v)_{\rho}=g_{\rho\mu}(\nabla\times\mathbf v)^{\mu}$.
	Applying the curl twice gives
	\begin{equation}
		\left[\nabla\times(\nabla\times\mathbf v)\right]^{\mu}
		=\frac{1}{g}\,\epsilon^{\mu\nu\rho}\,\epsilon_{\rho}{}^{\sigma\lambda}\,v_{\lambda;\sigma;\nu},
		\qquad
		\epsilon_{\rho}{}^{\sigma\lambda}=g_{\rho\alpha}\epsilon^{\alpha\sigma\lambda}.
	\end{equation}
	Since covariant differentiation is commutative in the Euclidean ambient space
	(see above), $v_{\lambda;\sigma;\nu}=v_{\lambda;\nu;\sigma}$; with the contraction
	$\epsilon^{\mu\nu\rho}\epsilon_{\rho}{}^{\sigma\lambda}
	=g\,(g^{\mu\sigma}g^{\nu\lambda}-g^{\mu\lambda}g^{\nu\sigma})$,
	the double curl then yields the familiar identity (the flat-space form of the
	Bochner--Weitzenb\"{o}ck identity, which in a curved space would contain an
	additional curvature term)
	\begin{equation}
		\Delta\mathbf{v}
		= \nabla(\nabla\cdot\mathbf{v})
		- \nabla\times(\nabla\times\mathbf{v}).
		\label{eq:bochner}
	\end{equation}
	In particular, for an incompressible field ($\nabla\cdot\mathbf{v}=0$),
	\begin{equation}
		\Delta\mathbf{v} = -\nabla\times(\nabla\times\mathbf{v}),
		\label{eq:laplacian_incompressible}
	\end{equation}
	a relation that will be used in the boundary-layer analysis of
	Sec.~\ref{sec6}.

	\subsection*{Near-wall coordinate patches}

	Only the coordinates specific to the vicinity of the walls, which will be used in the boundary-layer analysis, are recalled here. The pore surface $S_{p}$ is assumed to be of class $C^{\infty}$.
	Near $S_{p}$, we construct each patch by
	adapting the coordinates to the surface and its normal.
	Two coordinates $(q^{1},q^{2})$ are laid down directly on $S_{p}$
	as surface coordinates; Roman indices $i,j,k,\ldots$ run over
	$\{1,2\}$.
	Let $\mathbf{n}$ denote the \emph{outward} unit normal from $V_p$
	(pointing into the solid skeleton), and let $\zeta\geq 0$ denote the
	geodesic normal distance measured \emph{into the fluid}, so that
	$\zeta=0$ on $S_{p}$ and $\zeta>0$ in the interior of $V_p$.
	The coordinate map is the projection along normals:
	\begin{equation}
		\mathbf{r}(q,\zeta)
		= \mathbf{r}_{0}(q) - \zeta\,\mathbf{n}(q),
		\label{eq:parallel_surface}
	\end{equation}
	where $\mathbf{r}_{0}(q)$ is the wall point with surface coordinates
	$q=(q^{1},q^{2})$, and $\mathbf{r}(q,\zeta)$ is the fluid point
	at distance $\zeta$ from it along the inward normal.
	The Gauss coordinates of the patch are then $x^{i}=q^{i}$ and $x^{3}=-\zeta$, so that
	\begin{equation}
		\partial_{3} = -\partial_{\zeta},
		\qquad
		\partial_{n} = -\partial_{\zeta}.
		\label{eq:normal_deriv}
	\end{equation}
	Since $\partial\mathbf r/\partial x^3=\mathbf n$, the basis vector of the coordinate $x^3$ is the unit normal itself: $n^\mu=\delta^\mu_3$ and, by Eq.~\eqref{eq:metric_block2} below, $n_\mu=\delta_\mu^3$. In near-wall patches the index $3$ therefore takes the place of the normal: the normal derivative of a scalar is $f_{;3}=\partial_3f$, and that of a vector field $a^\mu$ is the covariant derivative $a^\mu{}_{;3}$, not the derivative $\partial_3a^\mu$ of its components.

	These coordinates are used in the \emph{tubular neighbourhood}
	\begin{equation}
		\mathcal V_S=\{\,\mathbf r_0-\zeta\,\mathbf n(\mathbf r_0):\ \mathbf r_0\in S_p,\ 0\le\zeta<\zeta_S\,\}\subset V_p ,
		\label{eq:tubular-neighbourhood}
	\end{equation}
	the layer of fluid of thickness $\zeta_S$ along the wall, whose part $\zeta=0$ is $S_p$ itself. The thickness $\zeta_S>0$ is chosen small enough for the map~\eqref{eq:parallel_surface} to be one-to-one on $\mathcal V_S$, i.e.\ for each point of $\mathcal V_S$ to have a unique nearest point on $S_p$, at distance $\zeta$, whose surface coordinates it takes: $\zeta_S$ must be smaller than the radii of curvature of $S_p$ whose centre lies in the fluid, and than half the smallest distance across a pore or a throat. Such a $\zeta_S$ exists because $S_p$ is smooth and the cell is bounded. The boundary-layer fields of Secs.~\ref{sec6} and~\ref{sec8} are confined to $\mathcal V_S$ up to exponentially small terms, since the viscous and thermal skin depths of Eq.~\eqref{eq:complex-skin-depths} satisfy $|\varepsilon|,|\varepsilon'|\ll\zeta_S$.

	\subsection*{Block metric and its normal dependence}

	Because the normal direction is everywhere orthogonal to the
	parallel surfaces $\{\zeta=\text{const}\}$, and because
	$\zeta$ measures a Euclidean length, the metric in the near-wall
	patch takes the block-diagonal form
	\begin{equation}
		\mathrm{d}s^{2}
		= g_{\mu\nu}(x)dx^{\mu}dx^{\nu}
		=g_{ij}(q,\zeta)\,\mathrm{d}q^{i}\mathrm{d}q^{j}
		+ \mathrm{d}\zeta^{2},
		\label{eq:metric_block}
	\end{equation}
	that is,
	\begin{equation}
		g_{i3}=g_{3i}=0,\qquad g_{33}=1\qquad\text{in }\mathcal V_S.
		\label{eq:metric_block2}
	\end{equation}
	The three-dimensional line element is thus controlled entirely by
	the two-dimensional surface metric $g_{ij}(q,\zeta)$, which varies
	with the normal distance $\zeta$ as the parallel surfaces expand
	or contract relative to $S_{p}$.

	\subsection*{Christoffel symbols in near-wall patches}

	In the adapted coordinates $(q^{i},x^{3}=-\zeta)$, the
	block structure \eqref{eq:metric_block2} implies that the
	Christoffel symbols \eqref{eq:christoffel_def} reduce to
	\begin{equation}
		\Gamma^{3}_{3\alpha}=0,\quad
		\Gamma^{i}_{33}=0,\quad
		\Gamma^{3}_{ij}=\tfrac{1}{2}\,\partial_{\zeta}\,g_{ij},
		\label{eq:christoffel_a}
	\end{equation}
	\begin{equation}
		\Gamma^{i}_{3j}
		= \Gamma^{i}_{j3}
		= -\tfrac{1}{2}\,g^{ik}\partial_{\zeta}\,g_{kj},
		\label{eq:christoffel_b}
	\end{equation}
	\begin{equation}
		\Gamma^{i}_{jk}
		= \tfrac{1}{2}\,g^{i\ell}
		\bigl(\partial_{j}g_{\ell k}+\partial_{k}g_{\ell j}
		-\partial_{\ell}g_{jk}\bigr),
		\label{eq:christoffel_c}
	\end{equation}
	where, in \eqref{eq:christoffel_c}, the derivatives $\partial_{i}$ are
	taken at fixed $\zeta$, and $\Gamma^i_{jk}$ are the intrinsic Christoffel symbols
	of the parallel surface at that distance. The nonvanishing symbols $\Gamma^3_{ij}$ and $\Gamma^i_{3j}$ of Eqs.~\eqref{eq:christoffel_a} and~\eqref{eq:christoffel_b} are related by
	\begin{equation}
		\Gamma^3_{ij}=-g_{ik}\,\Gamma^k_{3j},
		\label{eq:christoffel_lower}
	\end{equation}
	so that $g_{ik}\Gamma^k_{3j}$ and $g^{jk}\Gamma^i_{3k}$ are symmetric, and $\partial_\zeta g_{ij}=2\Gamma^3_{ij}$, $\partial_\zeta g^{ij}=2g^{jk}\Gamma^i_{3k}$. The area element of $S_p$ is $dS=\sqrt{g(q,0)}\,dq^{1}dq^{2}$, with $g(q,\zeta)=\det[g_{ij}(q,\zeta)]$, equal to $\det(g_{\mu\nu})$ by Eq.~\eqref{eq:metric_block2}, and that of the parallel surface at distance $\zeta$ is $\mathcal J_S\,dS$, with $\mathcal J_S(q,\zeta)=\sqrt{g(q,\zeta)/g(q,0)}$. Since $\Gamma^{\kappa}_{\kappa\mu}=\partial_{\mu}\ln\sqrt{g}$ and $\Gamma^3_{33}=0$,
	\begin{equation}
		\Gamma^k_{3k}=\partial_3\ln\mathcal J_S=-\partial_\zeta\ln\mathcal J_S .
		\label{eq:trace-Gamma}
	\end{equation}

	The coordinate changes that preserve the near-wall structure are $q^{i'}=q^{i'}(q)$, $x^{3'}=x^{3}$. In the transformation law of $\Gamma^\lambda_{\mu\nu}$, the inhomogeneous term vanishes as soon as one of the three indices is $3$. Under these changes, $\Gamma^i_{3j}$ and $\Gamma^3_{ij}$ are therefore true surface tensors, mixed and covariant respectively, whereas $\Gamma^k_{ij}$ is not a tensor: it is the connection of the parallel surfaces. We denote by a vertical bar the covariant derivative of the parallel surface passing through the point considered. For a scalar $f$, a surface vector $a^i$, a surface covector $b_i$ and a mixed surface tensor $T^i{}_j$, depending on $\zeta$ as a parameter,
	\begin{equation}
		f_{|j}=\partial_jf,\quad
		a^i{}_{|j}=\partial_ja^i+\Gamma^i_{jk}a^k,\quad
		b_{i|j}=\partial_jb_i-\Gamma^k_{ji}b_k,\quad
		T^i{}_{j|k}=\partial_kT^i{}_j+\Gamma^i_{k\ell}T^\ell{}_j-\Gamma^\ell_{kj}T^i{}_\ell,
		\label{eq:surface-operators}
	\end{equation}
	and $g_{ij|k}=0$. The bar derivative involves only the tangential components; the coupling with the normal direction is contained in the three-dimensional covariant derivative, through the Christoffel symbols having $3$ among their indices: for a vector field, $a^i{}_{;j}=a^i{}_{|j}+\Gamma^i_{3j}a^3$.

	A single symbol suffices for the wall and for the parallel surfaces: the level is that of the point where the quantity is evaluated. On the wall, Eq.~\eqref{eq:surface-operators} is the intrinsic covariant derivative of $S_p$: for a tangent field $s^i$ on $S_p$, $s^i{}_{|j}$ is its intrinsic gradient, $s^k{}_{|k}$ its surface divergence and $g^{jk}s^i{}_{|j|k}$ its connection Laplacian. We write $[X]_0$ for the value of a quantity $X$ on the wall, at $\zeta=0$.

	\subsection*{Curvatures of the wall}

	At the wall, $-[\Gamma^i_{3j}]_0$ is the shape (Weingarten) operator of $S_p$, and $[\Gamma^3_{ij}]_0=\tfrac12\partial_\zeta g_{ij}(q,0)$ its second fundamental form~\cite{WEIN}. Where matrix notation is convenient, we write $\Gamma_3$ for the $2\times2$ matrix of the mixed components at the level $\zeta$ considered,
	\begin{equation}
		(\Gamma_3)^i{}_j=\Gamma^i_{3j},
		\label{eq:Gamma3-matrix}
	\end{equation}
	and $\mathbf I$ for the identity, so that the shape operator is $-[\Gamma_3]_0$. The eigenvalues $\kappa_{1}(q)$ and $\kappa_{2}(q)$ of $-[\Gamma^i_{3j}]_0$ are the two
	principal curvatures, with the sign convention that $\kappa_{a}>0$
	when $S_{p}$ is \emph{convex} as seen from the fluid
	(centre of curvature on the solid side) and $\kappa_{a}<0$ when
	it is \emph{concave} (centre of curvature on the fluid side).
	The mean and Gaussian curvatures are
	\begin{equation}
		H = \tfrac{1}{2}(\kappa_{1}+\kappa_{2})
		= -\tfrac{1}{2}\big[\Gamma^k_{3k}\big]_0,
		\qquad
		K = \kappa_{1}\kappa_{2}=\tfrac12\big[\Gamma^k_{3k}\Gamma^\ell_{3\ell}-\Gamma^k_{3\ell}\Gamma^\ell_{3k}\big]_0 .
		\label{eq:curvatures}
	\end{equation}
	They are used below only where their classical meaning matters; the calculations are carried out with the Christoffel symbols. We also use the trace-free part of $\Gamma^i_{3j}$,
	\begin{equation}
		\mathring\Gamma^i_{3j}=\Gamma^i_{3j}-\tfrac12\,\delta^i_j\,\Gamma^k_{3k},
		\label{eq:Gamma-trace-free}
	\end{equation}
	which vanishes at the umbilic points of the wall, where $\kappa_1=\kappa_2$. The Cayley--Hamilton identity for $2\times2$ matrices gives, at every level,
	\begin{equation}
		\Gamma^i_{3k}\Gamma^k_{3j}=\Gamma^k_{3k}\,\Gamma^i_{3j}-\tfrac12\big(\Gamma^k_{3k}\Gamma^\ell_{3\ell}-\Gamma^k_{3\ell}\Gamma^\ell_{3k}\big)\delta^i_j,
		\qquad
		\mathring\Gamma^i_{3k}\mathring\Gamma^k_{3j}=\tfrac12\,\mathring\Gamma^k_{3\ell}\mathring\Gamma^\ell_{3k}\,\delta^i_j,
		\label{eq:cayley-hamilton}
	\end{equation}
	with $\Gamma^k_{3\ell}\Gamma^\ell_{3k}=\mathring\Gamma^k_{3\ell}\mathring\Gamma^\ell_{3k}+\tfrac12(\Gamma^k_{3k})^2$; on the wall, $\tfrac12[\mathring\Gamma^k_{3\ell}\mathring\Gamma^\ell_{3k}]_0=H^2-K=\tfrac14(\kappa_1-\kappa_2)^2$.

	\subsection*{Metric of parallel surfaces}

	The map $\mathbf{r}_{0}\mapsto\mathbf{r}_{0}-\zeta\mathbf{n}$ deforms
	each tangent vector $\partial_i\mathbf r_0$ into $\partial_i\mathbf r=(\delta^k_i-\zeta[\Gamma^k_{3i}]_0)\,\partial_k\mathbf r_0$, so
	the surface metric at distance $\zeta$ is \emph{exactly}
	\begin{equation}
		g_{ij}(q,\zeta)
		= \big[g_{ij}+2\zeta\,\Gamma^3_{ij}+\zeta^{2}\,g_{k\ell}\Gamma^k_{3i}\Gamma^\ell_{3j}\big]_0 ,
		\label{eq:metric_parallel}
	\end{equation}
	and
	\begin{equation}
		\mathcal{J}_{S}(q,\zeta)
		= \det\!\big(\delta^k_i-\zeta[\Gamma^k_{3i}]_0\big)
		= (1+\zeta\kappa_{1})(1+\zeta\kappa_{2})
		= 1 + 2H\zeta + K\zeta^{2}.
		\label{eq:jacobian_parallel}
	\end{equation}
	The formula \eqref{eq:metric_parallel} is exact within the
	regular tubular neighbourhood of $S_{p}$ in which the
	parametrization \eqref{eq:parallel_surface} is invertible;
	it is not merely a truncated Taylor expansion.

	\subsection*{Flatness relations}

	The normal dependence of the Christoffel symbols is constrained by the flatness of the ambient space. Since the metric \eqref{eq:metric_block2} describes Euclidean space in curvilinear coordinates, its Riemann tensor
	\begin{equation*}
		R^{\alpha}{}_{\beta\gamma\delta}=\partial_\gamma\Gamma^{\alpha}_{\beta\delta}-\partial_\delta\Gamma^{\alpha}_{\beta\gamma}+\Gamma^{\alpha}_{\gamma\kappa}\Gamma^{\kappa}_{\beta\delta}-\Gamma^{\alpha}_{\delta\kappa}\Gamma^{\kappa}_{\beta\gamma},
	\end{equation*}
	which measures the failure of second covariant derivatives to commute, vanishes identically. By Eq.~\eqref{eq:christoffel_a}, its components $R^i{}_{33j}$, $R^k{}_{i3j}$ and $R^3{}_{ijk}$ reduce to
	\begin{equation}
		R^i{}_{33j}=\partial_3\Gamma^i_{3j}+\Gamma^i_{3k}\Gamma^k_{3j},\qquad
		R^k{}_{i3j}=\partial_3\Gamma^k_{ij}-\Gamma^k_{3i|j},\qquad
		R^3{}_{ijk}=\Gamma^3_{ik|j}-\Gamma^3_{ij|k},
	\end{equation}
	the derivatives combining with the products of Christoffel symbols into bar derivatives. Their vanishing gives
	\begin{equation}
		\partial_\zeta\Gamma^i_{3j}=\Gamma^i_{3k}\Gamma^k_{3j},
		\qquad
		\partial_\zeta\Gamma^k_{ij}=-\Gamma^k_{3i|j},
		\qquad
		\Gamma^i_{3j|k}=\Gamma^i_{3k|j}\qquad\text{in }\mathcal V_S,
		\label{eq:riccati-codazzi}
	\end{equation}
	where the last one also uses Eq.~\eqref{eq:christoffel_lower} and $g_{ij|k}=0$. The first relation is a Riccati equation. In the matrix notation~\eqref{eq:Gamma3-matrix}, it integrates to $\Gamma_3=[\Gamma_3]_0(\mathbf I-\zeta[\Gamma_3]_0)^{-1}$, and Eq.~\eqref{eq:trace-Gamma} then gives Eq.~\eqref{eq:jacobian_parallel} again. The second relation shows that $\partial_\zeta$ does not commute with the bar derivative: for a surface vector depending on $\zeta$, $\partial_\zeta(a^i{}_{|j})=(\partial_\zeta a^i)_{|j}-\Gamma^i_{3j|k}a^k$. The third is the Codazzi--Mainardi equation of the parallel surface~\cite{DOC}: $\Gamma^3_{ij|k}$ is totally symmetric. (The remaining components, $R^{i}{}_{jkl}=0$, give the Gauss equation, which relates the Gaussian curvature to the intrinsic curvature of the surface.) The contractions used below are
	\begin{equation}
		\partial_\zeta\Gamma^k_{3k}=\Gamma^k_{3\ell}\Gamma^\ell_{3k},\qquad
		\partial_\zeta\Gamma^k_{kj}=-\partial_j\Gamma^k_{3k},\qquad
		\Gamma^k_{3j|k}=\partial_j\Gamma^k_{3k}\qquad\text{in }\mathcal V_S,
		\label{eq:codazzi-contracted}
	\end{equation}
	with $\Gamma^k_{3k}=-2H$ on $S_p$.

	\subsection*{Microscopic equations in Gauss coordinates: two conceptions of volume counting}

	We now write the microscopic equations of Section~\ref{sec2} in Gauss coordinates. The kinematic fields are defined without reference to any counting of volume. The velocity is $v^{\mu}=dx^{\mu}/dt$, where $x^{\mu}$ is the position of a fluid particle whose motion is followed; it is automatically a true contravariant vector.
	The condensation $b$ is defined kinematically by $\partial_t b+v^{\mu}{}_{;\mu}=0$, with $b=0$
	at equilibrium. Since $v^{\mu}$ is a true vector, $b$ is a true scalar. In the
	linearized limit, this is equivalent to defining $b$ as minus the divergence of
	the fluid displacement, $b=-a^{\mu}{}_{;\mu}$, with $\partial_t a^{\mu}=v^{\mu}$
	and $a^{\mu}=0$ at equilibrium. The displacement must then be infinitesimal: a
	finite difference $\Delta x^{\mu}$ between the coordinates of two distinct points is not a vector
	in curvilinear coordinates, since its transformation law involves the second
	derivatives of the coordinate change, whereas an infinitesimal displacement
	$dx^{\mu}$, like $v^{\mu}=dx^{\mu}/dt$, is a true vector. The linearization
	consists precisely in treating the displacement as infinitesimal, so that $a^{\mu}$
	is a true vector and $b=-a^{\mu}{}_{;\mu}$ a true scalar.
	The true vector nature of $a^{\mu}$ emphasizes the possibility of a deep analogy between the acoustic and electromagnetic equations, developed in Ref.~\cite{LAF1}, which is not our concern here but becomes especially enlightening in the nonlocal theory. In that analogy, the contravariant displacement $a^{\mu}$ takes the place of the covariant electromagnetic potential in the Weyl gauge, and the velocity $v^{\mu}=\partial_t a^{\mu}$ and the strain tensor $b_{\mu\nu}=-\tfrac12\bigl(a_{\mu;\nu}+a_{\nu;\mu}\bigr)$ play the roles of the electric field and of the magnetic tensor, the kinematic relation $\partial_t b_{\mu\nu}=-\tfrac12\bigl(v_{\mu;\nu}+v_{\nu;\mu}\bigr)$, minus the rate-of-strain tensor, playing that of Faraday's law. Since the antisymmetry of the magnetic tensor becomes the symmetry of $b_{\mu\nu}$, no correspondence of signs can be drawn; the minus sign is chosen so that the trace of $b_{\mu\nu}$ is the condensation, $b^{\mu}_{\mu}=b$.
	Up to this point, only kinematics is involved; physics enters with the conservation of mass: the continuity equation is the relation $b=(\rho-\rho_0)/\rho_0$.
	The excess temperature $\tau$ is the difference of two intensive state variables attached to a fluid particle, hence also a true scalar; so are $T_0$, $\beta_0$, $c_P$ and $\gamma$, which are defined per unit mass or through intensive derivatives (see \cite{LAF1} Eqs (6.18)).  These ``Lorentz'' fields and constants have weight zero in any conception of volume.

	The nature of the remaining quantities, namely the pressure $p$, the excess density $\rho'$ if used, and the constants $\rho_0$, $\chi_0$, $\eta$, $\zeta$, $\kappa$ (and of the force density $\mathbf{F}$ and heat-supply rate $\dot Q$ per unit volume), depends on how the mass, hence the volume, is counted:
	\begin{enumerate}
		\item[(i)] per unit \emph{physical volume}, $dm=\rho\sqrt{g}\,d^{3}x$: the density $\rho$ is a true scalar;
		\item[(ii)] per unit \emph{coordinate volume}, $dm=\rho\,d^{3}x$: the density $\rho$ is a scalar density of weight $-1$, since $\sqrt{g}$ is of weight $-1$ and $\rho=\sqrt{g}\,\rho_{\rm phys}$.
	\end{enumerate}
	The weights of all quantities, in both conceptions, are collected in Table~\ref{tab:weights}. Note that in this section the letter $\zeta$ has two different usages, namely
	the bulk (second) viscosity and the normal distance to the wall, with no
	ambiguity from the context.
	In conception (i), all fields and constants are true tensors (weight zero). Conception (ii) brings out the difference of nature between the kinematic and
	thermodynamic ``Lorentz'' fields, unchanged in both writings because they are
	not concerned by the counting of volume, and the ``Maxwell--Umov'' fields,
	which become densities because they are concerned by it (see \cite{LAF1} for the introduction
	of the notions of Lorentz true tensors and Maxwell density fields).
	\begin{table}[ht]
		\caption{Tensorial weights of the fields and constants under the two conceptions of volume counting.}
		\label{tab:weights}
		\begin{ruledtabular}
			\begin{tabular}{lcc}
				Quantity & (i) physical volume & (ii) coordinate volume\\
				\hline
				$v^{\mu},\ b,\ \tau$ & 0 & 0\\
				$T_0,\ \beta_0,\ c_P,\ \gamma$ & 0 & 0\\
				$g_{\mu\nu},\ g^{\mu\nu}$ & 0 & 0\\
				$\sqrt{g}$ & $-1$ & $-1$\\
				$\rho_0,\ p,\ \eta,\ \zeta,\ \kappa$ & 0 & $-1$\\
				$F^{\mu},\ \dot Q$ & 0 & $-1$\\
				$\chi_0$ & 0 & $+1$\\
			\end{tabular}
		\end{ruledtabular}
	\end{table}
	The combinations $\rho_0\chi_0$, $\gamma\chi_0 p$ and $T_0\beta_0^{2}/(\rho_0c_P\chi_0)$ have weight zero in both conceptions, so that the thermodynamic identity~\eqref{eq:thermodynamic-identity} and the equation of state are covariant.

	In both conceptions, the microscopic equations take the same appearance, provided that the semicolon is understood as the covariant derivative of the quantity with its own weight, Eq.~\eqref{eq:density_covariant}: $p_{;\nu}=\partial_{\nu}p$ in (i), and $p_{;\nu}=\partial_{\nu}p-\Gamma^{\kappa}_{\kappa\nu}p$ in (ii). The time derivative is taken at fixed coordinates, the solid being at rest. Equations~\eqref{eq:full-momentum}--\eqref{eq:isothermal-wall} become
	\begin{align}
		\rho_0\,\partial_t v^{\mu}
		&=-g^{\mu\nu}p_{;\nu}
		+\eta\,g^{\rho\sigma}v^{\mu}{}_{;\rho;\sigma}
		-\Bigl(\zeta+\frac{\eta}{3}\Bigr)g^{\mu\nu}\partial_{\nu}\partial_t b
		+F^{\mu}
		\notag\\
		&=-g^{\mu\nu}p_{;\nu}
		-\frac{\eta}{g}\,\epsilon^{\mu\nu\rho}\epsilon_{\rho\sigma\lambda}\,v^{\lambda;\sigma}{}_{;\nu}
		-\Bigl(\zeta+\frac{4\eta}{3}\Bigr)g^{\mu\nu}\partial_{\nu}\partial_t b
		+F^{\mu},
		&&\text{in }V_p,
		\label{eq:gauss-full-momentum}\\
		\partial_t b+v^{\mu}{}_{;\mu}&=0,
		&&\text{in }V_p,
		\label{eq:gauss-full-continuity}\\
		\gamma\chi_0\,p&=b+\beta_0\tau,
		&&\text{in }V_p,
		\label{eq:gauss-full-state}\\
		\rho_0c_P\,\partial_t\tau
		&=\beta_0T_0\,\partial_t p
		+\kappa\,g^{\rho\sigma}\tau_{;\rho;\sigma}+\dot Q,
		&&\text{in }V_p,
		\label{eq:gauss-full-energy}\\
		v^{\mu}&=0,
		&&\text{on }S_p,
		\label{eq:gauss-full-noslip}\\
		\tau&=0,
		&&\text{on }S_p.
		\label{eq:gauss-full-wall}
	\end{align}
	In conception (i) this is the simplest formal writing, with all quantities true tensors. In conception (ii), the only formal difference lies in the pressure gradient, through $p_{;\nu}$.
	In conception (ii), the covariance of the equations is ensured by
	the weights of Table~\ref{tab:weights}: all terms of
	Eqs.~\eqref{eq:gauss-full-momentum} and~\eqref{eq:gauss-full-energy} have
	weight $-1$, and those of Eqs.~\eqref{eq:gauss-full-continuity} and~\eqref{eq:gauss-full-state} weight zero; in conception (i), all the weights vanish. In another patch $x^{\mu'}$ the equations keep the same form, with primed quantities related to the unprimed ones, in conception (ii), by
	\begin{equation}
		\begin{gathered}
			v^{\mu'}=\frac{\partial x^{\mu'}}{\partial x^{\mu}}v^{\mu},\qquad
			F^{\mu'}=J^{-1}\frac{\partial x^{\mu'}}{\partial x^{\mu}}F^{\mu},\qquad
			\dot Q'=J^{-1}\dot Q,\\
			p'=J^{-1}p,\quad
			\rho_0'=J^{-1}\rho_0,\quad
			\eta'=J^{-1}\eta,\quad
			\zeta'=J^{-1}\zeta,\quad
			\kappa'=J^{-1}\kappa,\quad
			\chi_0'=J\,\chi_0,
		\end{gathered}
		\label{eq:weights-transformation}
	\end{equation}
	whereas $b$, $\tau$, $T_0$, $\beta_0$, $c_P$ and $\gamma$ are unchanged; in conception (i), all these quantities are unchanged apart from the tensorial transformation of vector components.

	As in Section~\ref{sec2}, the local ``divergence-free'' simplification separates Eqs.~\eqref{eq:gauss-full-momentum}--\eqref{eq:gauss-full-wall} into two decoupled problems, linked only through the macroscopic pressure $\langle p\rangle_p$, common to both. Let
	\begin{equation}
		f^{\mu}=-g^{\mu\nu}\langle p\rangle_{p\,;\nu},
	\end{equation}
	be the macroscopic force density, treated as uniform on the scale of the REV, in the covariant sense $f^{\mu}{}_{;\nu}=0$. The visco-inertial problem is
	\begin{align}
		-i\omega\rho_0\,v^{\mu}
		&=-g^{\mu\nu}\widetilde p_{;\nu}
		+\eta\,g^{\rho\sigma}v^{\mu}{}_{;\rho;\sigma}
		+f^{\mu},
		&&\text{in }V_p,
		\label{eq:gauss-local-viscous-momentum}\\
		v^{\mu}{}_{;\mu}&=0,
		&&\text{in }V_p,
		\label{eq:gauss-local-viscous-incompressibility}\\
		v^{\mu}&=0,
		&&\text{on }S_p,
		\label{eq:gauss-local-viscous-no-slip}
	\end{align}
	and the thermal problem is
	\begin{align}
		-i\omega\rho_0c_P\,\tau
		&=\kappa\,g^{\rho\sigma}\tau_{;\rho;\sigma}
		-i\omega\beta_0T_0\langle p\rangle_p,
		&&\text{in }V_p,
		\label{eq:gauss-local-thermal-energy}\\
		\tau&=0,
		&&\text{on }S_p,
		\label{eq:gauss-local-thermal-wall}
	\end{align}
	together with the conditions completing them, as in Section~\ref{sec2}. Only these two problems are needed in what follows. Their canonical normalization, which makes all the unknowns true tensors in both
	conceptions, is the object of the next section. From the resulting canonical
	problems, the section also derives exact representations of the inverse viscous
	and thermal tortuosities, $1/\alpha(\omega)$ and $1/\alpha'(\omega)$, each
	expressed as its high-frequency limit plus a boundary term, namely an integral
	over the pore surface $S_p$ involving the normal derivative of the corresponding
	canonical field.

	\section{Canonical Microscopic Problems Defining the Macroscopic Local Response Functions and Exact Boundary Representations of the Inverse Tortuosities}
	\label{sec4}

	We introduce the complex
	viscous and thermal skin depths $\varepsilon$ and $\varepsilon'$, defined through their squares
	\begin{equation}
		\varepsilon^2=\frac{\eta}{-i\omega\rho_0},
		\qquad
		\varepsilon'^2=\frac{\kappa}{-i\omega\rho_0c_P}.
		\label{eq:complex-skin-depths}
	\end{equation}
	For positive real $\omega$, the square roots are chosen with positive real parts:
	\begin{equation}
		\varepsilon=(1+i)\sqrt{\frac{\eta}{2\rho_0\omega}},
		\qquad
		\varepsilon'=(1+i)\sqrt{\frac{\kappa}{2\rho_0c_P\omega}}.
	\end{equation}
	Both quantities have dimensions of length and tend to zero in the high-frequency limit.

	For the visco-inertial problem, let
	\begin{equation}
		\mathbf{f}=-\nabla\langle p\rangle_p=f\,\mathbf{e}
	\end{equation}
	denote the prescribed macroscopic force density, treated as uniform within the microscopic problem, with amplitude $f$ and unit direction $\mathbf{e}$, taken arbitrarily. By linearity, we define a vector field $\mathbf{w}$ and a scalar field $\pi$ through
	\begin{equation}
		\mathbf{v}=\frac{f}{\eta}\,\mathbf{w},
		\qquad
		\widetilde p=f\,\pi.
		\label{eq:scaled-viscous-fields}
	\end{equation}
	The fields $\mathbf{w}$ and $\pi$ have dimensions of squared length and length, respectively. They depend linearly on the direction $\mathbf{e}$ of the driving, but the macroscopic response extracted from them below does not, by isotropy.

	Substitution into the microscopic momentum balance gives the canonical visco-inertial problem
	\begin{align}
		\frac{\mathbf{w}}{\varepsilon^2}
		&=-\nabla\pi
		+\nabla^2\mathbf{w}+\mathbf{e},
		&&\text{in }V_p,
		\label{eq:canonical-viscous-momentum}\\
		\nabla\cdot\mathbf{w}
		&=0,
		&&\text{in }V_p,
		\label{eq:canonical-viscous-divergence}\\
		\mathbf{w}
		&=\mathbf{0},
		&&\text{on }S_p.
		\label{eq:canonical-viscous-wall}
	\end{align}
	The zero-mean definition of the pressure fluctuation gives
	\begin{equation}
		\langle\pi\rangle_p=0.
	\end{equation}
	The conditions completing this problem are inherited from the geometrical and averaging formulation already specified.

	Averaging Eq.~\eqref{eq:scaled-viscous-fields} and comparing with Eq.~\eqref{eq:local-dynamic-tortuosity} gives $\langle\mathbf{w}\rangle_p=(\varepsilon^2/\alpha)\,\mathbf{e}$, the collinearity with $\mathbf{e}$ expressing the isotropy hypothesis. Hence
	\begin{equation}
		\alpha(\omega)
		=\frac{\varepsilon^2}{\mathbf{e}\cdot\langle\mathbf{w}\rangle_p}.
		\label{eq:tortuosity-scaled-viscous-field}
	\end{equation}
	Thus the dynamic tortuosity is obtained directly from the intrinsic fluid average of the canonical field, projected on the direction of the driving.

	For the thermal problem, we use the analogous normalization by the prescribed source and the thermal conductivity. We define a scalar field $\vartheta$ through
	\begin{equation}
		\tau
		=\frac{-i\omega\beta_0T_0\langle p\rangle_p}{\kappa}
		\,\vartheta.
		\label{eq:scaled-thermal-field}
	\end{equation}
	Like $\mathbf{w}$, the field $\vartheta$ has dimensions of squared length. Substitution into Eq.~\eqref{eq:local-thermal-energy} gives the canonical thermal problem
	\begin{align}
		\frac{\vartheta}{\varepsilon'^2}
		&=\nabla^2\vartheta+1,
		&&\text{in }V_p,
		\label{eq:canonical-thermal-equation}\\
		\vartheta
		&=0,
		&&\text{on }S_p.
		\label{eq:canonical-thermal-wall}
	\end{align}
	Here again, the conditions completing the problem are those associated with the chosen geometrical and averaging formulation. Comparison with Eq.~\eqref{eq:local-thermal-tortuosity} gives
	\begin{equation}
		\alpha'(\omega)
		=\frac{\varepsilon'^2}{\langle\vartheta\rangle_p}.
		\label{eq:tortuosity-scaled-thermal-field}
	\end{equation}

	The prescribed excitation amplitudes have now disappeared from both microscopic problems. The frequency dependence of the two problems enters only through $\varepsilon$ and $\varepsilon'$, while the pore geometry enters through the domain, the interface, and the conditions completing the problems.

	Before undertaking the high-frequency analysis of these canonical problems, we introduce the ideal velocity pattern and the visco-inertial deficit relative to it. These fields will allow us to separate the geometrical structure of the ideal flow from the modifications required by viscous coupling to the pore boundary. We also introduce the ideal temperature pattern, equal to unity, and the thermal deficit relative to it. These fields will allow us to separate the uniform ideal temperature pattern from the modifications required by thermal coupling to the pore boundary.

	We first consider the visco-inertial quantities.

	With the normalization adopted above, we define the dimensionless visco-inertial response vector
	\begin{equation}
		\bm{\mathcal{E}}
		=\frac{\mathbf{w}}{\varepsilon^2}.
		\label{eq:dimensionless-viscous-response}
	\end{equation}
	The dimensional velocity is therefore
	\begin{equation}
		\mathbf{v}
		=\frac{f}{-i\omega\rho_0}\,
		\bm{\mathcal{E}},
	\end{equation}
	and the canonical visco-inertial problem becomes
	\begin{align}
		\bm{\mathcal{E}}
		&=\mathbf{e}-\nabla\pi
		+\varepsilon^2\nabla^2\bm{\mathcal{E}},
		&&\text{in }V_p,
		\label{eq:normalized-viscous-momentum}\\
		\nabla\cdot\bm{\mathcal{E}}
		&=0,
		&&\text{in }V_p,\\
		\bm{\mathcal{E}}
		&=\mathbf{0},
		&&\text{on }S_p.
	\end{align}
	Its intrinsic fluid average determines the dynamic tortuosity through $\langle\bm{\mathcal{E}}\rangle_p=\mathbf{e}/\alpha$, that is,
	\begin{equation}
		\frac{1}{\alpha(\omega)}
		=\mathbf{e}\cdot\langle\bm{\mathcal{E}}\rangle_p.
		\label{eq:tortuosity-normalized-response}
	\end{equation}

	The ideal velocity pattern is represented by a dimensionless vector field $\mathbf{E}$. We introduce a scalar potential $\Phi$ and define
	\begin{equation}
		\mathbf{E}=\mathbf{e}-\nabla\Phi.
		\label{eq:ideal-velocity-pattern}
	\end{equation}
	The ideal microscopic problem is
	\begin{align}
		\mathbf{E}
		&=\mathbf{e}-\nabla\Phi,
		&&\text{in }V_p,\\
		\nabla\cdot\mathbf{E}
		&=0,
		&&\text{in }V_p,\\
		\mathbf{n}\cdot\mathbf{E}
		&=0,
		&&\text{on }S_p,
		\label{eq:ideal-impermeability}
	\end{align}
	where $\mathbf{n}$ is the outward unit normal of the fluid domain at the fluid--solid interface.

	Equivalently, the potential satisfies
	\begin{align}
		\nabla^2\Phi
		&=0,
		&&\text{in }V_p,\\
		(\mathbf{n}\cdot\nabla)\Phi
		&=\mathbf{n}\cdot\mathbf{e},
		&&\text{on }S_p.
	\end{align}
	The conditions completing this problem are inherited from the geometrical and averaging formulation, and the additive constant is chosen so that
	\begin{equation}
		\langle\Phi\rangle_p=0.
	\end{equation}
	In particular, the ideal field is harmonic:
	\begin{equation}
		\nabla^2\mathbf{E}=\mathbf{0}\qquad\text{in }V_p.
		\label{eq:ideal-field-harmonic}
	\end{equation}
	It satisfies impermeability but not no slip: it is tangent to the pore boundary and is generally nonzero there.

	Its intrinsic fluid average defines the ideal, high-frequency tortuosity, through $\langle\mathbf{E}\rangle_p=\mathbf{e}/\alpha_\infty$:
	\begin{equation}
		\frac{1}{\alpha_{\infty}}
		=\mathbf{e}\cdot\langle\mathbf{E}\rangle_p.
		\label{eq:ideal-tortuosity}
	\end{equation}

	We now define the dimensionless visco-inertial deficit by
	\begin{equation}
		\mathbf{w}_{\varepsilon}
		=\mathbf{E}-\bm{\mathcal{E}}.
		\label{eq:visco-inertial-deficit}
	\end{equation}
	The subscript distinguishes this deficit from the dimensional canonical response field $\mathbf{w}$. We also separate the ideal potential from the pressure response by writing
	\begin{equation}
		\pi
		=\Phi
		+\varepsilon^2 Q_{\varepsilon}.
		\label{eq:pressure-potential-decomposition}
	\end{equation}
	Here $Q_{\varepsilon}$, which we call the pressure potential, is a scalar field with dimensions of inverse length; the viscous part of the pore-scale pressure is $\varepsilon^2Q_\varepsilon$. This decomposition is an exact definition at each nonzero frequency; it does not assume that $Q_{\varepsilon}$ remains bounded in the high-frequency limit. In fact it is not bounded: Sec.~\ref{sec6} shows that $Q_\varepsilon$ grows as $\varepsilon^{-1}$ when $\varepsilon\to0$. The zero-mean conventions for $\pi$ and $\Phi$ give
	\begin{equation}
		\langle Q_{\varepsilon}\rangle_p=0.
	\end{equation}

	Substituting these definitions into
	Eq.~\eqref{eq:normalized-viscous-momentum} and using
	Eq.~\eqref{eq:ideal-field-harmonic} yields the exact deficit problem
	\begin{align}
		\varepsilon^2
		\left(
		\nabla Q_{\varepsilon}
		+\nabla^2\mathbf{w}_{\varepsilon}
		\right)
		-\mathbf{w}_{\varepsilon}
		&=\mathbf{0},
		&&\text{in }V_p,
		\label{eq:deficit-momentum}\\
		\nabla\cdot\mathbf{w}_{\varepsilon}
		&=0,
		&&\text{in }V_p,
		\label{eq:deficit-divergence}\\
		\mathbf{w}_{\varepsilon}
		&=\mathbf{E},
		&&\text{on }S_p.
		\label{eq:deficit-wall}
	\end{align}
	The deficit is therefore driven by the tangential ideal velocity at the pore boundary. Its normal component vanishes there. The field $Q_{\varepsilon}$ enforces the divergence-free constraint, while the deficit describes the departure of the actual velocity pattern from the ideal one, both near the boundary and within the pore volume.

	Averaging its definition gives the exact relation
	\begin{equation}
		\frac{1}{\alpha(\omega)}
		=\frac{1}{\alpha_{\infty}}
		-\mathbf{e}\cdot\langle\mathbf{w}_{\varepsilon}\rangle_p.
		\label{eq:tortuosity-deficit-average}
	\end{equation}
	The deficit also contains a potential part: the displacement flow induced by the boundary layer. This flow extends into the bulk of the pores, so that this direct average does not reduce to an integral over the boundary layer~\cite{CSGL}. Projecting the deficit onto the ideal field removes this contribution: the identities derived below lead to Eq.~\eqref{eq:tortuosity-vortical-projection}, in which only the vortical part of the deficit appears, and which is the form evaluated in the boundary-layer analysis of Sec.~\ref{sec6}.

	These identities follow from an orthogonality property of incompressible fields satisfying impermeability at the pore boundary.

	Throughout, we call microscopic fields and potentials \emph{stationary} if they satisfy the periodic or stationary-random conditions of the geometrical formulation, with the conditions of regularity, joint statistical properties, integrability and averaging needed to ensure cancellation or disappearance of exterior-boundary contributions in the integration-by-parts identities. Stationarity is understood spatially; the potentials contain no macroscopic drift.

	If $\mathbf{u}$ is an incompressible field satisfying $\mathbf{u}\cdot\mathbf{n}=0$ on $S_p$, stationary in the sense just defined, and $\psi$ is a single-valued stationary scalar potential, integration by parts gives
	\begin{equation}
		\langle\mathbf{u}\cdot\nabla\psi\rangle_p=0.
		\label{eq:gradient-orthogonality}
	\end{equation}
	In a periodic cell, the contributions from opposite cell faces cancel, while the fluid--solid boundary contribution vanishes because $\mathbf{u}\cdot\mathbf{n}=0$. In the statistically homogeneous formulation, the cancellation is not exact but rests on an idealization, which can be seen in three ways. (i)~In a single sample, the average is taken over a REV of size $L$ whose outer boundary cuts through the fluid; the corresponding surface contribution is not zero, but it is of relative order $\ell/L$, $\ell$ being the microstructural scale, and becomes negligible when the REV is large compared with the microstructure, provided $\psi\mathbf u$ does not grow from one region to another---whence the absence of macroscopic drift in the definition of stationarity. The statistical idealization removes this residual term. (ii)~If one imagines an ensemble of samples prepared in the same way, homogeneity makes the mean of $\chi\psi\mathbf u$, with $\chi$ the indicator of the fluid, the same everywhere, so that the mean of its divergence vanishes and no averaging volume is needed. (iii)~The mean of the REV boundary term over the ensemble is the flux of a uniform vector through a closed surface, hence zero even for a finite REV. A real material is neither exactly homogeneous nor sampled by infinitely large REVs, so that the identity, like the macroscopic description itself, holds up to corrections of order $\ell/L$, which we neglect.

	For either $\mathbf{A}=\mathbf{E}$ or $\mathbf{A}=\mathbf{w}_{\varepsilon}$, the field $\mathbf{A}$ is incompressible and satisfies impermeability. Since $\mathbf{E}=\mathbf{e}-\nabla\Phi$,
	\begin{equation}
		\langle\mathbf{E}\cdot\mathbf{A}\rangle_p
		=\mathbf{e}\cdot\langle\mathbf{A}\rangle_p
		-\langle\mathbf{A}\cdot\nabla\Phi\rangle_p
		=\mathbf{e}\cdot\langle\mathbf{A}\rangle_p,
	\end{equation}
	where the last equality follows from
	Eq.~\eqref{eq:gradient-orthogonality}. In particular,
	\begin{equation}
		\left\langle
		\mathbf{E}\cdot\mathbf{E}
		\right\rangle_p
		=\mathbf{e}\cdot\langle\mathbf{E}\rangle_p,
		\qquad
		\left\langle
		\mathbf{E}\cdot\mathbf{w}_{\varepsilon}
		\right\rangle_p
		=\mathbf{e}\cdot\langle\mathbf{w}_{\varepsilon}\rangle_p.
		\label{eq:tensor-projection-identities}
	\end{equation}

	The inverse dynamic tortuosity can consequently be written as
	\begin{equation}
		\frac{1}{\alpha(\omega)}
		=\frac{1}{\alpha_{\infty}}
		-\left\langle
		\mathbf{E}\cdot\mathbf{w}_{\varepsilon}
		\right\rangle_p,
		\qquad
		\frac{1}{\alpha_{\infty}}
		=\left\langle\mathbf{E}\cdot\mathbf{E}\right\rangle_p.
		\label{eq:tortuosity-deficit-projection}
	\end{equation}
	This representation replaces the direct average of the deficit by its projection onto the ideal velocity pattern.

	We next take the scalar product of the exact deficit equation with $\mathbf{E}$ and average over the fluid. The pressure term vanishes by the same orthogonality property:
	\begin{equation}
		\left\langle
		\mathbf{E}\cdot\nabla Q_{\varepsilon}
		\right\rangle_p=0.
	\end{equation}
	Thus
	\begin{equation}
		\left\langle
		\mathbf{E}\cdot\mathbf{w}_{\varepsilon}
		\right\rangle_p
		=\varepsilon^2
		\left\langle
		\mathbf{E}\cdot\nabla^2\mathbf{w}_{\varepsilon}
		\right\rangle_p.
		\label{eq:deficit-projected-laplacian}
	\end{equation}
	The pressure potential remains necessary in the local deficit problem, but no longer appears explicitly in this averaged identity.

	The second Green identity, applied componentwise in Cartesian coordinates to $\mathbf{E}$ and $\mathbf{w}_{\varepsilon}$, gives
	\begin{align}
		&\int_{V_p}
		\left[
		\mathbf{E}\cdot\nabla^2\mathbf{w}_{\varepsilon}
		-\nabla^2\mathbf{E}\cdot\mathbf{w}_{\varepsilon}
		\right]\,dV
		\notag\\
		&\qquad=
		\int_{S_p}
		\left[
		\mathbf{E}\cdot\partial_n\mathbf{w}_{\varepsilon}
		-\partial_n\mathbf{E}\cdot\mathbf{w}_{\varepsilon}
		\right]\,dS,
		\label{eq:tensor-second-green-identity}
	\end{align}
	where $\partial_n=\mathbf{n}\cdot\nabla$ and $\mathbf{n}$ is the outward normal of the fluid domain. In the periodic formulation, contributions from opposite artificial cell faces cancel. Likewise, under the statistical averaging assumptions stated above, the exterior-boundary contribution per unit fluid volume vanishes in the macroscopic spatial average.

	Since $\nabla^2\mathbf{E}=\mathbf{0}$ in the fluid and
	$\mathbf{w}_{\varepsilon}=\mathbf{E}$ on $S_p$, we obtain
	\begin{equation}
		\left\langle
		\mathbf{E}\cdot\mathbf{w}_{\varepsilon}
		\right\rangle_p
		=
		\frac{\varepsilon^2}{|V_p|}
		\int_{S_p}
		\left[
		\mathbf{E}\cdot\partial_n\mathbf{w}_{\varepsilon}
		-\partial_n\mathbf{E}\cdot\mathbf{E}
		\right]\,dS.
		\label{eq:deficit-boundary-identity}
	\end{equation}
	The first exact boundary representation of the inverse dynamic tortuosity is therefore
	\begin{equation}
		\frac{1}{\alpha(\omega)}
		=
		\frac{1}{\alpha_{\infty}}
		-\frac{\varepsilon^2}{|V_p|}
		\int_{S_p}
		\left[
		\mathbf{E}\cdot\partial_n\mathbf{w}_{\varepsilon}
		-\partial_n\mathbf{E}\cdot\mathbf{E}
		\right]\,dS.
		\label{eq:tortuosity-boundary-total-deficit}
	\end{equation}

	An equivalent representation separates the deficit into vortical and potential contributions:
	\begin{equation}
		\mathbf{w}_{\varepsilon}
		=\mathbf{w}_{\mathrm{vort},\varepsilon}
		+\mathbf{w}_{\mathrm{pot},\varepsilon},
		\qquad
		\mathbf{w}_{\mathrm{pot},\varepsilon}
		=\nabla\chi_{\varepsilon},
		\qquad
		\nabla^2\mathbf{w}_{\mathrm{pot},\varepsilon}
		=\mathbf{0}\qquad\text{in }V_p.
		\label{eq:deficit-vortical-potential}
	\end{equation}
	Here the potential $\chi_{\varepsilon}$ itself must be single-valued and stationary in the sense defined above. Its gradient is then orthogonal to the ideal field in the fluid average, by Eq.~\eqref{eq:gradient-orthogonality} with $\mathbf u=\mathbf E$. This property does not require the pore domain to be simply connected. The conditions imposed on the potential part do not fix the split: the gradient of a single-valued, stationary harmonic function satisfies them and is also divergence-free, so that it can be moved from one part to the other. Equations~\eqref{eq:tortuosity-vortical-projection} and~\eqref{eq:tortuosity-boundary-vortical-deficit} below hold for any such $\chi_\varepsilon$. Section~\ref{sec6-pressure} makes the natural choice $\chi_\varepsilon=\varepsilon^2Q_\varepsilon$ [Eq.~\eqref{eq:exact-splitting}], for which the vortical part satisfies Eq.~\eqref{eq:deficit-momentum} without its pressure term, $\varepsilon^2\nabla^2\mathbf w_{\mathrm{vort},\varepsilon}=\mathbf w_{\mathrm{vort},\varepsilon}$.

	Orthogonality to the ideal field gives
	\begin{equation}
		\left\langle
		\mathbf{E}\cdot\mathbf{w}_{\mathrm{pot},\varepsilon}
		\right\rangle_p=0,
	\end{equation}
	and hence
	\begin{equation}
		\frac{1}{\alpha(\omega)}
		=
		\frac{1}{\alpha_{\infty}}
		-\left\langle
		\mathbf{E}\cdot\mathbf{w}_{\mathrm{vort},\varepsilon}
		\right\rangle_p.
		\label{eq:tortuosity-vortical-projection}
	\end{equation}
	Although the potential contribution can affect the local velocity and its direct volume average, it disappears from the projection onto $\mathbf{E}$.

	Because the Laplacian of the potential contribution also vanishes, Eq.~\eqref{eq:deficit-projected-laplacian} holds with $\mathbf{w}_{\mathrm{vort},\varepsilon}$ in place of $\mathbf{w}_{\varepsilon}$, and the second Green identity now yields the second exact boundary representation:
	\begin{equation}
		\frac{1}{\alpha(\omega)}
		=
		\frac{1}{\alpha_{\infty}}
		-\frac{\varepsilon^2}{|V_p|}
		\int_{S_p}
		\left[
		\mathbf{E}\cdot
		\partial_n\mathbf{w}_{\mathrm{vort},\varepsilon}
		-\partial_n\mathbf{E}\cdot
		\mathbf{w}_{\mathrm{vort},\varepsilon}
		\right]\,dS.
		\label{eq:tortuosity-boundary-vortical-deficit}
	\end{equation}
	In this second form, the vortical deficit alone appears, but its boundary value must not be replaced by $\mathbf{E}$: it is the total deficit that satisfies that boundary condition.

	Equations~\eqref{eq:tortuosity-boundary-total-deficit}
	and~\eqref{eq:tortuosity-boundary-vortical-deficit}
	are equivalent exact identities, which express the averaged visco-inertial response through quantities evaluated at the pore boundary. Their meaning is clearest in the first form: since $\mathbf w_\varepsilon=\mathbf E-\bm{\mathcal E}$, its integrand is $-\mathbf E\cdot\partial_n\bm{\mathcal E}$, so that $1/\alpha(\omega)-1/\alpha_\infty=(\varepsilon^2/|V_p|)\int_{S_p}\mathbf E\cdot\partial_n\bm{\mathcal E}\,dS$. The departure from the ideal response is thus set by the viscous shear stress of the actual flow at the walls, weighted by the ideal slip. With the Stokes-layer profile, for which $\partial_n\bm{\mathcal E}\simeq-\mathbf E/\varepsilon$ on $S_p$, this expression gives the Johnson--Koplik--Dashen length at once. Beyond this order, the boundary forms require the normal derivatives of the layer profile at the wall; the boundary-layer analysis of Sec.~\ref{sec6} evaluates the volume form~\eqref{eq:tortuosity-vortical-projection}, which gives the same coefficients.

	We now turn to the corresponding thermal quantities. Their treatment is simpler because the ideal normalized temperature is uniform and no constraint analogous to microscopic incompressibility needs to be enforced.

	With the normalization adopted above, we define the dimensionless thermal response scalar
	\begin{equation}
		\mathcal{T}
		=\frac{\vartheta}{\varepsilon'^2}.
		\label{eq:dimensionless-thermal-response}
	\end{equation}
	The frequency dependence of $\mathcal{T}$, through $\varepsilon'$, is left implicit in the notation, as is that of $\bm{\mathcal{E}}$ through $\varepsilon$.
	The dimensional excess temperature is consequently
	\begin{equation}
		\tau
		=\frac{\beta_0T_0}{\rho_0c_P}
		\langle p\rangle_p\,\mathcal{T}.
		\label{eq:temperature-normalized-response}
	\end{equation}
	The canonical thermal problem becomes
	\begin{align}
		\mathcal{T}
		&=1+\varepsilon'^2\nabla^2\mathcal{T},
		&&\text{in }V_p,
		\label{eq:normalized-thermal-equation}\\
		\mathcal{T}
		&=0,
		&&\text{on }S_p.
		\label{eq:normalized-thermal-wall}
	\end{align}
	The field satisfies the same stationarity conditions as the other microscopic response fields. Its intrinsic fluid average determines the thermal tortuosity:
	\begin{equation}
		\frac{1}{\alpha'(\omega)}
		=\langle\mathcal{T}\rangle_p.
		\label{eq:thermal-tortuosity-normalized-response}
	\end{equation}

	In the absence of thermal diffusion, the normalized temperature is simply unity. This ideal temperature pattern is independent of the pore geometry, but it does not satisfy the isothermal condition at the fluid--solid interface. We therefore define the dimensionless thermal deficit by
	\begin{equation}
		d_{\varepsilon'}
		=1-\mathcal{T}.
		\label{eq:thermal-deficit}
	\end{equation}
	Substitution into Eq.~\eqref{eq:normalized-thermal-equation}
	gives the exact deficit problem
	\begin{align}
		\varepsilon'^2\nabla^2d_{\varepsilon'}
		-d_{\varepsilon'}
		&=0,
		&&\text{in }V_p,
		\label{eq:thermal-deficit-equation}\\
		d_{\varepsilon'}
		&=1,
		&&\text{on }S_p.
		\label{eq:thermal-deficit-wall}
	\end{align}
	Thus the thermal deficit is driven by the mismatch between the uniform ideal temperature and the zero excess temperature imposed at the pore boundary. Unlike the visco-inertial deficit, it requires no additional potential field enforcing a divergence-free constraint.

	Averaging its definition gives
	\begin{equation}
		\frac{1}{\alpha'(\omega)}
		=1-\langle d_{\varepsilon'}\rangle_p.
		\label{eq:thermal-tortuosity-deficit-average}
	\end{equation}
	Since the ideal temperature pattern is unity, this direct average is already the counterpart of the projection onto the ideal velocity pattern used in the visco-inertial problem.

	Integrating Eq.~\eqref{eq:thermal-deficit-equation}
	over the fluid and applying the divergence theorem yields
	\begin{equation}
		\langle d_{\varepsilon'}\rangle_p
		=
		\frac{\varepsilon'^2}{|V_p|}
		\int_{S_p}\partial_n d_{\varepsilon'}\,dS.
		\label{eq:thermal-deficit-boundary-identity}
	\end{equation}
	Here $\partial_n=\mathbf{n}\cdot\nabla$, with $\mathbf{n}$
	the outward normal of the fluid domain. Contributions from
	opposite artificial cell faces cancel in the periodic formulation.
	In the statistically homogeneous formulation, the exterior-boundary
	contribution per unit fluid volume vanishes in the macroscopic
	average under the stationarity and averaging conditions stated above.

	The exact boundary representation of the inverse thermal
	tortuosity is therefore
	\begin{equation}
		\frac{1}{\alpha'(\omega)}
		=
		1-\frac{\varepsilon'^2}{|V_p|}
		\int_{S_p}\partial_n d_{\varepsilon'}\,dS.
		\label{eq:thermal-tortuosity-boundary}
	\end{equation}
	Equivalently, because
	$\partial_n d_{\varepsilon'}=-\partial_n\mathcal{T}$,
	\begin{equation}
		\frac{1}{\alpha'(\omega)}
		=
		1+\frac{\varepsilon'^2}{|V_p|}
		\int_{S_p}\partial_n\mathcal{T}\,dS.
		\label{eq:thermal-tortuosity-boundary-response}
	\end{equation}

	Using Eq.~\eqref{eq:dynamic-compliability}, the dynamic
	compliability can likewise be expressed directly in terms
	of the averaged thermal deficit or its boundary derivative:
	\begin{equation}
		\beta(\omega)
		=
		1+(\gamma-1)\langle d_{\varepsilon'}\rangle_p
		=
		1+(\gamma-1)\frac{\varepsilon'^2}{|V_p|}
		\int_{S_p}\partial_n d_{\varepsilon'}\,dS.
		\label{eq:compliability-thermal-boundary}
	\end{equation}
	The thermal response is thus reduced to a scalar boundary quantity, the heat conducted into the walls; with the layer profile, for which $\partial_n\mathcal T\simeq-1/\varepsilon'$ on $S_p$, Eq.~\eqref{eq:thermal-tortuosity-boundary-response} gives the Champoux--Allard length at once.
	As in the visco-inertial case, the boundary-layer analysis of Sec.~\ref{sec8} uses the volume form~\eqref{eq:thermal-tortuosity-deficit-average} rather than these boundary forms.

	\subsection*{True tensor character of the canonical fields and general Gauss form}

	The results of Section~\ref{sec3} show that the scaled unknowns $w^{\mu}=\eta\,v^{\mu}/f$ and $\pi=\widetilde p/f$ are true tensor fields, not tensor densities, in both conceptions of volume counting. Indeed, by Table~\ref{tab:weights}, the force density $f^{\mu}=f\,e^{\mu}$ has the weight of the pressure, $0$ in (i) and $-1$ in (ii), and so does its amplitude $f$, the unit vector $\mathbf{e}$ being a true vector of weight zero ($g_{\mu\nu}e^{\mu}e^{\nu}=1$, $e^{\mu}{}_{;\nu}=0$). The amplitude $f$, like the viscosity $\eta$, is covariantly constant, $f_{;\nu}=\eta_{;\nu}=0$. The field $w^{\mu}$ then has weight $(W_\eta+0-W_f)=0$, and $\pi$ also has weight $0$. The skin depths $\varepsilon^2=\eta/(-i\omega\rho_0)$ and $\varepsilon'^2=\kappa/(-i\omega\rho_0c_P)$ are true scalars of weight zero, and so is the prefactor $\langle p\rangle_p/\kappa$ in Eq.~\eqref{eq:scaled-thermal-field}. In summary, $\mathbf{w}$ is a true vector field, $\pi$ and $\vartheta$ are true scalar fields, and the same geometrical character is inherited by $\bm{\mathcal{E}}$, $\mathbf{E}$, $\Phi$, $\mathbf{w}_\varepsilon$, $Q_\varepsilon$, $\mathcal{T}$ and $d_{\varepsilon'}$. Their coordinate transformations therefore involve no volume-Jacobian factors, and their definitions require no convention as to whether the metric-volume factor $\sqrt{g}$ is incorporated into the field components or retained in the volume element. The scalars $\alpha(\omega)$, $\alpha'(\omega)$ and $\beta(\omega)$ are likewise of weight zero. Note also that only the scalar projection $\mathbf{e}\cdot\mathbf{w}=g_{\mu\nu}e^{\mu}w^{\nu}$ is averaged in Eq.~\eqref{eq:tortuosity-scaled-viscous-field}, an operation that is meaningful in any patch, in contrast to an average of vector components taken in different curvilinear charts.

	Since the canonical fields are true tensors, the canonical equations of this section,
	written above in Cartesian coordinates, can be written directly in Gauss coordinates,
	identically in both conceptions, by covariant transformation, that is, by replacing the
	Cartesian differential operators by the covariant differential operators of
	Section~\ref{sec3}, without any further calculation. In particular, $\pi$ being a true
	scalar, its covariant derivative is simply the partial derivative,
	$\pi_{;\nu}=\partial_\nu\pi$, with no connection term, even in conception~(ii), where
	the physical pressure $p$ is a density of weight $-1$. They read
	\begin{subequations}
		\label{eq:canonical_gauss}
		\begin{align}
			\frac{w^{\mu}}{\varepsilon^{2}}
			&=-g^{\mu\nu}\pi_{;\nu}
			+g^{\rho\sigma}w^{\mu}{}_{;\rho;\sigma}
			+e^{\mu}
			&&\text{in }V_p,
			\label{eq:canonical_gauss_mom}\\
			w^{\mu}{}_{;\mu}
			&=0
			&&\text{in }V_p,
			\label{eq:canonical_gauss_div}\\
			w^{\mu}
			&=0
			&&\text{on }S_{p},
			\label{eq:canonical_gauss_bc}
		\end{align}
	\end{subequations}
and,
\begin{subequations}
	\label{eq:thermal_canonical_gauss}
	\begin{align}
		\frac{\vartheta}{\varepsilon'^{\,2}}
		&=g^{\mu\nu}\,\vartheta_{;\mu;\nu}+1
		&&\text{in }V_p,
		\label{eq:thermal_canonical_gauss_pde}\\
		\vartheta
		&=0
		&&\text{on }S_{p}.
		\label{eq:thermal_canonical_gauss_bc}
	\end{align}
\end{subequations}
	The tortuosities are recovered from
	$\alpha(\omega)=\varepsilon^{2}\langle g_{\mu\nu}e^{\mu}w^{\nu}\rangle_p^{-1}$ and
	$\alpha'(\omega)=\varepsilon'^{2}\langle\vartheta\rangle_p^{-1}$, where
	$\langle\cdot\rangle_p=|V_p|^{-1}\int_{V_p}\cdot\,\sqrt{g}\,d^{3}x$, the integration
	being subdivided over the patches covering $V_p$.

	The ideal problem and the deficit problem take, in turn, the covariant forms
	\begin{subequations}
	\label{eq:ideal_gauss}
	\begin{align}
		E^{\mu}=e^{\mu}-g^{\mu\nu}\Phi_{;\nu},
		\qquad
		E^{\mu}{}_{;\mu}
		&=0
		&&\text{in }V_p,
		\label{eq:ideal_gauss_pde}\\
		n_{\mu}E^{\mu}
		&=0
		&&\text{on }S_{p},
		\label{eq:ideal_gauss_bc}
	\end{align}
\end{subequations}
where $n^{\mu}$ are the components of the outward unit normal of $V_p$, and
	\begin{subequations}
		\label{eq:deficit_gauss}
		\begin{align}
			\varepsilon^{2}\bigl(
			g^{\mu\nu}Q_{\varepsilon;\nu}
			+ g^{\rho\sigma}(w_{\varepsilon})^{\mu}{}_{;\rho;\sigma}
			\bigr)
			- (w_{\varepsilon})^{\mu}
			&= 0
			&&\text{in }V_p,
			\label{eq:deficit_gauss_mom}\\
			(w_{\varepsilon})^{\mu}{}_{;\mu}
			&= 0
			&&\text{in }V_p,
			\label{eq:deficit_gauss_div}\\
			(w_{\varepsilon})^{\mu}
			&= E^{\mu}
			&&\text{on }S_{p},
			\label{eq:deficit_gauss_bc}
		\end{align}
	\end{subequations}
	with $\pi=\Phi+\varepsilon^{2}Q_{\varepsilon}$. Equivalently, the normalized velocity components
	$\mathcal{E}^{\mu}=E^{\mu}-(w_{\varepsilon})^{\mu}$
	satisfy
	\begin{subequations}
		\label{eq:vel_gauss}
		\begin{align}
			\mathcal{E}^{\mu}
			+ g^{\mu\nu}\pi_{;\nu}
			- \varepsilon^{2}
			g^{\rho\sigma}\mathcal{E}^{\mu}{}_{;\rho;\sigma}
			&= e^{\mu}
			&&\text{in }V_p,
			\label{eq:vel_gauss_mom}\\
			\mathcal{E}^{\mu}{}_{;\mu}
			&= 0
			&&\text{in }V_p,
			\label{eq:vel_gauss_div}\\
			\mathcal{E}^{\mu}
			&= 0
			&&\text{on }S_{p}.
			\label{eq:vel_gauss_bc}
		\end{align}
	\end{subequations}
	Finally, the thermal deficit $d_{\varepsilon'}=1-\mathcal{T}$
	satisfies
	\begin{subequations}
		\label{eq:thermal_deficit_gauss}
		\begin{align}
			\varepsilon'^{\,2}\,\Delta d_{\varepsilon'}
			- d_{\varepsilon'}
			&= 0
			&&\text{in }V_p,
			\label{eq:thermal_deficit_gauss_pde}\\
			d_{\varepsilon'}
			&= 1
			&&\text{on }S_{p},
			\label{eq:thermal_deficit_gauss_bc}
		\end{align}
	\end{subequations}
	where the scalar Laplacian is given by Eq.~\eqref{eq:scalar_laplacian}.

	The integral identities of this section involve only scalars built by contraction and true scalar integrals, hence they hold unchanged in Gauss coordinates. For instance, $\partial_n=n^{\mu}\partial_{\mu}$ on scalars (in a near-wall patch of Sec.~\ref{sec3}, $n^\mu=\delta^\mu_3$, so that $\partial_nf=f_{;3}$, and the normal derivative of a vector field is $a^\mu{}_{;3}$), the orthogonality property reads $\langle u^{\mu}\psi_{;\mu}\rangle_p=0$ for $u^{\mu}{}_{;\mu}=0$ in $V_p$ and $u^{\mu}n_{\mu}=0$ on $S_p$, and the two volume forms used in the high-frequency analysis, Eqs.~\eqref{eq:tortuosity-vortical-projection} and~\eqref{eq:thermal-tortuosity-deficit-average}, read
	\begin{equation}
		\frac1{\alpha(\omega)}=\frac1{\alpha_\infty}-\frac1{|V_p|}\int_{V_p}g_{\mu\nu}E^\mu(w_{\mathrm{vort},\varepsilon})^\nu\,dV,\qquad
		\frac1{\alpha'(\omega)}=1-\frac1{|V_p|}\int_{V_p}d_{\varepsilon'}\,dV,
	\end{equation}
	with $dV=\sqrt g\,d^3x$.
	The near-wall structure of equations
	\eqref{eq:deficit_gauss} and \eqref{eq:thermal_deficit_gauss} will be
	unfolded systematically in Sections~\ref{sec6} and \ref{sec8}
	using the block metric \eqref{eq:metric_block}, the Christoffel symbols
	\eqref{eq:christoffel_a}--\eqref{eq:christoffel_c} and the flatness relations \eqref{eq:riccati-codazzi}--\eqref{eq:codazzi-contracted}.
	The rapid normal variable $\varsigma=\zeta/\varepsilon$
	(or $\varsigma'=\zeta/\varepsilon'$ for the thermal problem), complex like $\varepsilon$, is
	introduced there.

	\section{Relaxation-Mode Representation of the Canonical Responses}\label{sec5}

	Before analysing the boundary layers, we express the canonical fields of Sec.~\ref{sec4} on the viscous relaxation modes introduced by Avellaneda and Torquato~\cite{AT} (see also Ref.~\cite{L-AT}). This representation is exact at every frequency. It is introduced at this stage for three reasons. First, it shows that the normalization $\pi=\Phi+\varepsilon^2Q_\varepsilon$ of Eq.~\eqref{eq:pressure-potential-decomposition} is the natural one: $Q_\varepsilon$ is to $\bm{\mathcal E}$ what the modal pressures $Q_n$ are to the modal velocities $\bm\Psi_n$ introduced below. Second, it attaches every macroscopic parameter---low-frequency as well as high-frequency---to a single spectral object, the distribution of relaxation times, and therefore indicates how the high-frequency parameters derived in Secs.~\ref{sec6}--\ref{sec9} constrain the full-frequency response. Third, it gives a time-domain reading of the canonical problems, which clarifies the structure of the pressure field and a number of subtle convergence properties of the modal sums. The consequences are drawn in Sec.~\ref{sec10}, once the high-frequency coefficients are known.

	\subsection{Viscous relaxation modes}

	The viscous relaxation modes $(\bm\Psi_n,Q_n,\sigma_n)$ are the solutions of the self-adjoint eigenvalue problem
	\begin{align}
		\nabla^2\bm\Psi_n-\nabla Q_n&=-\frac{1}{\sigma_n}\bm\Psi_n,
		&&\text{in }V_p,
		\label{eq:AT-modes}\\
		\nabla\cdot\bm\Psi_n&=0,
		&&\text{in }V_p,
		\label{eq:AT-modes-div}\\
		\bm\Psi_n&=\mathbf 0,
		&&\text{on }S_p,
		\label{eq:AT-modes-wall}
	\end{align}
	completed by the stationarity conditions of Sec.~\ref{sec4}; we follow the notations of Refs.~\cite{AT,LAF_PhD,CSGL}. The eigenvalues $\sigma_n>0$ have dimensions of squared length; the corresponding relaxation times are $\Theta_n=\sigma_n/\nu$, with $\nu=\eta/\rho_0$. Taking the divergence of Eq.~\eqref{eq:AT-modes} shows that the modal pressures are harmonic, $\nabla^2Q_n=0$: like $\Phi$, they are Lagrange multipliers enforcing incompressibility. Their additive constants are fixed by $\langle Q_n\rangle_p=0$. Only the modes excited by the uniform driving, $\langle\bm\Psi_n\rangle_p\propto\mathbf e$, contribute to the macroscopic response; for them, this normalization is equivalent to the Umov--Heaviside--Poynting condition $\langle Q_n\bm\Psi_n\rangle_p=\mathbf 0$ discussed in Sec.~\ref{sec2}. The modes are orthonormalized,
	\begin{equation}
		\langle\bm\Psi_n\cdot\bm\Psi_m\rangle_p=\delta_{nm}.
		\label{eq:AT-orthonormality}
	\end{equation}
	Multiplying Eq.~\eqref{eq:AT-modes} by $\bm\Psi_n$, averaging and using Eq.~\eqref{eq:gradient-orthogonality} gives the Rayleigh quotient
	\begin{equation}
		\frac{1}{\sigma_n}
		=\frac{\langle\nabla\bm\Psi_n:\nabla\bm\Psi_n\rangle_p}{\langle\bm\Psi_n\cdot\bm\Psi_n\rangle_p}=\frac{\langle\partial_i(\Psi_n)_j\,\partial_i(\Psi_n)_j\rangle_p}{\langle(\Psi_n)_i(\Psi_n)_i\rangle_p}.
		\label{eq:AT-Rayleigh}
	\end{equation}
	For a divergence-free velocity field $\mathbf v$ vanishing on $S_p$, the viscous dissipation rate per unit fluid volume is $2\eta\langle\dot b_{ij}\dot b_{ij}\rangle_p$, where $\dot b_{ij}=\partial_t b_{ij}=-\tfrac12(\partial_iv_j+\partial_jv_i)$ is the time derivative of the strain tensor of Sec.~\ref{sec3}; since $\partial_iv_j\,\partial_jv_i=\partial_i(v_j\partial_jv_i)$ averages to zero, this equals $\eta\langle\partial_iv_j\,\partial_iv_j\rangle_p$. Applied to the mode $\mathbf v=\bm\Psi_n e^{-t/\Theta_n}$, the numerator of Eq.~\eqref{eq:AT-Rayleigh} is thus, up to the factor $\eta$, the dissipation rate of the mode. Hence $1/\Theta_n=\nu/\sigma_n$ is one half of the ratio of the dissipation rate to the kinetic energy $\tfrac12\rho_0\langle\bm\Psi_n\cdot\bm\Psi_n\rangle_p$ of the mode: the modes diagonalize simultaneously the kinetic energy and the dissipation, and each of them decays independently, $\bm\Psi_n e^{-t/\Theta_n}$, once the forcing has ceased. Modes concentrated near the walls have small $\sigma_n$: they carry little kinetic energy for much dissipation.

	The coupling of the uniform excitation to the mode $n$ is measured by the amplitude
	\begin{equation}
		b_n=\mathbf e\cdot\langle\bm\Psi_n\rangle_p=\langle\bm\Psi_n\cdot\mathbf E\rangle_p ,
		\label{eq:AT-weights}
	\end{equation}
	the second form following from Eq.~\eqref{eq:gradient-orthogonality}. Isotropy is used here as in Sec.~\ref{sec2}: $\langle\bm\Psi_n\rangle_p=b_n\mathbf e$.

	The modes are complete in the Hilbert space $\mathcal S$ obtained by closing, in the norm $\langle\mathbf u\cdot\mathbf u\rangle_p^{1/2}$, the space of smooth stationary divergence-free fields that vanish on $S_p$. Since the closure does not control boundary traces, $\mathcal S$ contains fields with nonzero tangential slip, and is precisely the space of divergence-free fields with zero normal component on $S_p$. The ideal field belongs to it. Therefore
	\begin{equation}
		\mathbf E=\sum_n b_n\bm\Psi_n\quad\text{in }V_p ,
		\qquad
		\sum_nb_n^2=\langle\mathbf E\cdot\mathbf E\rangle_p=\frac{1}{\alpha_\infty},
		\label{eq:AT-E-expansion}
	\end{equation}
	the second relation being Parseval's identity together with Eq.~\eqref{eq:tortuosity-deficit-projection}. The first relation holds in the mean-square sense, not as an equality of traces: every partial sum vanishes on $S_p$, whereas $\mathbf E$ does not. Equivalently, if $\mathsf P$ denotes the Helmholtz--Leray projection onto $\mathcal S$, then $\mathbf E=\mathsf P\mathbf e$ and $b_n=\langle\bm\Psi_n\cdot\mathsf P\mathbf e\rangle_p$.

	\subsection{Modal form of the canonical fields and of the pressure potential}

	The solution of the canonical problem~\eqref{eq:normalized-viscous-momentum} has the modal form
	\begin{equation}
		\bm{\mathcal E}=\sum_na_n(\varepsilon)\,\bm\Psi_n,
		\qquad
		\pi=\Phi+\varepsilon^2\sum_na_n(\varepsilon)\,Q_n,
		\qquad
		a_n(\varepsilon)=\frac{b_n}{1+\varepsilon^2/\sigma_n}=\frac{b_n\,\sigma_n}{\sigma_n+\varepsilon^2},
		\label{eq:AT-canonical}
	\end{equation}
	the expansions of the fields holding in $V_p$. The incompressibility and no-slip conditions are satisfied term by term, since every $\bm\Psi_n$ satisfies them. In the momentum equation, $\bm{\mathcal E}=\mathbf e-\nabla\pi+\varepsilon^2\nabla^2\bm{\mathcal E}$, the right-hand side is the sum of the driving force $\mathbf e$, the pressure force $-\nabla\pi$ and the viscous force $\varepsilon^2\nabla^2\bm{\mathcal E}$, all normalized by the amplitude $f$ of the driving. Writing the mode equation~\eqref{eq:AT-modes} as $\nabla^2\bm\Psi_n=\nabla Q_n-\bm\Psi_n/\sigma_n$ and differentiating the series term by term, which is formal, the viscous force becomes
	\begin{equation*}
		\varepsilon^2\nabla^2\bm{\mathcal E}=\varepsilon^2\nabla\Big(\sum_na_nQ_n\Big)-\sum_n\frac{\varepsilon^2}{\sigma_n}\,a_n\bm\Psi_n .
	\end{equation*}
	Its gradient part cancels the modal part $-\varepsilon^2\nabla\sum_na_nQ_n$ of the pressure force, and the momentum equation reduces to
	\begin{equation*}
		\sum_n\Big(1+\frac{\varepsilon^2}{\sigma_n}\Big)a_n\bm\Psi_n=\mathbf e-\nabla\Phi=\mathbf E=\sum_nb_n\bm\Psi_n ,
	\end{equation*}
	by Eq.~\eqref{eq:AT-E-expansion}. It is satisfied mode by mode if $a_n(1+\varepsilon^2/\sigma_n)=b_n$. Conversely, these are the only possible coefficients, as the projection of the momentum equation on $\bm\Psi_n$ shows without differentiating the series term by term. Multiplying the momentum equation by $\bm\Psi_n$ and averaging, the left-hand side gives $\langle\bm\Psi_n\cdot\bm{\mathcal E}\rangle_p=a_n$ by orthonormality, and $\mathbf e$ gives $b_n$ by Eq.~\eqref{eq:AT-weights}; the pressure force gives zero by Eq.~\eqref{eq:gradient-orthogonality}; and in the viscous force, Green's identity, whose boundary terms vanish because $\bm\Psi_n$ and $\bm{\mathcal E}$ both vanish on $S_p$, moves the Laplacian from $\bm{\mathcal E}$ onto $\bm\Psi_n$:
	\begin{equation*}
		\langle\bm\Psi_n\cdot\nabla^2\bm{\mathcal E}\rangle_p=\langle\bm{\mathcal E}\cdot\nabla^2\bm\Psi_n\rangle_p=\langle\bm{\mathcal E}\cdot\nabla Q_n\rangle_p-\frac{a_n}{\sigma_n}=-\frac{a_n}{\sigma_n},
	\end{equation*}
	by Eqs.~\eqref{eq:AT-modes} and~\eqref{eq:gradient-orthogonality}. Hence $a_n=b_n-(\varepsilon^2/\sigma_n)\,a_n$. Comparing the pressure in Eq.~\eqref{eq:AT-canonical} with Eq.~\eqref{eq:pressure-potential-decomposition} gives
	\begin{equation}
		Q_\varepsilon=\sum_na_n(\varepsilon)\,Q_n\qquad\text{in }V_p .
		\label{eq:AT-Q}
	\end{equation}
	The pressure potential is thus built from the modal pressures with exactly the same coefficients as the velocity is built from the modal velocities. This is the reason for the factor $\varepsilon^2$ in Eq.~\eqref{eq:pressure-potential-decomposition}: in the momentum equation, which then reads $\bm{\mathcal E}=\mathbf E+\varepsilon^2(\nabla^2\bm{\mathcal E}-\nabla Q_\varepsilon)$, it is the factor multiplying the Stokes operator $\nabla^2\bm{\mathcal E}-\nabla Q_\varepsilon$, the operator of the mode equation~\eqref{eq:AT-modes}, and the pair $(\bm{\mathcal E},Q_\varepsilon)$ is the same superposition of the pairs $(\bm\Psi_n,Q_n)$. Accordingly, $Q_\varepsilon$ has the dimensions of an inverse length, like the $Q_n$, and $\langle Q_\varepsilon\rangle_p=0$ follows from $\langle Q_n\rangle_p=0$. The deficit reads
	\begin{equation}
		\mathbf w_\varepsilon=\mathbf E-\bm{\mathcal E}=\sum_nb_n\,\frac{\varepsilon^2}{\sigma_n+\varepsilon^2}\,\bm\Psi_n\qquad\text{in }V_p,
		\label{eq:AT-deficit}
	\end{equation}
	again in the mean-square sense, since $\mathbf w_\varepsilon=\mathbf E\neq\mathbf 0$ on $S_p$.

	Projecting on $\mathbf E$ with Eq.~\eqref{eq:tortuosity-normalized-response}, we obtain the Stieltjes representation of the inverse dynamic tortuosity,
	\begin{equation}
		\frac{1}{\alpha(\omega)}
		=\sum_n\frac{b_n^2}{1+\varepsilon^2/\sigma_n}
		=\sum_n b_n^2\,\frac{-i\omega\Theta_n}{1-i\omega\Theta_n},
		\label{eq:AT-alpha}
	\end{equation}
	since $\varepsilon^2/\sigma_n=1/(-i\omega\Theta_n)$. With the dynamic permeability $k(\omega)=\phi\,\varepsilon^2/\alpha(\omega)$ [Eq.~\eqref{eq:tortuosity-scaled-viscous-field}], this is equivalent to the Avellaneda--Torquato form $k(\omega)/\phi=\sum_nb_n^2\sigma_n/(1-i\omega\Theta_n)$~\cite{AT,L-AT}.

	The two frequency limits are instructive.

	\emph{Low frequencies} ($|\varepsilon|\to\infty$). Then $a_n\simeq b_n\sigma_n/\varepsilon^2$, and
	\begin{equation}
		\varepsilon^2\bm{\mathcal E}\to\mathbf w_0=\sum_nb_n\sigma_n\bm\Psi_n,
		\qquad
		\varepsilon^2Q_\varepsilon\to\sum_nb_n\sigma_nQ_n=\pi_0-\Phi,
		\label{eq:AT-Darcy}
	\end{equation}
	where $(\mathbf w_0,\pi_0)$ is the Darcy (Stokes) solution of $-\nabla^2\mathbf w_0=-\nabla\pi_0+\mathbf e$, $\nabla\cdot\mathbf w_0=0$, $\mathbf w_0=\mathbf 0$ on $S_p$, $\langle\pi_0\rangle_p=0$. Thus $\pi\to\pi_0$: the decomposition~\eqref{eq:pressure-potential-decomposition}, with Eq.~\eqref{eq:AT-Q}, connects the ideal pressure $\Phi$ and the Darcy pressure $\pi_0$ and gives, with Eq.~\eqref{eq:AT-canonical}, the identity $\pi_0=\Phi+\sum_nb_n\sigma_nQ_n$ already stated in Ref.~\cite{L-AT} or Ref.~\cite{LAF_PhD}.

	\emph{High frequencies} ($\varepsilon\to0$). Then $a_n\to b_n$ and $\bm{\mathcal E}\to\mathbf E$ in the mean-square sense, but $Q_\varepsilon\to\sum_nb_nQ_n$ is a series that does not converge in general. Section~\ref{sec6} shows that $Q_\varepsilon$ grows as $\varepsilon^{-1}$, its coefficient being a harmonic potential driven by the boundary layer. The growth of $Q_\varepsilon$ is therefore not an artefact of the normalization~\eqref{eq:pressure-potential-decomposition}; it is the frequency-domain expression of the divergence of the modal sum of pressures.

	\subsection{Relaxation-time distribution and spectral zeta function}

	All the information on the visco-inertial response contained in Eq.~\eqref{eq:AT-alpha} is carried by the nondecreasing spectral function
	\begin{equation}
		G(\sigma)=\sum_{\sigma_n\le\sigma}b_n^2,
		\qquad
		G(\infty)=\frac{1}{\alpha_\infty},
		\label{eq:AT-G}
	\end{equation}
	whose normalized form $\alpha_\infty G$ is the cumulative distribution of relaxation times of Avellaneda and Torquato, expressed in the geometrical variable $\sigma=\nu\Theta$~\cite{AT}. Equation~\eqref{eq:AT-alpha} becomes the Stieltjes integral
	\begin{equation}
		\frac{1}{\alpha(\omega)}=\int_0^\infty\frac{dG(\sigma)}{1+\varepsilon^2/\sigma}.
		\label{eq:AT-Stieltjes}
	\end{equation}
This representation is automatically causal, but, being tied to the divergence-free simplification, it expresses more than causality: the response is purely relaxational, as discussed in Sec.~\ref{sec5-time}. It is convenient to encode the moments of this distribution in the spectral zeta function
	\begin{equation}
		\mathcal Z(s)=\sum_nb_n^2\,\sigma_n^{\,s}=\int_0^\infty\sigma^{s}\,dG(\sigma),
		\label{eq:AT-zeta}
	\end{equation}
	which has the dimensions of a length to the power $2s$. For nonnegative $s$ the series converges, and its values at integers are related to the tortuosity and the familiar low-frequency parameters~\cite{RFL}. Indeed, $\mathcal Z(0)=1/\alpha_\infty$ by Eq.~\eqref{eq:AT-E-expansion}; $\mathcal Z(1)=\mathbf e\cdot\langle\mathbf w_0\rangle_p=k_0/\phi$, the static permeability, by Eq.~\eqref{eq:AT-Darcy}; and $\mathcal Z(2)=\langle\mathbf w_0\cdot\mathbf w_0\rangle_p$, so that the static viscous tortuosity $\alpha_0=\langle\mathbf w_0\cdot\mathbf w_0\rangle_p/\langle\mathbf w_0\rangle_p^2$~\cite{LAF1} is $\alpha_0=\mathcal Z(2)/\mathcal Z(1)^2$. More generally, the low-frequency expansion of Eq.~\eqref{eq:AT-alpha} reads
	\begin{equation}
		\frac{1}{\alpha(\omega)}=\sum_{j\ge1}(-1)^{j+1}\,\frac{\mathcal Z(j)}{\varepsilon^{2j}},
		\qquad
		|\varepsilon|^2>\max_n\sigma_n ,
		\label{eq:AT-LF}
	\end{equation}
	so that the successive low-frequency parameters are the positive integer moments of the relaxation-time distribution. Inverting this series gives the low-frequency expansion of the tortuosity itself, $\alpha(\omega)=\varepsilon^2/\mathcal Z(1)+\alpha_0-\alpha_1^2\,\mathcal Z(1)/\varepsilon^2+\alpha_2^3\,\mathcal Z(1)^2/\varepsilon^4-\cdots$, which defines the dimensionless parameters $\alpha_1$, $\alpha_2,\dots$ of Ref.~\cite{RFL}, whose quantities $S_m$ are our $\mathcal Z(m)$. With $\alpha_0=\mathcal Z(2)/\mathcal Z(1)^2$,
	\begin{equation}
		\alpha_1^2=\frac{\mathcal Z(3)}{\mathcal Z(1)^3}-\alpha_0^2,
		\qquad
		\alpha_2^3=\frac{\mathcal Z(4)}{\mathcal Z(1)^4}-2\alpha_0\alpha_1^2-\alpha_0^3 .
		\label{eq:RFL-alphas}
	\end{equation}
	All the $\alpha_p^{p+1}$ are positive, as anticipated in Ref.~\cite{RFL}. For $\alpha_1^2$ this is the Cauchy--Schwarz inequality $\mathcal Z(2)^2\le\mathcal Z(1)\mathcal Z(3)$; in general it follows from the Stieltjes character of Eq.~\eqref{eq:AT-Stieltjes}, as will be shown elsewhere.

	The high-frequency parameters, by contrast, involve the behaviour of $G(\sigma)$ as $\sigma\to0$, that is, the short relaxation times of the modes confined near the walls. We shall find in Sec.~\ref{sec10} that $G(\sigma)\propto\sigma^{1/2}$ there, so that $\mathcal Z(s)$ diverges for $\operatorname{Re}s\le-\tfrac12$. The high-frequency parameters are then attached to the analytic continuation of $\mathcal Z$ to negative $s$: some to its poles, others to its regular values. This dictionary is the main subject of Sec.~\ref{sec10}.

	\subsection{Thermal relaxation modes}

	The thermal problem admits the same representation, with scalar Dirichlet modes
	\begin{equation}
		-\nabla^2\Psi'_n=\frac{1}{\sigma'_n}\Psi'_n\ \text{ in }V_p,
		\qquad
		\Psi'_n=0\ \text{ on }S_p,
		\qquad
		\langle\Psi'_n\Psi'_m\rangle_p=\delta_{nm},
		\qquad
		b'_n=\langle\Psi'_n\rangle_p ,
		\label{eq:AT-thermal-modes}
	\end{equation}
	and relaxation times $\Theta'_n=\sigma'_n\rho_0c_P/\kappa$. The ideal temperature pattern $1$ belongs to the mean-square closure of the functions vanishing on $S_p$, so that $1=\sum_nb'_n\Psi'_n$ and $\sum_nb'^2_n=1$. Projecting Eq.~\eqref{eq:normalized-thermal-equation} on $\Psi'_n$ gives
	\begin{equation}
		\mathcal T=\sum_n\frac{b'_n}{1+\varepsilon'^2/\sigma'_n}\,\Psi'_n\quad\text{in }V_p,
		\qquad
		\frac{1}{\alpha'(\omega)}=\sum_n\frac{b'^2_n}{1+\varepsilon'^2/\sigma'_n}=\int_0^\infty\frac{dG'(\sigma')}{1+\varepsilon'^2/\sigma'},
		\label{eq:AT-thermal}
	\end{equation}
	with $G'(\sigma')=\sum_{\sigma'_n\le\sigma'}b'^2_n$; here and below, primes label thermal quantities and never denote derivatives. The moments of this distribution are encoded in the thermal zeta function $\mathcal Z'(s)=\sum_nb'^2_n\sigma'^{\,s}_n=\int_0^\infty\sigma'^{\,s}\,dG'(\sigma')$. Then $\mathcal Z'(0)=1$, $\mathcal Z'(1)=k'_0/\phi$, $k'_0$ being the static thermal permeability, and $\alpha'_0=\mathcal Z'(2)/\mathcal Z'(1)^2$ is the static thermal tortuosity~\cite{LAF_PhD}, \cite{LLAT}. The low-frequency expansion of $1/\alpha'(\omega)$ is Eq.~\eqref{eq:AT-LF} with $\varepsilon$, $\mathcal Z$ and $\sigma_n$ replaced by $\varepsilon'$, $\mathcal Z'$ and $\sigma'_n$. In the time domain, $\sum_nb'^2_ne^{-t/\Theta'_n}$ is the fraction of diffusing heat (or, equivalently, of Brownian walkers initially uniformly distributed) that has not yet been absorbed by the isothermal walls at time $t$.

	\subsection{Time-domain reading: impulse response, ideal and visco-inertial parts}\label{sec5-time}

	Since $\varepsilon^2/\sigma_n=1/(-i\omega\Theta_n)$, the representation~\eqref{eq:AT-canonical} is the Fourier transform of a superposition of exponential relaxations. For a uniform impulsive force density $\mathbf f(t)=\rho_0v_0\mathbf e\,\delta(t)$, where $v_0$ is a constant velocity, the solution of the time-dependent problem associated with Eqs.~\eqref{eq:local-viscous-momentum}--\eqref{eq:local-viscous-no-slip} is
	\begin{equation}
		\frac{\mathbf v(t)}{v_0}=H(t)\sum_nb_n\bm\Psi_n\,e^{-t/\Theta_n},
		\qquad
		\frac{\widetilde p(t)}{\rho_0v_0}=\Phi\,\delta(t)+\nu H(t)\sum_nb_nQ_n\,e^{-t/\Theta_n}\qquad\text{in }V_p,
		\label{eq:AT-impulse}
	\end{equation}
	with $H$ the Heaviside function. Substitution confirms it: the impulsive terms reproduce $\mathbf E=\mathbf e-\nabla\Phi$, and the damped terms reproduce Eq.~\eqref{eq:AT-modes}. The response to a time-harmonic force is recovered by convolution, and Eq.~\eqref{eq:AT-alpha} follows. Causality is thereby built into the representation: since the impulse response~\eqref{eq:AT-impulse} vanishes for $t<0$, its Fourier transform~\eqref{eq:AT-alpha} is analytic and bounded in the upper half $\omega$-plane, each mode contributing a single pole at $\omega=-i/\Theta_n$. The positivity of the weights $b_n^2$ and of the relaxation times $\Theta_n$ adds more than causality and passivity: all singularities lie on the negative imaginary axis, and the averaged impulse response is completely monotonic. These properties come from the divergence-free simplification: the response is purely relaxational, without resonance, because the incompressible fluid in a rigid frame stores kinetic energy only, and no elastic energy; energy cannot oscillate between two reservoirs and can only be dissipated. The same holds for the thermal response Eq.~\eqref{eq:AT-thermal}, whose singularities lie at $\omega=-i/\Theta'_n$ on the negative imaginary axis, with positive weights $b'^2_n$: within the same simplification, the pressure is uniform at the pore scale, so that compression acts only as a uniform heat source, and the excess temperature obeys a pure diffusion equation and can only relax. These features will have to be reconsidered in a nonlocal theory~\cite{LAF1,LAF2,LAF_HF}, where the compressibility of the fluid at the pore scale provides a second, elastic reservoir and resonances may appear.

	The impulse response separates exactly into two motions of different nature~\cite{L-AT}. The first one, $\mathbf v_\infty=v_0\mathbf E\,H(t)$, $\widetilde p_\infty=\rho_0v_0\Phi\,\delta(t)$, is the ideal motion created instantaneously in the volume: the impulsive pressure $\Phi\delta(t)$ projects the excitation onto the admissible fields, $\mathbf E=\mathsf P\mathbf e$, and then disappears. The second one, $\mathbf v-\mathbf v_\infty=v_0H(t)\sum_nb_n\bm\Psi_n(e^{-t/\Theta_n}-1)$, is a visco-inertial motion without volume forcing (since $\nabla^2\mathbf E=\mathbf 0$), driven by the wall datum $-v_0\mathbf E$ which restores the no-slip condition, and carrying all the vorticity. The averaged velocity defines the relaxation function
	\begin{equation}
		R(t)=\frac{\mathbf e\cdot\langle\mathbf v(t)\rangle_p}{v_0}=\sum_nb_n^2\,e^{-t/\Theta_n},
		\qquad
		\frac{1}{\alpha(\omega)}=-i\omega\int_0^\infty R(t)\,e^{i\omega t}\,dt ,
		\label{eq:AT-relaxation-function}
	\end{equation}
	which decreases from $R(0^+)=1/\alpha_\infty$ to zero. In momentum terms, the impulse $\rho_0v_0\mathbf e$ delivered to the fluid is shared at $t=0^+$ between the visible fluid momentum $\rho_0v_0\mathbf e/\alpha_\infty$ and the momentum $\rho_0v_0(1-1/\alpha_\infty)\mathbf e$ transmitted reversibly to the frame by the impulsive wall pressure; the visible part is then transferred irreversibly to the frame by the viscous wall stresses and the visco-inertial pressure, at the rate $-\rho_0v_0\,dR/dt$.

	\subsection{Meaning of the modal sums}\label{sec5-convergence}

	Several modal sums encountered above converge only in a weak sense, and their manipulation requires care. Three facts govern them.

	(i) \emph{Mean-square versus pointwise convergence.} Expansions such as~\eqref{eq:AT-E-expansion} and~\eqref{eq:AT-deficit} converge in the mean-square sense. Near the walls, the partial sums must pass from the nonzero value of $\mathbf E$ to zero on $S_p$ over a distance that shrinks as more modes are retained. The convergence is therefore not uniform, and the value of a sum at a wall point must be understood as the limit of its value at interior points.

	(ii) \emph{Summation kernels.} The factors $e^{-t/\Theta_n}$ and $(1+\varepsilon^2/\sigma_n)^{-1}$ act as summation kernels that suppress the modes with $\sigma_n\ll\nu t$ or $\sigma_n\ll|\varepsilon|^2$. Limits such as $t\to0^+$ or $\varepsilon\to0$ cannot in general be interchanged with the summation. A statement such as ``$\sum_nc_n=0$'' is meaningful only as a regularized (Abel-type) limit, either in the time domain, $\lim_{t\to0^+}\sum_nc_ne^{-t/\Theta_n}$, or in the frequency domain, $\lim_{\varepsilon\to0}\sum_nc_n(1+\varepsilon^2/\sigma_n)^{-1}$, and must be checked as such. The two forms are related: with $s=-i\omega=\nu/\varepsilon^2$, $(1+\varepsilon^2/\sigma_n)^{-1}=s\int_0^\infty e^{-st}e^{-t/\Theta_n}\,dt$, so the frequency form is the Laplace transform of the time form multiplied by $s$, and the existence of the time limit implies that of the frequency limit, with the same value, when $\varepsilon\to0$ with $\varepsilon^2>0$ (that is, along the positive imaginary $\omega$-axis).

	(iii) \emph{Divergent sums with finite regularized values.} Since $G(\sigma)\propto\sigma^{1/2}$ as $\sigma\to0$, sums such as $\sum_nb_n^2/\sigma_n$, which represents $-\nu^{-1}dR/dt$ at $t=0^+$, diverge. Section~\ref{sec10} shows that such sums nevertheless possess well-defined regularized values, which are precisely the even-order high-frequency coefficients, whereas their divergent parts are governed by the odd-order ones.

	These facts will allow us, in Sec.~\ref{sec10}, to determine the short-time behaviour of the modal sums of pressures and velocities, and to correct the expectation, expressed in Ref.~\cite{L-AT}, that $\sum_nb_nQ_n$ and $\sum_nb_n\bm\Psi_n/\sigma_n$ vanish in the pore volume.

\section{Visco-Inertial Boundary-Layer Analysis in Gauss Coordinates}\label{sec6}

In this section we construct the complete high-frequency asymptotic solution of the deficit problem~\eqref{eq:deficit_gauss}, and we use it to determine the successive coefficients of the high-frequency expansion of the dynamic tortuosity. Following the convention of Ref.~\cite{LBF} [Eq.~(10a) there], these coefficients are written
\begin{equation}
	\frac{\alpha(\omega)}{\alpha_\infty}=1+\frac{2\varepsilon}{\Lambda}+\frac{3\varepsilon^2}{\Sigma}+\frac{15\,\varepsilon^3}{4V}+O(\varepsilon^4),
	\label{eq:alpha-expansion-LSV}
\end{equation}
where the numerical factors are chosen so that, for straight circular pores of radius $R$, $\Lambda=R$, $\Sigma=R^2$ and $V=R^3$. The general expressions obtained below for $\Lambda$, $\Sigma$ and $V$ indeed reduce to these values in that case, as checked in Sec.~\ref{sec6-checks}. The characteristic length $\Lambda$ is that of Johnson, Koplik and Dashen; $\Sigma$ and $V$ are the viscous surface and volume parameters.

The response $\alpha(\omega)$ obeys five general properties: causality, reality, passivity, reciprocity, and, within the divergence-free simplification, a purely relaxational structure. Their consequences for the high-frequency parameters introduced above are limited but precise: these parameters are real, $\Lambda$ is positive, and all the following ones, starting with $\Sigma$ and $V$, may take either sign.

	Causality requires $\alpha(\omega)$, and equivalently its inverse or the dynamic permeability $k(\omega)=\phi\,\varepsilon^2/\alpha(\omega)$, to be analytic in the upper half $\omega$-plane (see Landau and Lifshitz~\cite[\S\,62, pp.~256--263]{LALIel} and~\cite[\S\,125, pp.~384--390]{LALIstp}). Reality, the fact that a real-valued excitation produces a real-valued response, implies the reflection symmetry $\alpha(-\omega^*)=\alpha^*(\omega)$, as emphasized by Johnson, Koplik and Dashen~\cite{JKD}, citing Landau and Lifshitz~\cite[p.~257]{LALIel}. Together, they make the coefficients real. On the positive imaginary axis, $\omega=iy$ with $y>0$, $\alpha$ is real by the reflection symmetry, and so is $\varepsilon=(\nu/y)^{1/2}$. Since $\alpha$ has no singularity in the upper half-plane, where $|\arg\varepsilon|<\pi/4$, and since the high-frequency solution constructed below does not require $\omega$ to be real but holds throughout this half-plane [Eq.~\eqref{eq:sector} and Sec.~\ref{sec6-twoscale}], the expansion~\eqref{eq:alpha-expansion-LSV} holds there, in particular on the positive imaginary axis; as the asymptotic expansion of a real function in powers of a real variable, it has real coefficients. Johnson, Koplik and Dashen drew this conclusion for $\alpha_\infty$ and $\Lambda$; it extends to all orders.

	Passivity, that is, positive dissipation, fixes the sign of the leading dissipative term. For real $\omega>0$, $\varepsilon=(1+i)(\nu/2\omega)^{1/2}$ and $\mathrm{Im}\,\alpha\simeq(2\alpha_\infty/\Lambda)(\nu/2\omega)^{1/2}$, so that positive absorption requires $\Lambda>0$. The following parameters, starting with $\Sigma$ and $V$, enter at higher orders, where the sign of the absorption is already fixed by the $\Lambda$ term; passivity does not constrain them.

	The divergence-free simplification, together with passivity, restricts the singularities much further. As shown in Appendix~A of Ref.~\cite{JKD}, a nontrivial free motion of the incompressible fluid in the rigid frame, which stores kinetic energy only and dissipates it by viscosity, can only decay without oscillating, so that all poles, zeros and branch points of $\alpha(\omega)$ lie on the negative imaginary axis. The same argument, applied to heat diffusion with a pore-scale uniform source, places those of $\alpha'(\omega)$ on the same axis~\cite{LAF_PhD}. Reciprocity makes the corresponding weights positive. In the divergence-free setting, it takes the form of the reciprocal theorem of Lorentz for the unsteady Stokes equations~\cite{LOR96,HB}: if two force densities $\mathbf f_1$ and $\mathbf f_2$ acting on the fluid produce the velocities $\mathbf v_1$ and $\mathbf v_2$, then $\langle\mathbf f_1\cdot\mathbf v_2\rangle_p=\langle\mathbf f_2\cdot\mathbf v_1\rangle_p$. It holds because the kinetic energy and the viscous dissipation are both symmetric quadratic forms of the divergence-free velocity, while the pore-scale pressure correction $\widetilde p$, which only enforces the constraint, does no net work on divergence-free fields with zero normal component on the walls and periodic (or stationary) over the REV. As a consequence, the uniform force excites the mode $n$ with the amplitude $b_n=\mathbf e\cdot\langle\bm\Psi_n\rangle_p$, and the mean velocity along $\mathbf e$ reads this mode with the same amplitude $b_n$, so that its weight in the response is $b_n\cdot b_n=b_n^2\geq0$; without reciprocity, the two amplitudes would differ and their product could take either sign. The relaxation-mode representation of Sec.~\ref{sec5} makes these properties explicit: a sum of simple poles at $\omega=-i/\Theta_n$ with positive weights $b_n^2$.

	The divergence-free simplification, passivity and reciprocity thus fix the signs of the relaxation times and of the weights, but not those of the parameters beyond $\Lambda$; what they give the parameters is a common status. Expanded in powers of $\varepsilon$, the inverse tortuosity $1/\alpha$ has odd-order coefficients that are data of the short-time end, $\sigma\to0$, of a single positive relaxation-time distribution $G(\sigma)$ [Eq.~\eqref{eq:AT-G}], and even-order coefficients that are sum rules over it (Sec.~\ref{sec10}). This confirms $\Lambda>0$: since $G$ vanishes at $\sigma=0$ and is nondecreasing, the coefficient of its leading behaviour $G\simeq[4/(\pi\alpha_\infty\Lambda)]\,\sigma^{1/2}$ [Eq.~\eqref{eq:G-tail}] is positive. It leaves, however, the signs of $\Sigma$, $V$ and the following parameters free. The subleading tail coefficients are not constrained in sign, and the even-order coefficients are finite parts of divergent sums of positive terms; such finite parts may be negative (e.g., $\mathcal Z(-1)=-1/R^2$ for the circular pore). For instance, for an in-parallel bundle of disjoint straight circular pores aligned with $\mathbf e$, of radii $R_i$ occupying the fractions $p_i$ of the pore volume, $\mathbf E=\mathbf e$, $\alpha_\infty=1$, and $1/\alpha$ is the volume-weighted sum of the responses of the tubes. For a single tube, for which $\Lambda=R$, $\Sigma=R^2$ and $V=R^3$, inverting Eq.~\eqref{eq:alpha-expansion-LSV} gives $1/\alpha=1-2\varepsilon/R+\varepsilon^2/R^2+\varepsilon^3/(4R^3)+O(\varepsilon^4)$. With $m_k=\langle R^{-k}\rangle=\sum_ip_iR_i^{-k}$, the bundle thus has $1/\alpha=1-2m_1\varepsilon+m_2\varepsilon^2+m_3\varepsilon^3/4+O(\varepsilon^4)$, and inverting this series gives
	\begin{equation}
		\Lambda=\frac1{m_1},\qquad
		\frac3\Sigma=4m_1^2-m_2,\qquad
		\frac{15}{4V}=8m_1^3-4m_1m_2-\frac{m_3}4 ;
		\label{eq:bundle}
	\end{equation}
	$\Sigma$ becomes negative as soon as $m_2>4m_1^2$, that is, for a sufficiently broad in-parallel distribution of radii, consistently with the in-parallel interpretation of the negative values reported in Ref.~\cite{LBF} (Sec.~\ref{sec7}).

The deficit problem~\eqref{eq:deficit_gauss} is solved by expanding its two unknown fields, the deficit $\mathbf w_\varepsilon$ and the pressure potential $Q_\varepsilon$, in powers of $\varepsilon$. The two fields do not behave in the same way. Since the wall is smooth, nothing varies rapidly along it, and the only rapid variable is the stretched normal variable
\begin{equation}
\varsigma=\zeta/\varepsilon .
\label{eq:stretched-variable}
\end{equation}
This variable is complex. For $\omega$ in the closed upper half-plane, the root of $\varepsilon^2=\nu/(-i\omega)$ [Eq.~\eqref{eq:complex-skin-depths}] with positive real part satisfies $|\arg\varepsilon|\le\pi/4$, so that, for $0\le\zeta<\zeta_S$, $\varsigma$ lies in the closed sector
\begin{equation}
\mathcal C=\{\varsigma\in\mathbb C:\ |\arg\varsigma|\le\pi/4\}
\label{eq:sector}
\end{equation}
of the complex plane. This sector contains every value of $\zeta/\varepsilon$ that occurs for these frequencies: $\varsigma$ is real and positive for $\omega$ on the positive imaginary axis, where $\varepsilon$ is real and positive, and lies on the edge $\arg\varsigma=-\pi/4$ for real $\omega>0$, where $\varepsilon=|\varepsilon|\,e^{i\pi/4}$. In $\mathcal C$, $\mathrm{Re}\,\varsigma\ge|\varsigma|/\sqrt2$, so that $|e^{-\varsigma}|\le e^{-|\varsigma|/\sqrt2}$: a function that decays like $e^{-\varsigma}$ tends to zero as $|\varsigma|\to\infty$ in $\mathcal C$, uniformly in the direction of $\varsigma$. All the limits $|\varsigma|\to\infty$ below are taken in this sense.

The momentum equation of the deficit contains the operator $\varepsilon^2\Delta-1$, and with it a length of its own, the skin depth $|\varepsilon|$: near a plane wall along which the slip $\mathbf E$ is uniform, the deficit is the Stokes-layer profile $\mathbf E\,e^{-\zeta/\varepsilon}$. In a general pore, it varies in the normal direction on two scales at once, over the skin depth and on the geometric scales, through the curvature of the wall and the nonuniformity of the wall slip. The pressure potential, on the contrary, is harmonic at every frequency [Eq.~\eqref{eq:Q-harmonic}], and Laplace's equation has no length of its own: its wall data varying on the geometric scales only, it varies on these scales alone. Accordingly, the deficit is sought as a two-scale series, whose coefficients depend on the position and on $\varsigma$, and the pressure potential as an ordinary power series, whose coefficients depend on the position only:
\begin{equation}
(w_\varepsilon)^\mu=\sum_{m\ge0}\varepsilon^m\,W^\mu_{(m)}(x^\nu;\zeta/\varepsilon),\qquad
Q_\varepsilon=\sum_{m\ge0}\varepsilon^{m-1}\,Q^{(m)}(x^\nu).
\label{eq:expansion-principle}
\end{equation}
Section~\ref{sec6-pressure} justifies the form of the second series, explains why it begins at $\varepsilon^{-1}$, and splits each coefficient $W^\mu_{(m)}$ into a gradient flow, which extends over the whole pore volume, and a part confined to a boundary layer of thickness $|\varepsilon|$. This part is constructed by the two-scale method of Burridge and Keller~\cite{BK} (Sec.~\ref{sec6-twoscale}), and the coefficients of the two series, then those of $1/\alpha$, are determined in Secs.~\ref{sec6-hierarchy}--\ref{sec6-reduction}.

\subsection{Near-wall operators}\label{sec6-operators}

The boundary-layer analysis is carried out in the near-wall patches of Sec.~\ref{sec3}. There, the components $a^\mu{}_{;\nu}$ of the covariant derivative of a vector field $a^\mu$ are, by Eqs.~\eqref{eq:christoffel_a}--\eqref{eq:christoffel_c} and~\eqref{eq:surface-operators},
\begin{equation}
a^i{}_{;j}=a^i{}_{|j}+\Gamma^i_{3j}a^3,\qquad
a^3{}_{;j}=\partial_ja^3+\Gamma^3_{jk}a^k,\qquad
a^i{}_{;3}=-\partial_\zeta a^i+\Gamma^i_{3k}a^k,\qquad
a^3{}_{;3}=-\partial_\zeta a^3 .
\label{eq:cov-components}
\end{equation}
Here the vertical bar denotes the covariant derivative of the parallel surface through the point considered [Eq.~\eqref{eq:surface-operators}]: $a^i{}_{|j}=\partial_ja^i+\Gamma^i_{jk}a^k$ and, at second order, $a^i{}_{|j|k}=\partial_k(a^i{}_{|j})+\Gamma^i_{k\ell}\,a^\ell{}_{|j}-\Gamma^\ell_{kj}\,a^i{}_{|\ell}$, while the normal component $a^3$, like any scalar field $f$, is a scalar on the parallel surfaces, so that $a^3{}_{|j}=\partial_ja^3$ and $f_{|j|k}=\partial_k\partial_jf-\Gamma^\ell_{kj}\,\partial_\ell f$. The bar derivatives collect the Christoffel symbols whose indices are all tangential; the others, which carry a $3$, couple the tangential and normal components. For a tangent field $s^i$ on $S_p$, for instance, $s^i{}_{;j}=s^i{}_{|j}$, but $s^3{}_{;j}=\Gamma^3_{jk}s^k$, which does not vanish in general. A direct computation of the vector Laplacian~\eqref{eq:vector_laplacian}, $(\Delta a)^\mu=g^{jk}a^\mu{}_{;j;k}+a^\mu{}_{;3;3}$, using the flatness relations~\eqref{eq:riccati-codazzi}--\eqref{eq:codazzi-contracted} and the symmetry of $g^{jk}\Gamma^i_{3k}$, then yields, for any vector field $a^\mu$ and any scalar field $f$ in $\mathcal V_S$,
\begin{align}
(\Delta a)^i&=\partial_\zeta^2a^i-\big(2\Gamma^i_{3\ell}+\delta^i_\ell\Gamma^k_{3k}\big)\partial_\zeta a^\ell+g^{jk}a^i{}_{|j|k}+\big(\Gamma^k_{3k}\Gamma^i_{3\ell}-\Gamma^i_{3k}\Gamma^k_{3\ell}\big)a^\ell\nonumber\\
&\qquad+2g^{jk}\Gamma^i_{3j}\,\partial_ka^3+g^{ij}(\partial_j\Gamma^k_{3k})\,a^3,\label{eq:lap-tan}\\
(\Delta a)^3&=\partial_\zeta^2a^3-\Gamma^k_{3k}\,\partial_\zeta a^3+g^{jk}a^3{}_{|j|k}-2\Gamma^k_{3\ell}\,a^\ell{}_{|k}-(\partial_\ell\Gamma^k_{3k})\,a^\ell-\Gamma^k_{3\ell}\Gamma^\ell_{3k}\,a^3,\label{eq:lap-nor}\\
a^\mu{}_{;\mu}&=a^k{}_{|k}+\Gamma^k_{3k}\,a^3-\partial_\zeta a^3,\qquad
\Delta f=\partial_\zeta^2f-\Gamma^k_{3k}\,\partial_\zeta f+g^{jk}f_{|j|k}.\label{eq:div-grad}
\end{align}

\subsection{The pressure potential has no boundary layer}\label{sec6-pressure}

We keep the decomposition $\pi=\Phi+\varepsilon^2Q_\varepsilon$ of Eq.~\eqref{eq:pressure-potential-decomposition}. The deficit problem is Eq.~\eqref{eq:deficit_gauss}. Taking the divergence of Eq.~\eqref{eq:deficit_gauss_mom}, and using the incompressibility of $w_\varepsilon$ together with the commutation of divergence and Laplacian in Euclidean space, we find that the pressure potential is harmonic at every frequency:
\begin{equation}
\Delta Q_\varepsilon=0\qquad\text{in }V_p .
\label{eq:Q-harmonic}
\end{equation}

This exact property provides the natural choice of the two parts of the decomposition~\eqref{eq:deficit-vortical-potential}. The gradient $\varepsilon^2\nabla Q_\varepsilon$ meets both requirements imposed there on the potential part: it derives from a single-valued, stationary potential, $\varepsilon^2Q_\varepsilon$, and its vector Laplacian vanishes, since $\nabla^2\nabla Q_\varepsilon=\nabla\nabla^2Q_\varepsilon=\mathbf 0$. We therefore set
\begin{equation}
\chi_\varepsilon=\varepsilon^2Q_\varepsilon,\qquad
\mathbf w_{\mathrm{pot},\varepsilon}=\varepsilon^2\nabla Q_\varepsilon,\qquad
\mathbf w_{\mathrm{vort},\varepsilon}=\mathbf w_\varepsilon-\varepsilon^2\nabla Q_\varepsilon .
\label{eq:exact-splitting}
\end{equation}
Substituting into Eq.~\eqref{eq:deficit_gauss}, the pressure terms cancel, and the vortical part satisfies exactly
\begin{subequations}\label{eq:vort-problem}
\begin{align}
\varepsilon^2g^{\rho\sigma}(w_{\mathrm{vort},\varepsilon})^{\mu}{}_{;\rho;\sigma}-(w_{\mathrm{vort},\varepsilon})^{\mu}&=0 &&\text{in }V_p,\label{eq:vort-problem-mom}\\
(w_{\mathrm{vort},\varepsilon})^{\mu}{}_{;\mu}&=0 &&\text{in }V_p,\\
(w_{\mathrm{vort},\varepsilon})^{\mu}&=E^{\mu}-\varepsilon^2g^{\mu\nu}Q_{\varepsilon;\nu} &&\text{on }S_p .\label{eq:vort-problem-bc}
\end{align}
\end{subequations}
The two parts are coupled only at the wall. The normal component of Eq.~\eqref{eq:vort-problem-bc}, with $E^3=0$ on $S_p$, gives the Neumann datum of the pressure potential,
\begin{equation}
Q_{\varepsilon;3}=-\varepsilon^{-2}\,(w_{\mathrm{vort},\varepsilon})^{3}\qquad\text{on }S_p ,
\label{eq:Q-neumann}
\end{equation}
while its tangential components give the wall slip seen by the vortical part, $\mathbf E$ corrected by the tangential gradient of the pressure.

Two conclusions follow at once. First, $\mathbf w_{\mathrm{vort},\varepsilon}$ is a boundary-layer field. Equation~\eqref{eq:vort-problem-mom} has the same form as the thermal deficit equation~\eqref{eq:thermal_deficit_gauss_pde}, with a vector field in place of a scalar. In both, the operator $\varepsilon^2\Delta-1$ \emph{screens} the wall data: a value imposed on the wall is not transmitted through the pore, as it would be by Laplace's equation, but decays away from it like $e^{-\zeta/\varepsilon}$, $\zeta$ being the distance to the wall, that is, over a distance of the order of $|\varepsilon|$. Hence $\mathbf w_{\mathrm{vort},\varepsilon}$ is exponentially small at distances from $S_p$ large compared with $|\varepsilon|$. It carries all the vorticity of the deficit. Second, the pressure potential has \emph{no} boundary layer. It is harmonic, and its Neumann datum~\eqref{eq:Q-neumann} is a smooth function on $S_p$ which, the wall being smooth, varies along the wall only on geometric scales. A harmonic function cannot vary rapidly in the normal direction while varying slowly along the wall, since, in $\Delta Q_\varepsilon=\partial_\zeta^2Q_\varepsilon-\Gamma^k_{3k}\,\partial_\zeta Q_\varepsilon+g^{jk}Q_{\varepsilon|j|k}=0$ [Eq.~\eqref{eq:div-grad}], $\partial_\zeta^2Q_\varepsilon$ is balanced by tangential derivatives and curvature terms. Hence $Q_\varepsilon$ varies on the geometric scales only, throughout $V_p$ and up to the wall, and its expansion in powers of $\varepsilon$ is an ordinary power series, whose coefficients are functions of position.

\begin{figure}[t]
		\centering
		\includegraphics[width=0.55\textwidth]{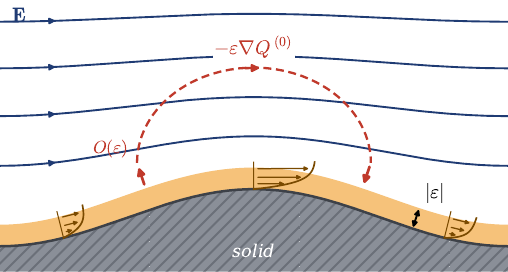}
		\caption{Visco-inertial boundary layer on a curved wall (schematic). The ideal flow $\mathbf E$ (blue) slips along the wall; the boundary layer, of thickness $|\varepsilon|$ (shaded), restores no slip, as shown by the velocity profiles. Where the slip increases or decreases along the wall ($E^k{}_{|k}\neq0$), the flux deficit carried by the layer changes, and mass conservation forces a transpiration velocity of order $\varepsilon$ through its outer edge (red arrows). This normal velocity is the wall datum of the potential displacement flow, which perturbs the core velocity by $-\varepsilon\nabla Q^{(0)}$ (dashed) over the whole pore volume; the pressure potential that drives it has no boundary layer.}
		\label{fig:bl}
	\end{figure}

The size of $Q_\varepsilon$ follows from Eq.~\eqref{eq:Q-neumann}. Within a layer of thickness $|\varepsilon|$, incompressibility makes the normal component of $\mathbf w_{\mathrm{vort},\varepsilon}$ of order $\varepsilon$. This point, on which the whole construction rests, is a simple mass balance (Fig.~\ref{fig:bl}). On the wall, the tangential part of $\mathbf w_{\mathrm{vort},\varepsilon}$ is the slip $\mathbf E$, of order unity, and it varies along the wall only on geometric scales, so that its tangential divergence is of order unity as well. Since $\mathbf w_{\mathrm{vort},\varepsilon}$ is divergence-free, this tangential divergence must be balanced by the normal derivative of the normal component, which is therefore of order unity; integrated across the layer, outside which $\mathbf w_{\mathrm{vort},\varepsilon}$ vanishes, it yields a normal component of order $\varepsilon$ at the wall. Physically, where the deficit flux carried by the layer converges or diverges along the wall, fluid must leave or enter the layer through its outer edge; the layer being thin, the corresponding normal velocity, the \emph{transpiration}, is of order $\varepsilon$. As the total deficit has zero normal component on the wall, the transpiration must be compensated there by the potential part, and Eq.~\eqref{eq:Q-neumann} gives $Q_{\varepsilon;3}=O(\varepsilon^{-1})$. In the core, this wall datum drives the potential displacement flow discussed below. As anticipated in Sec.~\ref{sec5}, $Q_\varepsilon$ is in general not bounded as $\varepsilon\to0$, and its ordinary power series~\eqref{eq:expansion-principle} begins at order $\varepsilon^{-1}$, $Q_\varepsilon=\varepsilon^{-1}Q^{(0)}+Q^{(1)}+\varepsilon\,Q^{(2)}+\cdots$, with
\begin{equation}
\Delta Q^{(m)}=0\ \text{in }V_p,\qquad\langle Q^{(m)}\rangle_p=0,
\label{eq:Q-conditions}
\end{equation}
where the \emph{pressure potentials} $Q^{(m)}(x^\mu)$ are ordinary functions of position, whose Neumann data are determined below by the boundary layer. The product $\varepsilon^2Q_\varepsilon$, which is what enters the pressure, is of order $\varepsilon$. The superscript $(m)$ distinguishes these potentials from the modal pressures $Q_n$ of Sec.~\ref{sec5}; in modal terms, $Q^{(0)}$ is the coefficient of $\varepsilon^{-1}$ in the high-frequency expansion of the sum $\sum_na_n(\varepsilon)Q_n$ of Eq.~\eqref{eq:AT-Q}.

At fixed distance from $S_p$, the vortical part is exponentially small, and \emph{the deficit reduces to the gradient of $\varepsilon^2Q_\varepsilon$},
\begin{equation}
\mathbf w_\varepsilon=\varepsilon^2\nabla Q_\varepsilon+O(\varepsilon^\infty)\qquad\text{at fixed distance from }S_p .
\label{eq:core-deficit}
\end{equation}
This is the inviscid balance $-i\omega\rho_0\mathbf v=-\nabla\widetilde p-\nabla\langle p\rangle_p$ of the core flow, in which the viscous term vanishes because $\nabla^2\mathbf E=\mathbf 0$ and $\nabla^2\nabla Q_\varepsilon=\mathbf 0$. The deficit flow in the core is the potential \emph{displacement flow} induced by the boundary layer. This flow was identified in Ref.~\cite{CSGL}. It was obtained there through an electrical analogy that makes the mechanism transparent. Within the boundary layer the fluid behaves as a conductor of reduced conductivity, $\sigma=\varepsilon^2(1-e^{-\zeta/\varepsilon})$ [their Eq.~(28c)], so that the walls are lined by a thin layer of different conductivity. Where the tangential current carried by this layer varies along the wall, current conservation forces a normal current into the bulk, as at an interface carrying a surface conductivity; this normal current is the transpiration, and the resulting perturbation of the bulk field is the displacement flow. At leading order it appears there as a perturbation $\varepsilon\mathbf N$ of the electric field, with $\mathbf N$ harmonic, orthogonal to $\mathbf E$ in the mean, and with wall datum $\mathbf N\cdot\mathbf n=\partial E_\zeta/\partial\zeta$, where $\zeta$ is the distance to the wall (denoted $\beta$ there) and $E_\zeta$ the component of $\mathbf E$ along the inward normal. Since $E_\zeta=0$ on $S_p$ and $\nabla\cdot\mathbf E=0$, this datum equals $-E^k{}_{|k}$, and $\mathbf N=-\nabla Q^{(0)}$, the first pressure potential being given by Eq.~\eqref{eq:Qp0} below.

In the linear average $\langle\mathbf w_\varepsilon\cdot\mathbf e\rangle_p$, the bulk contribution of this flow is of the same order as that of the boundary layer; its omission is the origin of the incomplete expression of $\Lambda$ found in Ref.~\cite{SZ}. Avellaneda and Torquato [Ref.~\cite{AT}, Appendix~D] observed that the boundary-layer analysis of Ref.~\cite{SZ} had to be carried to the next order, and found that the normal velocity of order $\varepsilon^3$ (in their normalization) driven by $E^k{}_{|k}$ does not decay across the layer. Their analysis is nevertheless not consistent with the present solution. The velocity and the total pressure are expanded there with coefficients depending only on the tangential coordinates and on the stretched normal variable, the metric being evaluated on the wall, so that any slow normal dependence is excluded. The non-decaying normal velocity is in fact the wall trace of the displacement flow $\varepsilon\mathbf N=-\varepsilon\nabla Q^{(0)}$, a harmonic field that varies on the geometric scale throughout the pore volume and cannot be represented in that form. In their analysis, this velocity is extended as a constant across the slab of finite thickness over which the average is taken, and produces in the linear average a volume term that does not reduce to a wall integral [Eq.~(D34) there]. This term was eliminated by projecting on $\mathbf E$, which is legitimate and yields the correct $\Lambda$, because the displacement flow is orthogonal to $\mathbf E$ in the mean; but the bulk nature of the displacement flow, and hence the origin of the incomplete expression of $\Lambda$ of Ref.~\cite{SZ}, was established only in Ref.~\cite{CSGL}.

In the present solution, the displacement flow and the boundary layer are separated from the outset: the pressure potential has no boundary layer and carries the displacement flow in the whole pore volume, whereas the vortical deficit is confined to the layer and varies in the normal direction on both the skin depth and the geometric scales, the latter through the transport equations~\eqref{eq:transport}.

In the series~\eqref{eq:expansion-principle}, the limit of each coefficient $W^\mu_{(m)}$ as $|\varsigma|\to\infty$ is therefore $g^{\mu\nu}Q^{(m-1)}_{;\nu}$, with $Q^{(-1)}=0$, and what remains, which decays as $|\varsigma|\to\infty$, is the contribution of the vortical deficit. The problem is thereby reduced to a boundary-layer analysis of $\mathbf w_{\mathrm{vort},\varepsilon}$ alone, the pressure entering only through its wall values.

\subsection{Two-scale expansion of the vortical deficit}\label{sec6-twoscale}

\paragraph{Two-scale form.} It remains to construct the contribution of the vortical deficit to the coefficients $W^\mu_{(m)}$ of the two-scale series~\eqref{eq:expansion-principle}. In this series, the parameter $\varepsilon$ enters the coefficients through their fast argument $\zeta/\varepsilon$, and not only through the powers $\varepsilon^m$: for the deficit, Eq.~\eqref{eq:expansion-principle} is a \emph{generalized} asymptotic expansion, in the sense of Erd\'elyi and Wyman~\cite{EW}. This raises two difficulties. First, unlike an expansion whose coefficients are independent of $\varepsilon$, it is not unique. Since $\zeta=\varepsilon\varsigma$, a dependence on $\zeta$ can be transferred from the slow argument of a coefficient to its fast argument at the cost of one power of $\varepsilon$, so that a given field admits infinitely many two-scale series of this form: a contribution $\zeta f(q)\,e^{-\zeta/\varepsilon}$ to a component, for instance, may be represented either by the term $\zeta f(q)\,e^{-\varsigma}$ in the coefficient of $\varepsilon^0$, or by the term $\varsigma f(q)\,e^{-\varsigma}$ in the coefficient of $\varepsilon^1$. Second, substituted into the field equations, the two-scale form does not give one equation per power of $\varepsilon$: a derivative with respect to $\zeta$ acts on both arguments at once, on the fast one with a factor $\varepsilon^{-1}$, and a sum $\sum_m\varepsilon^mR_{(m)}(x^\nu;\zeta/\varepsilon)$ may vanish without its coefficients vanishing, as the example just given shows. The two-scale principle of Burridge and Keller, introduced below, removes the second difficulty, but not the first, which is removed by a closure (Sec.~\ref{sec6-closure}).

\paragraph{Abstract space and abstract field.} In the two-scale series~\eqref{eq:expansion-principle}, the first argument of each coefficient accounts for the slow variations of the deficit, and the second for its rapid variations. Following Burridge and Keller~\cite{BK}, we consider the vortical deficit itself to be a function of these two sets of variables, the position $x^\nu$ and the stretched normal variable $\varsigma$, which, however, we consider to be independent. We are thus extending the definition of the field from the tubular neighbourhood $\mathcal V_S$ of Eq.~\eqref{eq:tubular-neighbourhood}, a region of the three-dimensional physical space in which the position is represented by $(q,\zeta)$, to the product $\mathcal V_S\times\mathcal C$ of the pairs $(x^\nu;\varsigma)$, which we call the \emph{abstract space}. Its additional variable $\varsigma$ is complex and ranges over the sector $\mathcal C$ of Eq.~\eqref{eq:sector}, which contains all the values of $\zeta/\varepsilon$ for $\omega$ in the closed upper half-plane: the construction applies directly at all these frequencies, the real ones included, without analytic continuation. The abstract fields are taken to be analytic functions of $\varsigma$, so that $\partial/\partial\varsigma$ is a complex derivative and the chain rule~\eqref{eq:chain-rule} below holds on the diagonal, along which $\varsigma=\zeta/\varepsilon$ runs on the ray $\arg\varsigma=-\arg\varepsilon$; the coefficients obtained below, products of $e^{-\varsigma}$ and of amplitudes independent of $\varsigma$ [Eq.~\eqref{eq:C-form}], are entire functions of $\varsigma$. The extended field, the \emph{abstract field}, is denoted by an overline, $\overline w^{\,\mu}(x^\nu;\varsigma)$; its coefficients, the overlined coefficients $\overline w^{\,\mu}_{(m)}$, are functions of the independent variables $(x^\nu;\varsigma)$. The actual physical field is then obtained by restricting the abstract field to the \emph{physical diagonal}, the subset of the abstract space on which $\varsigma$ takes the value $\zeta/\varepsilon$:
\begin{equation}
(w_{\mathrm{vort},\varepsilon})^\mu(x^\nu)=\overline w^{\,\mu}(x^\nu;\zeta/\varepsilon),\qquad
\overline w^{\,\mu}=\sum_{m\ge0}\varepsilon^m\,\overline w^{\,\mu}_{(m)}(x^\nu;\varsigma),\qquad
\overline w^{\,\mu}_{(m)}\to0\ \text{as }|\varsigma|\to\infty .
\label{eq:two-scale-substitution}
\end{equation}
The order $m$ is written in parentheses, so that it cannot be mistaken for a tensor index, and the subscripts of $\mathbf w_{\mathrm{vort},\varepsilon}$ are omitted, the vortical deficit being the only field represented in the abstract space. The position is confined to $\mathcal V_S$ because the diagonal is defined through the normal coordinate $\zeta$ of the near-wall patches; on the diagonal, the abstract field therefore gives the physical field in $\mathcal V_S$ only. This is immaterial: the vortical deficit is exponentially small at distances from $S_p$ large compared with $|\varepsilon|$ (Sec.~\ref{sec6-pressure}), so that the physical field so obtained may be multiplied by a smooth function of position, equal to one near the wall and vanishing near the outer boundary $\zeta=\zeta_S$ of $\mathcal V_S$, and extended by zero to the rest of $V_p$, with an error $O(\varepsilon^\infty)$. The pressure potential, which has no boundary layer (Sec.~\ref{sec6-pressure}), is not represented in the abstract space: it enters only through the wall condition, as its ordinary series~\eqref{eq:expansion-principle}.

\paragraph{Overlined operators.} The equations are extended in the same way, each derivative being replaced by its overlined version~\eqref{eq:chain-rule}. This replacement is dictated by the chain rule. On the physical diagonal, $\varsigma=\zeta/\varepsilon$, and $\partial_\nu(\zeta/\varepsilon)=-\delta^3_\nu/\varepsilon$. For any abstract field $\overline F(x^\nu;\varsigma)$ and its restriction $F(x^\nu)=\overline F(x^\nu;\zeta/\varepsilon)$ to the diagonal, the chain rule therefore gives $\partial_\nu F=\big(\overline\partial_\nu\overline F\big)\big|_{\varsigma=\zeta/\varepsilon}$, with the \emph{overlined} derivatives
\begin{equation}
\overline\partial_\nu=\partial_\nu-\frac{\delta^3_\nu}{\varepsilon}\,\frac{\partial}{\partial\varsigma}
\quad\Big(\text{i.e. }\overline\partial_i=\partial_i,\ \ \overline\partial_\zeta=\partial_\zeta+\frac1\varepsilon\frac{\partial}{\partial\varsigma}\Big),
\qquad
(\cdot)_{\ocd{\nu}}=(\cdot)_{;\nu}-\frac{\delta^3_\nu}{\varepsilon}\,\frac{\partial}{\partial\varsigma}.
\label{eq:chain-rule}
\end{equation}
Every differential operator acting on the physical field has an overlined counterpart acting on abstract fields, the \emph{overlined operator}, obtained by replacing each derivative by its overlined version. For a vector field, the overlined covariant derivative is written with an overlined semicolon, $\overline w^{\,\mu}{}_{\ocd{\nu}}$ [Eq.~\eqref{eq:chain-rule}], to distinguish it from the ordinary covariant derivative $\overline w^{\,\mu}{}_{;\nu}$, taken at fixed $\varsigma$; the overlined divergence and Laplacian are $\overline w^{\,\mu}{}_{\ocd{\mu}}$ and $(\overline\Delta\,\overline w)^\mu=g^{\rho\sigma}\,\overline w^{\,\mu}{}_{\ocd{\rho}\ocd{\sigma}}$. The metric coefficients and Christoffel symbols depend on $x^\mu$ only; they are kept exact and are \emph{not} expanded in powers of $\zeta=\varepsilon\varsigma$. Because $\partial_\varsigma$ commutes with $\partial_\zeta$, $\partial_i$ and with these coefficients, the overlined operators compose like the ordinary ones, and each of them, applied to an abstract field and restricted to the diagonal, gives the ordinary operator applied to the physical field. Their near-wall forms follow from Eqs.~\eqref{eq:lap-tan}--\eqref{eq:div-grad} by applying Eq.~\eqref{eq:chain-rule} once for the divergence and twice for the Laplacian. In the second derivative, the components $\delta^3_\nu$ are themselves differentiated: their covariant derivative is $-\Gamma^3_{\nu\rho}$ (since $\delta^3_\nu=\partial_\nu x^3$ has constant components, $\delta^3_{\nu;\rho}=\partial_\rho\delta^3_\nu-\Gamma^\sigma_{\rho\nu}\delta^3_\sigma=-\Gamma^3_{\rho\nu}$), whose contraction with $g^{\nu\rho}$ is $-g^{jk}\Gamma^3_{jk}=\Gamma^k_{3k}$ [Eq.~\eqref{eq:christoffel_lower}]. The overlined divergence is $\overline w^{\,\mu}{}_{\ocd{\mu}}=\overline w^{\,\mu}{}_{;\mu}-\varepsilon^{-1}\partial_\varsigma\overline w^{\,3}$, and the overlined Laplacian contains \emph{exactly three powers of $\varepsilon$}, the same for all components:
\begin{equation}
\varepsilon^2\big(\overline\Delta\,\overline w\big)^\mu=\varepsilon^2g^{\rho\sigma}\,\overline w^{\,\mu}{}_{\ocd{\rho}\ocd{\sigma}}=\partial_\varsigma^2\overline w^{\,\mu}-\varepsilon\big(2\,\partial_\varsigma\overline w^{\,\mu}{}_{;3}+\Gamma^k_{3k}\,\partial_\varsigma\overline w^{\,\mu}\big)+\varepsilon^2(\Delta\overline w)^\mu .
\label{eq:three-term}
\end{equation}
The coefficient $\Gamma^k_{3k}$ is the divergence of the field of unit normals $\delta^\mu_3$, equal to $-2H$ on the wall.

\paragraph{Identification of equal powers.} In the abstract space, the momentum equation and the incompressibility condition of Eqs.~\eqref{eq:vort-problem} become
\begin{equation}
\varepsilon^2\big(\overline\Delta\,\overline w\big)^\mu-\overline w^{\,\mu}=0,\qquad
\overline w^{\,\mu}{}_{\ocd{\mu}}=0\qquad x^\nu\in\mathcal V_S,\ \varsigma\in\mathcal C,
\label{eq:overlined-problem}
\end{equation}
with the wall condition~\eqref{eq:vort-problem-bc} imposed at $\zeta=0$, $\varsigma=0$, and decay as $|\varsigma|\to\infty$. The solutions of this abstract problem give, on the diagonal, solutions of Eqs.~\eqref{eq:vort-problem}. The abstract problem is, however, more demanding than the physical one, since it must hold identically in the independent variables $(x^\nu;\varsigma)$, and not only on the diagonal: an abstract field whose restriction to the diagonal satisfies Eqs.~\eqref{eq:vort-problem} need not satisfy Eqs.~\eqref{eq:overlined-problem}. Requiring Eqs.~\eqref{eq:overlined-problem} identically amounts to requiring that the restriction of the abstract field to every surface $\varsigma=(\zeta-u)/\varepsilon$ of the abstract space, parallel to the diagonal ($u=0$), satisfy the field equations of Eqs.~\eqref{eq:vort-problem}: since $\partial_\nu\big((\zeta-u)/\varepsilon\big)=-\delta^3_\nu/\varepsilon$ for every $u$, the overlined operators, applied to an abstract field and restricted to any of these surfaces, give the ordinary operators applied to the restricted field. It is this additional requirement that gives one equation per power of $\varepsilon$. The series~\eqref{eq:two-scale-substitution} being substituted into Eqs.~\eqref{eq:overlined-problem}, the coefficient of each power of $\varepsilon$ is a function of the independent variables $(x^\nu;\varsigma)$ that does not depend on $\varepsilon$; at a given point of the abstract space, a series in powers of $\varepsilon$ with such coefficients is asymptotic to zero as $\varepsilon\to0$ only if all its coefficients vanish, and each of them is therefore required to vanish identically. On the diagonal, where $\varsigma=\zeta/\varepsilon$ depends on $\varepsilon$, this argument fails: this is the second difficulty noted above. By Eq.~\eqref{eq:three-term}, the leading part of $\varepsilon^2\overline\Delta-1$ is $\partial_\varsigma^2-1$, so that the equation obtained for each power of $\varepsilon$ is an ordinary differential problem in $\varsigma$, in which the position $x^\nu$ is a parameter, and whose source involves the coefficients of the lower powers. In the wall condition, $\varepsilon^2g^{\mu\nu}Q_{\varepsilon;\nu}$ contributes $\sum_m\varepsilon^{m+1}g^{\mu\nu}Q^{(m)}_{;\nu}$: the overlined coefficients and the pressure potentials are coupled through the wall conditions, and the local problems in $\varsigma$ and the global potential problems, of ideal-fluid type, are solved alternately (Sec.~\ref{sec6-hierarchy}).

\paragraph{Return to the diagonal.} The abstract space is finally left. Restricted to the diagonal, the abstract solution satisfies Eqs.~\eqref{eq:vort-problem} and, combined with the slow harmonic field $\varepsilon^2g^{\mu\nu}Q_{\varepsilon;\nu}$, it gives a uniformly valid composite representation of the deficit, without matching:
\begin{equation}
(w_\varepsilon)^\mu=\sum_{m\ge0}\varepsilon^{m+1}g^{\mu\nu}Q^{(m)}_{;\nu}+\sum_{m\ge0}\varepsilon^m\,\overline w^{\,\mu}_{(m)}\big|_{\varsigma=\zeta/\varepsilon} .
\label{eq:deficit-expansion}
\end{equation}
The first series carries the displacement flow and extends over the whole pore volume; the second, cut off as explained above, is confined to the boundary layer, and its decay in $\varsigma$ takes the place of matching. Together, the two series are the expansion~\eqref{eq:expansion-principle} of the deficit, whose coefficients are now identified: $W^\mu_{(m)}(x^\nu;\zeta/\varepsilon)=g^{\mu\nu}Q^{(m-1)}_{;\nu}+\overline w^{\,\mu}_{(m)}(x^\nu;\zeta/\varepsilon)$, with $Q^{(-1)}=0$. The overlined coefficients, and with them the $W_{(m)}$, are fixed only once a closure is chosen (Sec.~\ref{sec6-closure}), and depend on that choice, as do the partial sums of the series; uniqueness is recovered only for the field that the whole series represents, in the asymptotic sense made precise in Sec.~\ref{sec6-closure}.

\paragraph{A remark on the two-scale method and homogenization.} A word of caution is due here. Burridge and Keller~\cite{BK}, from whom we have just borrowed the two-scale principle, used it for a different purpose: to deduce the macroscopic equations of porous media by homogenization. In our opinion, that use lies beyond the real reach of the method, and we do not endorse it; ours stays within that reach. This remark is not needed for what follows, but it makes clear in what sense we follow Burridge and Keller, and in what sense we do not. The difference lies in the small parameter. Here the rapid variable is attached to a single length, the skin depth $|\varepsilon|$, which is small compared with all the geometrical lengths of the pore space. These lengths---radii of curvature, pore sizes, distances between walls---all belong to the slow description, in which they enter exactly through the metric and the curvatures, without any hierarchy being assumed between them. This is what the smoothness of the walls guarantees, and what fails at edges, where the distance to the edge introduces a further scale~\cite{FGL,AA,CSGL}. In homogenization, by contrast, the small parameter is the ratio of a microscopic length to the wavelength, whereas the microgeometry generally involves several lengths of different sizes, none of which is singled out. We consider that only the leading-order outcome of that procedure is meaningful. It expresses the divergence-free simplification, from which the macroscopic equations can be obtained directly~\cite{LAF_HF,LAF0,LAF2}, rather than deduced from the microstructure in a strict sense. The higher-order corrections given by the nested cell problems of this ``homogenization theory'', although widely studied and used, have in our view no physical meaning; the nonlocal theory~\cite{LAF1,LAF2}, which does not fit into that framework, illustrates this point. We therefore borrow from Burridge and Keller only the two-scale principle, as a tool for the boundary-layer problem, where it is justified, and not their ``homogenization'' as a method for deducing the macroscopic properties of porous media.

\subsection{The hierarchy}\label{sec6-hierarchy}

With $\overline w_{(m)}\equiv0$ for $m<0$, the coefficient of $\varepsilon^m$ in Eqs.~\eqref{eq:overlined-problem}, written with Eq.~\eqref{eq:three-term} and the overlined divergence, gives
\begin{subequations}\label{eq:hierarchy}
\begin{align}
&(\partial_\varsigma^2-1)\,\overline w^{\,\mu}_{(m)}=2\,\partial_\varsigma\overline w^{\,\mu}_{(m-1);3}+\Gamma^k_{3k}\,\partial_\varsigma\overline w^{\,\mu}_{(m-1)}-(\Delta\overline w_{(m-2)})^\mu,\label{eq:hier-T}\\
&\partial_\varsigma\overline w^{\,3}_{(m+1)}=\overline w^{\,\mu}_{(m);\mu},\label{eq:hier-D}
\end{align}
\end{subequations}
for $x^\nu\in\mathcal V_S$ and $\varsigma\in\mathcal C$, with decay as $|\varsigma|\to\infty$. The tangential components of Eq.~\eqref{eq:hier-T} govern $\overline w^{\,i}_{(m)}$, the incompressibility condition~\eqref{eq:hier-D} gives the normal component, and the normal component of Eq.~\eqref{eq:hier-T} is then identically satisfied. This consistency identity relies on the flatness of the ambient space. Substituted into the overlined momentum equation, the field so constructed leaves a residual, $\varepsilon^2(\overline\Delta\,\overline w)^\mu-\overline w^{\,\mu}$, which would vanish if the equation were satisfied. The tangential components of this residual vanish, because the tangential components of the equation were used to determine $\overline w^{\,i}$: the residual reduces to a normal component, $N\delta^\mu_3$. Because divergence and Laplacian commute, the overlined operators compose like the ordinary ones, and the overlined divergence of $\overline w$ vanishes, this residual has a vanishing overlined divergence. With $(N\delta^\mu_3)_{\ocd{\mu}}=\partial_3N+\Gamma^k_{3k}N-\varepsilon^{-1}\partial_\varsigma N$, the coefficients $N_{(m)}$ of its expansion therefore obey $\partial_\varsigma N_{(m+1)}=\partial_3N_{(m)}+\Gamma^k_{3k}N_{(m)}$, with $N_{(-1)}=0$; since they decay as $|\varsigma|\to\infty$, they vanish order by order. With the series of $Q_\varepsilon$ in Eq.~\eqref{eq:expansion-principle}, the wall condition~\eqref{eq:vort-problem-bc} and the Neumann datum~\eqref{eq:Q-neumann} read, on $S_p$ and at $\varsigma=0$,
\begin{equation}
\begin{aligned}
&\overline w^{\,i}_{(0)}=E^i,\qquad \overline w^{\,i}_{(m)}=-g^{ij}\partial_jQ^{(m-1)}\ \ (m\ge1),\\
&Q^{(m)}_{;3}=-\overline w^{\,3}_{(m+1)}(q,0;0)\ \ (m\ge0),
\end{aligned}
\qquad\text{on }S_p\ (\zeta=\varsigma=0).
\label{eq:wall-conditions}
\end{equation}
The last condition is the Neumann datum of the potential problem for $Q^{(m)}$: it is produced by the normal velocity that incompressibility forces inside the boundary layer (transpiration). The problems are thus solved in the alternating sequence
\[
\begin{aligned}
&\overline w_{(0)}\to\overline w^{\,3}_{(1)}\to Q^{(0)}\to\overline w^{\,i}_{(1)}\to\overline w^{\,3}_{(2)}\to Q^{(1)}\\
&\to\overline w^{\,i}_{(2)}\to\overline w^{\,3}_{(3)}\to Q^{(2)}\to\cdots.
\end{aligned}
\]

\subsection{Separation of the slow and rapid dependences: non-secularity}\label{sec6-closure}

For each $m$, the tangential components of Eq.~\eqref{eq:hier-T} form an ordinary differential equation in $\varsigma$ for $\overline w^{\,i}_{(m)}$, in which the position $x^\nu$ is a parameter: the operator is $\partial_\varsigma^2-1$, and the source is formed with the coefficients of the powers $\varepsilon^{m-1}$ and $\varepsilon^{m-2}$. The decaying solutions are a particular solution, forced by the source, plus a homogeneous term proportional to $e^{-\varsigma}$. A component of the source proportional to $e^{-\varsigma}$, which lies in the kernel of $\partial_\varsigma^2-1$, is \emph{resonant}: since $(\partial_\varsigma^2-1)(\varsigma e^{-\varsigma})=-2e^{-\varsigma}$, it forces a term $\varsigma e^{-\varsigma}$, in which the factor of $e^{-\varsigma}$ depends on $\varsigma$. The factor of the homogeneous term, a constant of integration of the equation in $\varsigma$, may be any function of $(q,\zeta)$; the wall condition~\eqref{eq:wall-conditions}, imposed at $\zeta=0$, $\varsigma=0$, fixes only its value on the wall. This is the freedom left by the hierarchy; it is the non-uniqueness of the two-scale series~\eqref{eq:expansion-principle} noted in Sec.~\ref{sec6-twoscale}, as it appears in the abstract space. At the power $\varepsilon^0$, for instance, the source vanishes, and $\overline w^{\,i}_{(0)}=\overline w^{\,i}_{(0)}(q,\zeta;0)\,e^{-\varsigma}$, where the amplitude $\overline w^{\,i}_{(0)}(q,\zeta;0)$ may be any function of $(q,\zeta)$ equal to $E^i$ on the wall. At the power $\varepsilon^1$, the source of Eq.~\eqref{eq:hier-T}, $-\big(2\,\overline w^{\,\mu}_{(0);3}+\Gamma^k_{3k}\,\overline w^{\,\mu}_{(0)}\big)$, is then proportional to $e^{-\varsigma}$, hence resonant, and $\overline w^{\,i}_{(1)}$ contains the term $\tfrac12\varsigma\big(2\,\overline w^{\,i}_{(0);3}+\Gamma^k_{3k}\,\overline w^{\,i}_{(0)}\big)$, which depends on the choice of that amplitude. A \emph{closure} removes this freedom, by fixing how the normal dependence of the coefficients is shared between $\zeta$ and $\varsigma$. Two closures are natural. In the \emph{non-secular} closure, all the rapid dependence is carried by the factor $e^{-\varsigma}$, each coefficient being this factor multiplied by an amplitude that depends on the position $(q,\zeta)$ only. In the \emph{inner} closure, the coefficients do not depend on $\zeta$, but on $(q;\varsigma)$ only; the geometric coefficients must then be expanded in powers of $\varepsilon\varsigma$, which modifies the hierarchy. Together with the wall conditions, each closure determines the coefficients completely.

We adopt the non-secular closure. Since a resonant source forces a term in which the factor of $e^{-\varsigma}$ depends on $\varsigma$, this closure is imposed by the \emph{non-secularity condition}: the resonant component of the source of Eq.~\eqref{eq:hier-T} is required to vanish at every power of $\varepsilon$. The terms so excluded decay, so that their elimination is a choice, not a condition of existence. They are called secular because they grow relative to $e^{-\varsigma}$: the restriction of $\varepsilon\varsigma e^{-\varsigma}$ to the diagonal is $\zeta e^{-\zeta/\varepsilon}$, so that, at a given point, a term $\varsigma e^{-\varsigma}$ in the coefficient of $\varepsilon$ contributes as much as the coefficient of $\varepsilon^0$. The reason for the choice lies in what is held fixed as $\varepsilon\to0$. At fixed $\varsigma$, that is, at a point kept at a fixed number of skin depths from the wall, the vortical deficit has an expansion in powers of $\varepsilon$, the classical inner expansion of the boundary layer, whose coefficients, functions of $(q;\varsigma)$, are those of the inner closure; its terms are ordered in absolute size. At a fixed position $x^\nu$ in $\mathcal V_S$, the vortical deficit is exponentially small, and all the coefficients of its expansion in powers of $\varepsilon$ vanish [Eq.~\eqref{eq:core-deficit}]. It has instead an expansion in the functions $\varepsilon^me^{-\zeta/\varepsilon}$, $\zeta$ being the distance from the point to the wall, whose coefficients, functions of position, are the amplitudes of the non-secular closure; its terms are ordered relative to the common factor $e^{-\zeta/\varepsilon}$, and the geometric coefficients enter it exactly, without being expanded about the wall. Each of these two expansions is unique, as is any asymptotic expansion with respect to a given sequence of functions of $\varepsilon$, with coefficients independent of $\varepsilon$~\cite{POI,ERD}, so that each closure gives the expansion of the field in one of the two limits. In an experiment, the geometry and the point of observation are fixed, and it is the frequency that increases: the physical limit is $\varepsilon\to0$ at fixed position, and the non-secular closure gives the expansion of the field at a given point. The expansion at fixed $\varsigma$ describes the layer at its own scale, as it tightens against the wall with increasing frequency. With the non-secularity condition, every abstract coefficient takes the form
\begin{equation}
\overline w_{(m)}=C_{(m)}(q,\zeta)\,e^{-\varsigma}\qquad x^\nu\in\mathcal V_S,\ \varsigma\in\mathcal C .
\label{eq:C-form}
\end{equation}
Indeed, if this holds at the powers $\varepsilon^{m-1}$ and $\varepsilon^{m-2}$, then $\partial_\varsigma\overline w_{(m-1)}=-C_{(m-1)}e^{-\varsigma}$ and $\Delta\overline w_{(m-2)}=(\Delta C_{(m-2)})\,e^{-\varsigma}$, since $\Delta$ acts on the slow variables only: the whole source of Eq.~\eqref{eq:hier-T} is proportional to $e^{-\varsigma}$, hence resonant, and the non-secularity condition requires it to vanish; the decaying solution of $(\partial_\varsigma^2-1)\overline w^{\,i}_{(m)}=0$ is then of the form~\eqref{eq:C-form}. The vanishing of the source gives the transport equations of the tangential amplitudes, and Eq.~\eqref{eq:hier-D} gives the normal amplitudes algebraically: for $m\ge0$,
\begin{equation}
\begin{aligned}
&2\,C^{\,\mu}_{(m);3}+\Gamma^k_{3k}\,C^{\,\mu}_{(m)}=-\big(\Delta C_{(m-1)}\big)^\mu,\\
&C^{\,3}_{(m+1)}=-C^{\,\mu}_{(m);\mu},
\end{aligned}
\qquad\text{in }\mathcal V_S,\qquad C_{(-1)}=0,\quad C^{\,3}_{(0)}=0,
\label{eq:transport}
\end{equation}
where $\Delta$ is the vector Laplacian~\eqref{eq:lap-tan}--\eqref{eq:lap-nor}, acting on the whole field $C_{(m-1)}$, tangential and normal components. The first relation, written for all components, determines the tangential amplitudes, its normal component being then satisfied identically; by Eq.~\eqref{eq:trace-Gamma}, it also reads $(\mathcal J_S^{1/2}C^{\,\mu}_{(m)})_{;3}=-\tfrac12\mathcal J_S^{1/2}(\Delta C_{(m-1)})^\mu$, $\mathcal J_S^{1/2}$ being a true scalar. The initial data of the tangential amplitudes on $S_p$ are given by the wall conditions~\eqref{eq:wall-conditions}: $C^{\,i}_{(0)}(q,0)=E^i$ and $C^{\,i}_{(m)}(q,0)=-g^{ij}\partial_jQ^{(m-1)}$ for $m\ge1$; in turn, the Neumann datum of the potential problem for $Q^{(m)}$ is $Q^{(m)}_{;3}=-C^{\,3}_{(m+1)}(q,0)=[C^{\,\mu}_{(m);\mu}]_0$. At leading order, $\mathcal J_S^{1/2}C_{(0)}$ is parallel along the normals, $(\mathcal J_S^{1/2}C^{\,\mu}_{(0)})_{;3}=0$, so that $C^{\,3}_{(0)}\equiv0$ and $\partial_\zeta C^{\,i}_{(0)}=(\Gamma^i_{3j}+\tfrac12\delta^i_j\Gamma^k_{3k})\,C^{\,j}_{(0)}$; with the solution $\Gamma_3=[\Gamma_3]_0(\mathbf I-\zeta[\Gamma_3]_0)^{-1}$ of the Riccati equation [first of Eqs.~\eqref{eq:riccati-codazzi}],
\begin{equation}
\overline w^{\,i}_{(0)}=\mathcal J_S^{-1/2}\big[(\mathbf I-\zeta[\Gamma_3]_0)^{-1}\big]^i{}_j\,E^j(q)\,e^{-\varsigma}\qquad x^\nu\in\mathcal V_S,\ \varsigma\in\mathcal C:
\label{eq:W0-WKB}
\end{equation}
the wall slip is transported parallel to itself along the normal [the factor $(\mathbf I-\zeta[\Gamma_3]_0)^{-1}$ in Gauss components, $\Gamma_3$ being the matrix~\eqref{eq:Gamma3-matrix}; in Cartesian components, the transported slip is constant along each normal], with the geometric spreading amplitude $\mathcal J_S^{-1/2}$. The construction is thus identical to a WKB (Wentzel--Kramers--Brillouin, or Liouville--Green) expansion $\mathbf w_{\mathrm{vort},\varepsilon}=e^{-\phi/\varepsilon}\sum_m\varepsilon^m\mathbf C_{(m)}$ of the evanescent shear field. Its eikonal equation $|\nabla\phi|^2=1$, with $\phi=0$ on $S_p$, gives the phase $\phi=\zeta$; its rays are the wall normals; its transport equations are Eqs.~\eqref{eq:transport}. This is a general fact: a two-scale expansion whose fast phase is exactly linear in the fast variable, closed by non-secularity, reproduces WKB. The factor $\mathcal J_S^{-1/2}$ can be checked on the exact solutions of the screened equation $\varepsilon^2\Delta f-f=0$ with unit datum $f=1$ on the wall and decay away from it, which is the problem satisfied by the axial component of the vortical deficit for a flow along a straight circular pore (where $\mathbf E=\mathbf e_z$ and $Q_\varepsilon=0$), and by the thermal deficit of Sec.~\ref{sec8}. Inside a circular tube of radius $R$, $f=I_0(r/\varepsilon)/I_0(R/\varepsilon)$, and outside a circular cylinder of radius $R$, $f=K_0(r/\varepsilon)/K_0(R/\varepsilon)$; for $|\varepsilon|\ll R$ both behave as $\sqrt{R/r}\,e^{-|r-R|/\varepsilon}$, while $\zeta=|r-R|$ and $\mathcal J_S=r/R$. Outside a sphere of radius $R$, $f=(R/r)\,e^{-(r-R)/\varepsilon}$ exactly, while $\mathcal J_S=(r/R)^2$. In each case the amplitude is precisely $\mathcal J_S^{-1/2}$.

The coefficients of the inner closure are polynomials in $\varsigma$ multiplied by $e^{-\varsigma}$, and the two closures distribute the same contributions differently between the powers of $\varepsilon$. At fixed $\varsigma$, the point considered is at the distance $\zeta=\varepsilon\varsigma$ from the wall, and the Taylor expansion of the amplitudes at the wall turns the non-secular terms into the inner ones: each term $\varepsilon^mC_{(m)}(q,\zeta)\,e^{-\zeta/\varepsilon}$ contributes to its own power of $\varepsilon$ and to all the higher ones, and the coefficient of $\varepsilon^m$ in the expansion at fixed $\varsigma$ is $e^{-\varsigma}\sum_{k=0}^m(\varsigma^k/k!)\,[\partial_\zeta^kC_{(m-k)}]_0$. For the circular tube considered above, for instance, $I_0(r/\varepsilon)/I_0(R/\varepsilon)$, with $r=R-\zeta$, is $e^{-\zeta/\varepsilon}(R/r)^{1/2}\big[1+\varepsilon\big(\tfrac1{8r}-\tfrac1{8R}\big)+O(\varepsilon^2)\big]$ at fixed $r$, and $e^{-\varsigma}\big[1+\varepsilon\varsigma/(2R)+O(\varepsilon^2)\big]$ at fixed $\varsigma$; the term $\varepsilon\varsigma/(2R)$ is the first Taylor term of the leading amplitude $(R/r)^{1/2}=(1-\varepsilon\varsigma/R)^{-1/2}$, the next amplitude vanishing on the wall. The overlined coefficients, and the coefficients $W_{(m)}$ of the series~\eqref{eq:expansion-principle}, therefore depend on the closure: the $W_{(m)}$ depend on $\varepsilon$ through $\zeta/\varepsilon$, and the uniqueness recalled above does not apply to them. What does not depend on the closure is, besides the exact fields, everything that is defined by an expansion with respect to a given sequence of functions of $\varepsilon$, with coefficients independent of $\varepsilon$, and is therefore unique: the two expansions of the vortical deficit, at fixed $\varsigma$ and at fixed position, of which each closure provides one; the value of each term on the wall, where $\zeta$ and $\varsigma$ vanish together and the two expansions coincide, and hence all the wall data; the pressure potentials, coefficients of the ordinary power series of $Q_\varepsilon$ in Eq.~\eqref{eq:expansion-principle}; and the tortuosity coefficients obtained below, coefficients of the ordinary power series of $1/\alpha$. The results, therefore, are the same with either closure.

Uniqueness is thus found in the field, and not in the two-scale series that represent it. Stopped at the same power $\varepsilon^m$, the non-secular and the inner series give two different approximations of the vortical deficit. Each differs from the exact deficit by $O(\varepsilon^{m+1})$, uniformly in $\mathcal V_S$, so that they differ from each other by $O(\varepsilon^{m+1})$ as well. The two closures give the same field in this sense only: not because their terms coincide, but because both approximate the same exact solution, more and more closely as further powers of $\varepsilon$ are retained.

This improvement has a limit, because the series are in general divergent; this does not affect the uniqueness of the expansions at fixed $\varsigma$ and at fixed position, which concerns their coefficients, not the sum of the series. In the circular tube, the coefficients of the bracket in the expansion at fixed $r$ given above grow for large $k$ like those of the asymptotic expansion of $I_0(r/\varepsilon)$, as $(k-1)!/[\pi(2r)^k]$. At a given frequency, the terms of the series therefore decrease only up to $k\simeq2r/|\varepsilon|$, and grow beyond: adding more terms then makes the approximation worse, and the best attainable error, relative to the leading term, is of the order of $|e^{-2r/\varepsilon}|$, up to an algebraic factor. This limit has a simple physical meaning. The boundary layer reaches a point of the tube mainly from the nearest part of the wall, at the distance $R-r$, but also from the opposite side, at the distance $R+r$, with an amplitude smaller by a factor of order $e^{-2r/\varepsilon}$. This second contribution is smaller than any power of $\varepsilon$ times $e^{-(R-r)/\varepsilon}$, so that no term of the series can contain it: the series describes the field to every power of $\varepsilon$, but two fields that differ only by such a contribution have the same series.

\subsection{The first three potential problems}

By Sec.~\ref{sec6-closure}, the Neumann datum of the problem for $Q^{(m)}$ is the wall value of the divergence of the amplitude $C_{(m)}$, $Q^{(m)}_{;3}=[C^{\,\mu}_{(m);\mu}]_0$, and the same divergence gives the next normal amplitude, $C^{\,3}_{(m+1)}=-C^{\,\mu}_{(m);\mu}$, with $C^{\,3}_{(0)}=0$ [Eq.~\eqref{eq:transport}]. By Eq.~\eqref{eq:div-grad}, this divergence is
\begin{equation}
C^{\,\mu}_{(m);\mu}=C^{\,k}_{(m)|k}+\Gamma^\ell_{3\ell}\,C^{\,3}_{(m)}-\partial_\zeta C^{\,3}_{(m)} .
\label{eq:div-Cm}
\end{equation}
For $m=0$, the normal amplitude vanishes, and the divergence reduces to the surface divergence $C^{\,k}_{(0)|k}$, equal to $E^k{}_{|k}$ on $S_p$. This gives the first potential problem
\begin{equation}
\Delta Q^{(0)}=0\ \text{in }V_p,\qquad
Q^{(0)}_{;3}=E^k{}_{|k}\ \text{on }S_p,\qquad
\langle Q^{(0)}\rangle_p=0 ,
\label{eq:Qp0}
\end{equation}
which is compatible because $\oint_{S_p}E^k{}_{|k}\,dS=0$, the integral of a surface divergence, the contributions of the faces of a periodic cell cancelling.

The divergences at the next powers are surface divergences as well, at every level of $\mathcal V_S$. Suppose that $C^{\,\mu}_{(m-1);\mu}=\mathcal F^{\,k}_{(m-1)|k}$ for some surface vector $\mathcal F_{(m-1)}$, as for $m=1$ with $\mathcal F_{(0)}=C_{(0)}$. Then $C^{\,3}_{(m)}=-\mathcal F^{\,k}_{(m-1)|k}$, and, since the normal derivative does not commute with the surface divergence, $\partial_\zeta(\mathcal F^{\,k}_{(m-1)|k})=(\partial_\zeta\mathcal F^{\,k}_{(m-1)})_{|k}-\mathcal F^{\,k}_{(m-1)}\,\partial_k\Gamma^\ell_{3\ell}$ [second relation~\eqref{eq:codazzi-contracted}]. The terms of Eq.~\eqref{eq:div-Cm} in the normal amplitude are therefore
\begin{align*}
\Gamma^\ell_{3\ell}\,C^{\,3}_{(m)}-\partial_\zeta C^{\,3}_{(m)}
&=-\Gamma^\ell_{3\ell}\,\mathcal F^{\,k}_{(m-1)|k}+\big(\partial_\zeta\mathcal F^{\,k}_{(m-1)}\big)_{|k}-\mathcal F^{\,k}_{(m-1)}\,\partial_k\Gamma^\ell_{3\ell}\\
&=\big(\partial_\zeta\mathcal F^{\,k}_{(m-1)}-\Gamma^\ell_{3\ell}\,\mathcal F^{\,k}_{(m-1)}\big)_{|k}
=\Big(\mathcal J_S^{-1}\partial_\zeta\big(\mathcal J_S\,\mathcal F^{\,k}_{(m-1)}\big)\Big)_{|k},
\end{align*}
the last form by Eq.~\eqref{eq:trace-Gamma}. Hence, at every level,
\begin{equation}
C^{\,\mu}_{(m);\mu}=\mathcal F^{\,k}_{(m)|k},\qquad
\mathcal F^{\,k}_{(0)}=C^{\,k}_{(0)},\qquad
\mathcal F^{\,k}_{(m)}=C^{\,k}_{(m)}+\mathcal J_S^{-1}\partial_\zeta\big(\mathcal J_S\,\mathcal F^{\,k}_{(m-1)}\big)\qquad\text{in }\mathcal V_S .
\label{eq:flux-recurrence}
\end{equation}
The trace $\Gamma^\ell_{3\ell}$, which measures the spreading of the parallel surfaces, enters only through their area ratio $\mathcal J_S$. The vectors $\mathcal F_{(m)}$ have a simple meaning. By Eq.~\eqref{eq:flux-recurrence}, $\mathcal J_S\,\mathcal F_{(m)}=\sum_{n=0}^m\partial_\zeta^n(\mathcal J_S\,C_{(m-n)})$, while repeated integrations by parts give $\int_\zeta^\infty\mathcal J_S\,C_{(m)}\,e^{-\zeta'/\varepsilon}d\zeta'\sim e^{-\zeta/\varepsilon}\sum_n\varepsilon^{n+1}\partial_\zeta^n(\mathcal J_S\,C_{(m)})$. With $\mathbf w_{\mathrm{vort},\varepsilon}\sim\sum_m\varepsilon^mC_{(m)}\,e^{-\zeta/\varepsilon}$ [Eq.~\eqref{eq:C-form} on the diagonal], it follows that
\begin{equation*}
\int_\zeta^\infty\mathcal J_S\,w^k_{\mathrm{vort},\varepsilon}\,d\zeta'\sim e^{-\zeta/\varepsilon}\sum_{m\ge0}\varepsilon^{m+1}\mathcal J_S\,\mathcal F^{\,k}_{(m)} .
\end{equation*}
The left-hand side is the tangential flux carried by the part of the layer beyond the level $\zeta$, and the relation $C^{\,3}_{(m+1)}=-\mathcal F^{\,k}_{(m)|k}$ is the mass balance of this part of the layer, term by term: the fluid that crosses the parallel surface at $\zeta$ feeds the divergence of this flux.

With $C^{\,i}_{(m);3}=-\partial_\zeta C^{\,i}_{(m)}+\Gamma^i_{3j}C^{\,j}_{(m)}$, $\Gamma^i_{3j}=\mathring\Gamma^i_{3j}+\tfrac12\delta^i_j\Gamma^\ell_{3\ell}$ and Eq.~\eqref{eq:trace-Gamma}, the tangential components of the transport equations~\eqref{eq:transport} take a similar form,
\begin{equation}
\mathcal J_S^{-1}\partial_\zeta\big(\mathcal J_S\,C^{\,i}_{(m)}\big)=\mathring\Gamma^i_{3j}C^{\,j}_{(m)}+\tfrac12\big(\Delta C_{(m-1)}\big)^i .
\label{eq:transport-J}
\end{equation}
For $m=1$, Eq.~\eqref{eq:flux-recurrence} and Eq.~\eqref{eq:transport-J} with $m=0$, which has no source, give
\begin{equation}
\mathcal F^{\,k}_{(1)}=C^{\,k}_{(1)}+\mathring\Gamma^k_{3j}C^{\,j}_{(0)} :
\label{eq:F1}
\end{equation}
of $\Gamma^i_{3j}$, only the trace-free part remains, the trace part having been absorbed into $\mathcal J_S$ both in the divergence and in the transport. For $m=2$, the normal derivative of $\mathring\Gamma^k_{3j}$ is needed. By the Riccati equation~\eqref{eq:riccati-codazzi}, its trace~\eqref{eq:codazzi-contracted} and the first identity~\eqref{eq:cayley-hamilton}, this derivative is proportional to $\mathring\Gamma^k_{3j}$ itself; with Eq.~\eqref{eq:trace-Gamma}, the relation integrates along the normals:
\begin{equation}
\partial_\zeta\mathring\Gamma^k_{3j}=\Gamma^\ell_{3\ell}\,\mathring\Gamma^k_{3j},\qquad
\mathcal J_S\,\mathring\Gamma^k_{3j}=\big[\mathring\Gamma^k_{3j}\big]_0\qquad\text{in }\mathcal V_S .
\label{eq:dz-ring}
\end{equation}
Hence $\mathcal J_S^{-1}\partial_\zeta(\mathcal J_S\,\mathring\Gamma^k_{3j}C^{\,j}_{(0)})=\mathring\Gamma^k_{3\ell}\,\partial_\zeta C^{\,\ell}_{(0)}$, and, with Eq.~\eqref{eq:transport-J} for $m=1$ and the leading-order transport,
\begin{equation*}
\mathcal J_S^{-1}\partial_\zeta\big(\mathcal J_S\,\mathcal F^{\,k}_{(1)}\big)=\mathring\Gamma^k_{3j}C^{\,j}_{(1)}+\tfrac12\big(\Delta C_{(0)}\big)^k+\mathring\Gamma^k_{3\ell}\big(\mathring\Gamma^\ell_{3j}+\delta^\ell_j\,\Gamma^i_{3i}\big)C^{\,j}_{(0)} .
\end{equation*}
By the second identity~\eqref{eq:cayley-hamilton}, $\mathring\Gamma^k_{3\ell}\mathring\Gamma^\ell_{3j}=\tfrac12\mathring\Gamma^i_{3\ell}\mathring\Gamma^\ell_{3i}\,\delta^k_j$, and Eq.~\eqref{eq:flux-recurrence} gives
\begin{equation}
\mathcal F^{\,k}_{(2)}=C^{\,k}_{(2)}+\mathring\Gamma^k_{3j}C^{\,j}_{(1)}+\Gamma^\ell_{3\ell}\,\mathring\Gamma^k_{3j}C^{\,j}_{(0)}+\tfrac12\mathring\Gamma^i_{3\ell}\mathring\Gamma^\ell_{3i}\,C^{\,k}_{(0)}+\tfrac12\big(\Delta C_{(0)}\big)^k,
\label{eq:F2}
\end{equation}
where Eq.~\eqref{eq:lap-tan}, with the leading-order transport, the Riccati equation and Eq.~\eqref{eq:cayley-hamilton}, gives in $\mathcal V_S$
\begin{equation}
(\Delta C_{(0)})^i=g^{jk}C^{\,i}_{(0)|j|k}+\Big(\tfrac14(\Gamma^k_{3k})^2\delta^i_\ell-\Gamma^k_{3k}\Gamma^i_{3\ell}\Big)C^{\,\ell}_{(0)} .
\label{eq:DeltaC0}
\end{equation}
On $S_p$, where $C^{\,k}_{(0)}=E^k$ and $C^{\,k}_{(m)}=-g^{kj}\partial_jQ^{(m-1)}$ for $m\ge1$, Eqs.~\eqref{eq:F1}--\eqref{eq:DeltaC0} give the Neumann data $Q^{(m)}_{;3}=\mathcal F^{\,k}_{(m)|k}$ of the next two problems,
\begin{align}
Q^{(1)}_{;3}&=\Big[\mathring\Gamma^k_{3j}E^j-g^{kj}\partial_jQ^{(0)}\Big]_{|k}\qquad\text{on }S_p,\label{eq:Qp1}\\
Q^{(2)}_{;3}&=\Big[\tfrac12g^{j\ell}E^k{}_{|j|\ell}+\tfrac12\Gamma^\ell_{3\ell}\,\mathring\Gamma^k_{3j}E^j+\Big(\tfrac12\mathring\Gamma^i_{3\ell}\mathring\Gamma^\ell_{3i}-\tfrac18(\Gamma^\ell_{3\ell})^2\Big)E^k\nonumber\\
&\qquad\qquad-\mathring\Gamma^k_{3j}\,g^{j\ell}\partial_\ell Q^{(0)}-g^{kj}\partial_jQ^{(1)}\Big]_{|k}\qquad\text{on }S_p,\label{eq:Qp2}
\end{align}
each potential being harmonic in $V_p$ with zero mean. These data are surface divergences, so that, like the first one, the two problems are compatible.

At $\zeta=0$, the flux beyond the level $\zeta$ is the flux across the whole layer, $\int_0^\infty\mathcal J_S\,w^k_{\mathrm{vort},\varepsilon}\,d\zeta$, and $Q^{(m)}_{;3}$ is the surface divergence of its coefficient of $\varepsilon^{m+1}$. The slips $s^i_{(0)}=E^i$ and $s^i_{(k)}=-g^{ij}\partial_jQ^{(k-1)}$ enter this flux at the orders $\varepsilon^k$. For the response of the layer to a single wall slip $s^i$, whose amplitudes have the wall values $C^{\,i}_{(0)}=s^i$ and $C^{\,i}_{(m)}=0$ for $m\ge1$, Eqs.~\eqref{eq:F1}--\eqref{eq:DeltaC0} give the flux across the layer, on $S_p$:
\begin{equation}
\varepsilon\,s^i+\varepsilon^2\,\mathring\Gamma^i_{3j}s^j+\varepsilon^3\Big[\tfrac12g^{j\ell}s^i{}_{|j|\ell}+\tfrac12\Gamma^\ell_{3\ell}\,\mathring\Gamma^i_{3j}s^j+\Big(\tfrac12\mathring\Gamma^k_{3\ell}\mathring\Gamma^\ell_{3k}-\tfrac18(\Gamma^k_{3k})^2\Big)s^i\Big]+O(\varepsilon^4).
\label{eq:deficit-flux}
\end{equation}
The term in $\varepsilon^2$ is the curvature-induced part of the flux. It involves only the trace-free part of $\Gamma^i_{3j}$ on the wall, which in principal axes is $[\mathring\Gamma^i_{3j}]_0=\operatorname{diag}\big(\tfrac{\kappa_2-\kappa_1}2,\tfrac{\kappa_1-\kappa_2}2\big)$, and vanishes on umbilic walls (planes, spheres).

\subsection{Boundary-layer moments}

With the exact splitting~\eqref{eq:exact-splitting}, the vortical part in Eq.~\eqref{eq:tortuosity-vortical-projection} is the boundary-layer field just constructed. We evaluate this volume form, which is equivalent to the boundary form of Eq.~\eqref{eq:tortuosity-boundary-vortical-deficit}: the two give the same coefficients, their integrands on $S_p$ differing by surface divergences. Since $\mathbf w_{\mathrm{vort},\varepsilon}$ is exponentially small outside the layer, the volume average may be restricted, up to $O(\varepsilon^\infty)$, to the tubular neighbourhood of the wall, and written in Gauss coordinates, with $dV=\mathcal J_S\,d\zeta\,dS$, as a wall integral of normal moments:
\begin{equation}
\frac1{\alpha(\omega)}=\frac1{\alpha_\infty}-\frac1{|V_p|}\int_{S_p}\!\!dS\int_0^\infty g_{\mu\nu}E^\mu\,(w_{\mathrm{vort},\varepsilon})^\nu\,\mathcal J_S\,d\zeta+O(\varepsilon^\infty).
\end{equation}
The inner integral is expanded by Watson's lemma~\cite{BH}, $\int_0^\infty f(\zeta)\,e^{-\zeta/\varepsilon}d\zeta\sim\sum_k\varepsilon^{k+1}f^{(k)}(0)$. The problem being linear, the vortical deficit is the sum $\sum_{k\ge0}\varepsilon^k\,w_{\mathrm{vort},\varepsilon}[s_{(k)}]$ of the responses of the layer to the successive wall slips of Eq.~\eqref{eq:wall-conditions}, $s^i_{(0)}=E^i$ and $s^i_{(k)}=-g^{ij}\partial_jQ^{(k-1)}$ for $k\ge1$, each one entering at its own order; $w_{\mathrm{vort},\varepsilon}[s]$ denotes the response to the slip $s^i$ alone. For a given slip, the layer moment is, up to its own third order,
\begin{multline}
\int_0^\infty\mathcal J_S\,E_\mu\,w^\mu_{\mathrm{vort},\varepsilon}[s]\,d\zeta=\varepsilon\,E_is^i+\varepsilon^2\Big[-\tfrac12\Gamma^k_{3k}E_is^i-E_{i;3}s^i\Big]\\
+\varepsilon^3\Big[\Big(-\tfrac12\mathring\Gamma^k_{3\ell}\mathring\Gamma^\ell_{3k}+\tfrac18(\Gamma^k_{3k})^2\Big)E_is^i-\tfrac12\Gamma^k_{3k}E_i\Gamma^i_{3j}s^j+\Gamma^k_{3k}E_{i;3}s^i+E_{i;3;3}s^i\\
+\tfrac12E_ig^{jk}s^i{}_{|j|k}-E^k{}_{|k}\,s^\ell{}_{|\ell}\Big]+O(\varepsilon^4)\qquad\text{on }S_p,
\label{eq:layer-moment}
\end{multline}
where all quantities are evaluated on $S_p$. Here $E_{i;3}$ and $E_{i;3;3}$ are components of the first and second covariant derivatives of the vector field $E^\mu$ along the normal, $E^\mu{}_{;3}$ and $E^\mu{}_{;3;3}$, and not derivatives of its components; the last term of the bracket of order $\varepsilon^3$, $-E^k{}_{|k}\,s^\ell{}_{|\ell}$, comes from the normal components: that of $E^\mu$, which vanishes on $S_p$ but grows as $\zeta E^k{}_{|k}$, and the transpiration velocity $-\varepsilon s^k{}_{|k}\,e^{-\zeta/\varepsilon}$ of the vortical deficit. All indices are contracted, and $\Gamma^i_{3j}$ and $\mathring\Gamma^i_{3j}$ are surface tensors, so that Eq.~\eqref{eq:layer-moment} and those derived from it are invariant under the changes of surface coordinates.

The derivation of Eq.~\eqref{eq:layer-moment} is short in this notation. Let $f_{(m)}(\zeta)=\mathcal J_S\,E_\mu C^{\,\mu}_{(m)}[s]$, where $C_{(m)}[s]$ are the amplitudes of $w_{\mathrm{vort},\varepsilon}[s]$: they obey Eqs.~\eqref{eq:transport} with $C^{\,i}_{(0)}[s]=s^i$ and $C^{\,i}_{(m)}[s]=0$ ($m\ge1$) on $S_p$, the later wall data belonging to the next slips. Since $f_{(m)}(0)=0$ for $m\ge1$ ($C^{\,i}_{(m)}[s]$ and $E_3$ vanish on $S_p$), the moment is $\varepsilon f_{(0)}(0)+\varepsilon^2f_{(0)}'(0)+\varepsilon^3[f_{(0)}''(0)+f_{(1)}'(0)]+O(\varepsilon^4)$. The functions $f_{(m)}$ are scalars and $\Gamma^\lambda_{33}=0$, so that $d/d\zeta=-(\cdot)_{;3}$ and $d^2/d\zeta^2=(\cdot)_{;3;3}$, with the Leibniz rule of the covariant derivative: the normal derivatives of the products are covariant derivatives of each factor. By Eq.~\eqref{eq:transport}, $f_{(0)}=\mathcal J_S^{1/2}E_\mu P^\mu$, with $P^\mu=\mathcal J_S^{1/2}C^{\,\mu}_{(0)}[s]$ and $P^\mu{}_{;3}=0$. By Eqs.~\eqref{eq:trace-Gamma}, \eqref{eq:codazzi-contracted} and~\eqref{eq:cayley-hamilton}, $(\mathcal J_S^{1/2})_{;3}=\tfrac12\Gamma^k_{3k}\mathcal J_S^{1/2}$ and $(\mathcal J_S^{1/2})_{;3;3}=-\tfrac12\mathring\Gamma^k_{3\ell}\mathring\Gamma^\ell_{3k}\mathcal J_S^{1/2}$. The value and the first two derivatives of $f_{(0)}$ at $\zeta=0$ thus give the terms of orders $\varepsilon$ and $\varepsilon^2$, and, at order $\varepsilon^3$, the terms in $E_{i;3;3}s^i$ and $\Gamma^k_{3k}E_{i;3}s^i$ and the part $-\tfrac12\mathring\Gamma^k_{3\ell}\mathring\Gamma^\ell_{3k}E_is^i$ of the term in $E_is^i$. On $S_p$, $C^{\,i}_{(1)}[s]=0$, $C^{\,3}_{(1)}[s]=-s^k{}_{|k}$ and $C^{\,i}_{(1);3}[s]=-\tfrac12(\Delta C_{(0)}[s])^i$, while $E^3=0$ and $E^3{}_{;3}=-E^k{}_{|k}$ by incompressibility [Eq.~\eqref{eq:div-grad}]; hence $f_{(1)}'(0)=\tfrac12E_i(\Delta C_{(0)}[s])^i-E^k{}_{|k}\,s^\ell{}_{|\ell}$. Finally, on $S_p$, $(\Delta C_{(0)}[s])^i$ is given by Eq.~\eqref{eq:DeltaC0} with $s^i$ in place of $C^{\,i}_{(0)}$, so that $f_{(1)}'(0)$ gives the other terms of order $\varepsilon^3$, including the part $\tfrac18(\Gamma^k_{3k})^2E_is^i$ of the term in $E_is^i$.
Summing the contributions of the slips $s^i_{(0)}=E^i$, $\varepsilon s^i_{(1)}$ and $\varepsilon^2s^i_{(2)}$, we obtain
\begin{equation}
\frac1{\alpha(\omega)}=\frac1{\alpha_\infty}-\big(d_1\varepsilon+d_2\varepsilon^2+d_3\varepsilon^3\big)+O(\varepsilon^4),
\label{eq:inverse-tortuosity-d}
\end{equation}
\begin{subequations}\label{eq:d123}
\begin{align}
|V_p|\,d_1&=\int_{S_p}E_iE^i\,dS,\\
|V_p|\,d_2&=\int_{S_p}\Big[-\tfrac12\Gamma^k_{3k}E_iE^i-E^iE_{i;3}-E^j\partial_jQ^{(0)}\Big]dS,\\
|V_p|\,d_3&=\int_{S_p}\Big[\Big(-\tfrac12\mathring\Gamma^k_{3\ell}\mathring\Gamma^\ell_{3k}+\tfrac18(\Gamma^k_{3k})^2\Big)E_iE^i-\tfrac12\Gamma^k_{3k}E_i\Gamma^i_{3j}E^j+\tfrac12E_ig^{jk}E^i{}_{|j|k}\nonumber\\
&\hspace{3.2em}+E^iE_{i;3;3}+\Gamma^k_{3k}E^iE_{i;3}-\big(E^k{}_{|k}\big)^2+\Big(E^j{}_{;3}+\tfrac12\Gamma^k_{3k}E^j\Big)\partial_jQ^{(0)}-E^j\partial_jQ^{(1)}\Big]dS .
\end{align}
\end{subequations}
Apart from the term in $\partial_jQ^{(0)}$ of $d_3$, which requires only $Q^{(0)}$, the potentials enter Eqs.~\eqref{eq:d123} through two wall integrals of the same form, the first potential in $d_2$ and the second in $d_3$:
\begin{equation*}
-\int_{S_p}E^j\partial_jQ^{(0)}\,dS\qquad\text{and}\qquad-\int_{S_p}E^j\partial_jQ^{(1)}\,dS ;
\end{equation*}
the second one would require solving a second Neumann problem. Both are transformed by two identities of potential theory. For two harmonic, single-valued and stationary functions $\psi$ and $\chi$, the divergence theorem applied to $\psi\,g^{\mu\nu}\chi_{;\nu}$ in $V_p$, the contributions of the faces of a periodic cell cancelling as for Eq.~\eqref{eq:gradient-orthogonality}, gives Green's first identity
\begin{equation*}
\int_{S_p}\psi\,\chi_{;3}\,dS=\int_{V_p}g^{\mu\nu}\psi_{;\mu}\chi_{;\nu}\,dV ,
\end{equation*}
the normal derivative $(\cdot)_{;3}$ being taken out of the fluid. Its right-hand side is symmetric in $\psi$ and $\chi$; hence the reciprocity relation
\begin{equation*}
\int_{S_p}\psi\,\chi_{;3}\,dS=\int_{S_p}\chi\,\psi_{;3}\,dS :
\end{equation*}
the integral over the wall of one potential weighted by the Neumann datum of the other does not change when the two potentials are exchanged. This is Green's reciprocity theorem of potential theory, distinct from the reciprocity of the response functions met at the beginning of this section. On the wall, $E^j$ being tangent,
\begin{equation*}
-\int_{S_p}E^j\partial_j\psi\,dS
=
\int_{S_p}E^k{}_{|k}\,\psi\,dS
=
\int_{S_p}Q^{(0)}_{;3}\,\psi\,dS .
\end{equation*}
The first equality is an integration by parts along $S_p$, in which the contributions of opposite faces of the periodic cell again cancel; the second uses the Neumann condition in Eq.~\eqref{eq:Qp0}.

For the first integral, setting $\psi=\chi=Q^{(0)}$ and using Green's first identity yields
\begin{equation*}
-\int_{S_p}E^j\partial_jQ^{(0)}\,dS
=
\int_{S_p}Q^{(0)}Q^{(0)}_{;3}\,dS
=
\int_{V_p}g^{\mu\nu}Q^{(0)}_{;\mu}Q^{(0)}_{;\nu}\,dV .
\end{equation*}
The contribution of $Q^{(0)}$ to $d_2$ is therefore the volume integral of the squared gradient of $Q^{(0)}$.

For the second integral, setting $\psi=Q^{(1)}$, $\chi=Q^{(0)}$ and using Green's reciprocity relation yields
\begin{equation*}
-\int_{S_p}E^j\partial_jQ^{(1)}\,dS
=
\int_{S_p}Q^{(1)}Q^{(0)}_{;3}\,dS
=
\int_{S_p}Q^{(0)}Q^{(1)}_{;3}\,dS .
\end{equation*}
The Neumann datum $Q^{(1)}_{;3}$ is known in terms of $E^j$ and $Q^{(0)}$ from Eq.~\eqref{eq:Qp1}. Substituting it and integrating by parts along the wall gives
\begin{equation*}
-\int_{S_p}E^j\partial_jQ^{(1)}\,dS
=
-\int_{S_p}\partial_kQ^{(0)}
\big[\mathring\Gamma^k_{3j}E^j-g^{kj}\partial_jQ^{(0)}\big]\,dS .
\end{equation*}
This eliminates $Q^{(1)}$ without solving its Neumann problem.

Finally, an integration by parts along $S_p$ gives
\begin{equation*}
\int_{S_p}E_i g^{jk}E^i{}_{|j|k}\,dS
=
-\int_{S_p}g_{ik}g^{j\ell}E^i{}_{|j}E^k{}_{|\ell}\,dS .
\end{equation*}
The right-hand side is minus the integral of the squared intrinsic gradient~\eqref{eq:surface-operators} of the tangential wall slip.

Combining these results, we obtain
\begin{align}
|V_p|\,d_2
&=\int_{S_p}\Big[
-\tfrac12\Gamma^k_{3k}E_iE^i-E^iE_{i;3}
\Big]dS
+\int_{V_p}g^{\mu\nu}Q^{(0)}_{;\mu}Q^{(0)}_{;\nu}\,dV,
\label{eq:d2-final}\\
|V_p|\,d_3
&=\int_{S_p}\Big[
\Big(-\tfrac12\mathring\Gamma^k_{3\ell}\mathring\Gamma^\ell_{3k}
+\tfrac18(\Gamma^k_{3k})^2\Big)E_iE^i
-\tfrac12\Gamma^k_{3k}E_i\Gamma^i_{3j}E^j
-\tfrac12g_{ik}g^{j\ell}E^i{}_{|j}E^k{}_{|\ell}
+E^iE_{i;3;3}\nonumber\\
&\hspace{3.2em}
+\Gamma^k_{3k}E^iE_{i;3}
-\big(E^k{}_{|k}\big)^2
+\Big(E^j{}_{;3}+\tfrac12\Gamma^k_{3k}E^j
-\mathring\Gamma^j_{3\ell}E^\ell\Big)\partial_jQ^{(0)}
+g^{jk}\partial_jQ^{(0)}\partial_kQ^{(0)}
\Big]dS .
\label{eq:d3-final}
\end{align}
Thus, evaluating $d_3$ requires neither the second potential nor the second-order boundary-layer profile.

\subsection{Reduction to the wall trace of the ideal field}\label{sec6-reduction}

The normal derivatives of the ideal field at the wall can be eliminated altogether, because it is the gradient of a harmonic function with zero normal derivative on $S_p$. Write $E_\mu=u_{;\mu}$, with $u=\mathbf e\cdot\mathbf x-\Phi$, $\Delta u=0$ and $u_{;3}=E_3=0$ on $S_p$; since $u_{;3}$ vanishes along $S_p$, so do its tangential derivatives. First, $E_{i;3}=u_{;i;3}=u_{;3;i}=\partial_i(u_{;3})-\Gamma^k_{3i}E_k$. Second, $E_{3;3}=u_{;3;3}=\partial_\zeta^2u$, and the Laplace equation, written with Eq.~\eqref{eq:div-grad}, gives $u_{;3;3}=-g^{jk}u_{|j|k}=-E^k{}_{|k}$ on $S_p$. Third, since the second covariant derivatives of a scalar are symmetric and covariant derivatives commute in Euclidean space, $E_{i;3;3}=u_{;i;3;3}=u_{;3;3;i}$; and $u_{;3;3}$, a component of the tensor $u_{;\mu;\nu}$ and not a scalar, has the derivative $u_{;3;3;i}=\partial_i(u_{;3;3})-2\Gamma^k_{3i}u_{;k;3}$ along the wall. Hence, on $S_p$,
\begin{equation}
E_{i;3}=-\Gamma^k_{3i}E_k,\qquad
E_{3;3}=-E^k{}_{|k},\qquad
E_{i;3;3}=-\partial_i\big(E^k{}_{|k}\big)+2\Gamma^k_{3i}\Gamma^\ell_{3k}E_\ell
\qquad\text{on }S_p .
\label{eq:wall-identities}
\end{equation}
(For instance, for the potential flow of speed $U$ around a circular cylinder of radius $R$, with $\theta$ the polar angle measured from the direction of the flow, $E_\mu E^\mu$ decreases away from the wall as $(E_\mu E^\mu)_{;3}=2E^iE_{i;3}=-2E_i\Gamma^i_{3j}E^j=8U^2\sin^2\theta/R$, which the exact solution confirms.) Substituting Eq.~\eqref{eq:wall-identities} into Eqs.~\eqref{eq:d2-final}--\eqref{eq:d3-final}, integrating $E^i\partial_i(E^k{}_{|k})$ by parts, and using Eq.~\eqref{eq:cayley-hamilton}, we obtain the compact forms
\begin{align}
|V_p|\,d_2&=\int_{S_p}E_i\mathring\Gamma^i_{3j}E^j\,dS+\int_{V_p}g^{\mu\nu}Q^{(0)}_{;\mu}Q^{(0)}_{;\nu}\,dV,\label{eq:d2-compact}\\
|V_p|\,d_3&=\int_{S_p}\Big[-\tfrac18(\Gamma^k_{3k})^2E_iE^i+\tfrac12\Gamma^k_{3k}E_i\mathring\Gamma^i_{3j}E^j-\tfrac12g_{ik}g^{j\ell}E^i{}_{|j}E^k{}_{|\ell}+g_{ij}V^iV^j\Big]dS,\label{eq:d3-compact}
\end{align}
with
\begin{equation}
V^i=g^{ij}\partial_jQ^{(0)}-\mathring\Gamma^i_{3j}E^j .
\label{eq:Vi-def}
\end{equation}
By Eq.~\eqref{eq:wall-identities}, the coefficient of $\partial_jQ^{(0)}$ in Eq.~\eqref{eq:d3-final} becomes $-2\mathring\Gamma^j_{3\ell}E^\ell$, and the last term of Eq.~\eqref{eq:d3-compact} results from completing the square: $-2\mathring\Gamma^j_{3\ell}E^\ell\partial_jQ^{(0)}+g^{jk}\partial_jQ^{(0)}\partial_kQ^{(0)}=g_{ij}V^iV^j-g_{ij}\mathring\Gamma^i_{3k}\mathring\Gamma^j_{3\ell}E^kE^\ell$, with $g_{ij}\mathring\Gamma^i_{3k}\mathring\Gamma^j_{3\ell}E^kE^\ell=\tfrac12\mathring\Gamma^k_{3\ell}\mathring\Gamma^\ell_{3k}E_iE^i$ by the symmetry of $g_{ij}\mathring\Gamma^j_{3k}$ [Eq.~\eqref{eq:christoffel_lower}] and Eq.~\eqref{eq:cayley-hamilton}. With the term $-\tfrac12\mathring\Gamma^k_{3\ell}\mathring\Gamma^\ell_{3k}E_iE^i$ so produced, the terms in $\mathring\Gamma^k_{3\ell}\mathring\Gamma^\ell_{3k}E_iE^i$ cancel in $d_3$. In principal axes, with principal curvatures $\kappa_1,\kappa_2$ and components $E_1,E_2$ of the wall slip on an orthonormal basis,
\begin{equation}
E_i\mathring\Gamma^i_{3j}E^j=\tfrac12(\kappa_2-\kappa_1)\,(E_1^2-E_2^2),
\qquad
\tfrac12\big[\mathring\Gamma^k_{3\ell}\mathring\Gamma^\ell_{3k}\big]_0=\tfrac14(\kappa_1-\kappa_2)^2 .
\label{eq:Gamma-principal}
\end{equation}
\emph{All three coefficients are therefore determined by the wall trace of the ideal field, its tangential gradient, the Christoffel symbols $\Gamma^i_{3j}$ of the wall, and the single harmonic potential $Q^{(0)}$, itself driven by $E^k{}_{|k}$.} No normal derivative of the ideal field and no higher boundary-layer profile is needed. The slip being tangent to the wall, these compact forms can also be written with Greek indices only, with the gradient $n^\mu{}_{;\nu}=\Gamma^\mu_{3\nu}$ of the unit normal and the tangential projector; Appendix~\ref{app:greek} gives these forms, on which the readings of Secs.~\ref{sec7} and~\ref{sec9} rest.

\subsection{The parameters $\Lambda$, $\Sigma$, $V$}

Inverting the series gives $\alpha/\alpha_\infty=1+\alpha_\infty d_1\varepsilon+\big[(\alpha_\infty d_1)^2+\alpha_\infty d_2\big]\varepsilon^2+\big[(\alpha_\infty d_1)^3+2\alpha_\infty^2d_1d_2+\alpha_\infty d_3\big]\varepsilon^3+O(\varepsilon^4)$. Comparison with Eq.~\eqref{eq:alpha-expansion-LSV} yields the three parameters:
\begin{align}
\frac2\Lambda&=\frac{\alpha_\infty}{|V_p|}\int_{S_p}E_iE^i\,dS,\label{eq:Lambda-def}\\
\frac3\Sigma&=\frac4{\Lambda^2}+\frac{\alpha_\infty}{|V_p|}\Big\{\int_{S_p}E_i\mathring\Gamma^i_{3j}E^j\,dS+\int_{V_p}g^{\mu\nu}Q^{(0)}_{;\mu}Q^{(0)}_{;\nu}\,dV\Big\},\label{eq:Sigma-def}\\
\frac{15}{4V}&=\frac{12}{\Lambda\Sigma}-\frac8{\Lambda^3}+\frac{\alpha_\infty}{|V_p|}\int_{S_p}\Big[-\tfrac18(\Gamma^k_{3k})^2E_iE^i+\tfrac12\Gamma^k_{3k}E_i\mathring\Gamma^i_{3j}E^j\nonumber\\
&\qquad-\tfrac12g_{ik}g^{j\ell}E^i{}_{|j}E^k{}_{|\ell}+g_{ij}V^iV^j\Big]dS ,\label{eq:V-def}
\end{align}
with $V^i$ given by Eq.~\eqref{eq:Vi-def}, all wall quantities being taken at $\zeta=0$. Equation \eqref{eq:Lambda-def} is the Johnson--Koplik--Dashen length~\cite{JKD}. Indeed, since $\langle\mathbf E\rangle_p=\mathbf e/\alpha_\infty$ [Eq.~\eqref{eq:ideal-tortuosity}] and $\langle\mathbf E\cdot\mathbf E\rangle_p=1/\alpha_\infty$ [Eq.~\eqref{eq:tortuosity-deficit-projection}], the ideal tortuosity is the ratio of the mean squared ideal velocity to the squared mean ideal velocity, $\alpha_\infty=\langle|\mathbf E|^2\rangle_p/|\langle\mathbf E\rangle_p|^2$. Hence $\alpha_\infty/|V_p|=1/\int_{V_p}E_\mu E^\mu dV$, and Eq.~\eqref{eq:Lambda-def} takes the original form $2/\Lambda=\int_{S_p}E_iE^i\,dS\big/\int_{V_p}E_\mu E^\mu dV$. It was rederived in Ref.~\cite{CSGL} in three ways: from the projection on $\mathbf E$ used here, from the energetic representation of the dynamic tortuosity, and from the linear average $\langle\mathbf w_\varepsilon\cdot\mathbf e\rangle_p$, in which the bulk contribution of the displacement flow must be retained. The extension of this last route to higher orders would provide independent checks of Eqs.~\eqref{eq:Sigma-def}--\eqref{eq:V-def}; it is not pursued here. Equations \eqref{eq:Sigma-def} and \eqref{eq:V-def} give the general definitions of the surface and volume parameters for an arbitrary smooth pore geometry. Their microgeometric content is discussed in Sec.~\ref{sec7}, and their spectral meaning in Sec.~\ref{sec10}. In both, the relevant quantities are not $\Sigma$ and $V$ themselves but the combinations
\begin{equation}
\frac3\Sigma-\frac4{\Lambda^2}=\alpha_\infty d_2,
\qquad
\frac{15}{4V}-\frac{12}{\Lambda\Sigma}+\frac8{\Lambda^3}=\alpha_\infty d_3 .
\label{eq:combinations}
\end{equation}
The reason is simple. The boundary layer removes, order by order, a part of the flow, and these contributions add up in the inverse tortuosity $1/\alpha$, which measures the flow [Eq.~\eqref{eq:inverse-tortuosity-d}], not in $\alpha$, which is its inverse. The coefficients of $\alpha$ therefore mix the orders: $3/\Sigma$ contains, besides the genuine second-order contribution, the square $4/\Lambda^2$ of the first-order one, and $15/(4V)$ similarly contains products of lower-order terms. The combinations~\eqref{eq:combinations} remove these lower-order echoes and retain only the genuine contribution of each order. In the electrical analogy, in which $\alpha_\infty=\phi F$ with $F$ the formation factor~\cite{JKD}, $1/\alpha$ plays the role of an effective conductivity and $\alpha$ that of an effective resistivity: the boundary layer acts as a poorly conducting sheath along the walls, and its successive contributions are additive in the conductivity, not in the resistivity.

\subsection{Checks on exact solutions}\label{sec6-checks}

\paragraph{Straight circular pore of radius $R$.} In Gauss coordinates $(\theta,z,\zeta=R-r)$, $g_{\theta\theta}=(R-\zeta)^2$ and $g_{zz}=1$; the surface Christoffel symbols vanish, and those with an index $3$ are $\Gamma^\theta_{3\theta}=1/(R-\zeta)$ and $\Gamma^3_{\theta\theta}=-(R-\zeta)$, so that, on the wall, $\Gamma^k_{3k}=1/R$ ($H=-1/(2R)$, $K=0$) and $\mathring\Gamma^\theta_{3\theta}=-\mathring\Gamma^z_{3z}=1/(2R)$. The ideal field $E^z=1$ is uniform, $E^\mu{}_{;\nu}=0$, $E^k{}_{|k}=0$ and $Q^{(0)}=0$, so that $E_i\mathring\Gamma^i_{3j}E^j=-1/(2R)$ and $V^z=1/(2R)$. Equations \eqref{eq:d123}, with $Q^{(1)}=0$ (its Neumann datum~\eqref{eq:Qp1} vanishes), give $1/\alpha=1-2\varepsilon/R+\varepsilon^2/R^2+\varepsilon^3/(4R^3)+O(\varepsilon^4)$, which coincides with the expansion of the exact result $1-2I_1(R/\varepsilon)/[(R/\varepsilon)I_0(R/\varepsilon)]$. Equations \eqref{eq:Lambda-def}--\eqref{eq:V-def} then give
\begin{equation}
\Lambda=R,\qquad\Sigma=R^2,\qquad V=R^3,
\end{equation}
which is the normalization underlying Eq.~\eqref{eq:alpha-expansion-LSV}.

\paragraph{Curved, non-umbilic wall.} For the exterior of a circular cylinder ($\kappa_1=1/R$, $\kappa_2=0$) with an arbitrary wall slip $(s_\theta,s_z)\,e^{i(m\theta+kz)}$, the exact solution of $\varepsilon^2\Delta\mathbf w_{\mathrm{vort},\varepsilon}-\mathbf w_{\mathrm{vort},\varepsilon}=\mathbf 0$, $\nabla\!\cdot\mathbf w_{\mathrm{vort},\varepsilon}=0$ (modified Bessel functions $K_m$) was evaluated numerically. Computed with this solution, the tangential flux across the layer agrees with Eq.~\eqref{eq:deficit-flux} component by component to the three orders shown, and the normal velocity at the wall with the surface divergence of Eq.~\eqref{eq:deficit-flux}; the layer moment \eqref{eq:layer-moment} computed with a generic test field satisfying $E^3=0$ and $E^\mu{}_{;\mu}=0$ on $S_p$ reproduces the coefficients of $\varepsilon$, $\varepsilon^2$ and $\varepsilon^3$, with a residual of order $\varepsilon^4$. For the exterior of a sphere, the exact solution in $k_l(r/\varepsilon)$ with $s^i=g^{ij}\partial_jY_l$ reproduces the deficit flux as well, its term in $\varepsilon^2$ vanishing on this umbilic wall.

\paragraph{Dilute array of spheres.} The preceding checks do not involve the potentials $Q^{(0)}$ and $Q^{(1)}$. A dilute array of fixed obstacles provides a check that does: the ideal slip varies along the wall of each obstacle, $E^k{}_{|k}\neq0$, and neither potential vanishes. At first order in the solid volume fraction $c$, and for $|\varepsilon|$ small compared with the distance between obstacles, the fields near each obstacle are those of the same obstacle isolated in the uniform field $\mathbf e$, up to corrections of relative order $c$; each coefficient $d_k$ is then $c/V_s$ times the value of the integrals of Eqs.~\eqref{eq:d123} for one obstacle, $V_s$ being its volume. For spheres of radius $a$, with $\theta$ the polar angle measured from $\mathbf e$, $\Gamma^i_{3j}=-\delta^i_j/a$ on the wall, $E^\theta=-3\sin\theta/(2a)$, $E^k{}_{|k}=-3\cos\theta/a$, and the potentials are the dipoles $Q^{(0)}=-3a^2\cos\theta/(2r^2)$ and $Q^{(1)}=-3a\cos\theta/(2r^2)$. Equations~\eqref{eq:d123} give $d_1=9c/(2a)$, $d_2=9c/(2a^2)$ and $d_3=0$. On this umbilic wall the surface terms of $d_2$ cancel, and $d_2$ comes entirely from the volume integral of Eq.~\eqref{eq:d2-final}. The eight terms of $d_3$ do not vanish separately; the one in $Q^{(1)}$, computed with the explicit second potential, equals the term $g^{jk}\partial_jQ^{(0)}\partial_kQ^{(0)}$ that replaces it in Eq.~\eqref{eq:d3-final}, and in Eq.~\eqref{eq:d3-compact} the terms $-\tfrac18(\Gamma^k_{3k})^2E_iE^i$ and $-\tfrac12g_{ik}g^{j\ell}E^i{}_{|j}E^k{}_{|\ell}$, each equal to $-3\pi$ per sphere, are compensated by $g_{ij}V^iV^j$, equal to $6\pi$. The exact solution of Eq.~\eqref{eq:normalized-viscous-momentum} around a sphere is $\bm{\mathcal E}=\mathbf e+\nabla\times\nabla\times[f(r)\mathbf e]$, with $f'(r)=A\,e^{-r/\varepsilon}(1/r+\varepsilon/r^2)+B/r^2$, the constants $A$ and $B$ being fixed by the no-slip condition. Inserted in Eq.~\eqref{eq:tortuosity-boundary-total-deficit}, it gives $1/\alpha=1/\alpha_\infty-\tfrac92c\,[\varepsilon/a+(\varepsilon/a)^2]+O(c^2)$: the three coefficients agree.

\paragraph{Dilute array of fibres.} For circular fibres of radius $a$ across the flow, $V_s$ being their cross-sectional area and the integrals being taken per unit length, the wall is not umbilic, $\mathring\Gamma^\theta_{3\theta}=-\mathring\Gamma^z_{3z}=-1/(2a)$, and the terms in $\mathring\Gamma^i_{3j}$ enter together with the potentials, in the Neumann datum~\eqref{eq:Qp1} and in $V^i$. With $E^\theta=-2\sin\theta/a$, $Q^{(0)}=-2a\cos\theta/r$ and $Q^{(1)}=-\cos\theta/r$, Eqs.~\eqref{eq:d123} give $d_1=4c/a$, $d_2=2c/a^2$ and $d_3=-c/(2a^3)$, the coefficients of the expansion of the exact result $1/\alpha=1/\alpha_\infty-4c\,(\varepsilon/a)\,K_1(a/\varepsilon)/K_0(a/\varepsilon)+O(c^2)$. These checks involve constant curvatures only; the terms involving tangential variations of the curvature have not yet been tested on an exact solution.

	\section{Microgeometric Significance of the First Three Hierarchical Dimensional Visco-Inertial HF Parameters}\label{sec7}

	The compact forms~\eqref{eq:d2-compact}--\eqref{eq:d3-compact} make the microgeometric content of the parameters explicit. Apart from the prefactor $\alpha_\infty$, which only fixes the normalization, all the quantities involved are evaluated on the wall, with a single exception: the volume integral $\int_{V_p}g^{\mu\nu}Q^{(0)}_{;\mu}Q^{(0)}_{;\nu}\,dV$ in $d_2$, which is (twice) the kinetic energy of the displacement flow. By Green's identity, with $Q^{(0)}_{;3}=E^k{}_{|k}$, it equals the wall integral $-\int_{S_p}E^j\partial_jQ^{(0)}\,dS$; the volume form shows that it is nonnegative. All the terms depend on the whole pore space through the ideal field $\mathbf E$, which solves a potential problem in $V_p$. Once the wall trace of $\mathbf E$ is known, however, the other terms follow point by point on the wall, whereas the volume integral requires the solution of a second potential problem in the whole pore space, for $Q^{(0)}$, driven by $E^k{}_{|k}$. Whereas $\mathbf E$ expresses the response of the whole pore space to the macroscopic driving, $Q^{(0)}$ expresses its response to the boundary layer itself: this term is the back-reaction of the boundary layer on the flow in the entire pore space. We denote by $\kappa_E=-E_i\Gamma^i_{3j}E^j/(E_kE^k)$, taken at $\zeta=0$, the normal curvature of the wall in the direction of the slip, so that
	\begin{equation}
		E_i\mathring\Gamma^i_{3j}E^j=(H-\kappa_E)\,E_iE^i .
	\end{equation}

	\paragraph{The length $\Lambda$.} Equation~\eqref{eq:Lambda-def} weights the interface by the squared ideal slip: the wall regions where the potential flow is fast dominate the boundary-layer dissipation at leading order. Unlike the steady viscous flow, which concentrates in the widest channels, the ideal flow is fast in narrow passages as well. In an in-parallel bundle it has the same speed in all the tubes, so that $\Lambda=1/m_1$, with $m_k=\langle R^{-k}\rangle$ [Eq.~\eqref{eq:bundle}], is controlled by the narrowest ones, whereas $k_0=\phi\langle R^2\rangle/8$ is controlled by the widest, the averages being weighted by the volume fractions $p_i$; the form factor $M=8\alpha_\infty k_0/(\phi\Lambda^2)=\langle R^2\rangle m_1^2$ then becomes large. This is the mechanism invoked in Refs.~\cite{LBF,BLF} for the foams of Ref.~\cite{LBF}, where $\Lambda\simeq29$--$36\,\mu$m and $M\simeq9$--$19$. Sharp edges, such as the sharp corners observed on the membranes of these foams~\cite{LBF}, lower $\Lambda$ further, the measured value being then an apparent one (Appendix~\ref{app:wedge}).

	\paragraph{The surface parameter $\Sigma$.} The combination $3/\Sigma-4/\Lambda^2=\alpha_\infty d_2$ is the sum of two contributions of different nature.

	The first one, $\int_{S_p}(H-\kappa_E)E_iE^i\,dS$, follows point by point on the wall from the trace of the ideal field. It compares the mean curvature of the wall with its curvature along the slip. Both terms involve the unit normal field $n^\mu$. On the wall, $2H=-n^\nu{}_{;\nu}$ is minus its divergence [Eqs.~\eqref{eq:curvatures} and~\eqref{eq:B-normal-gradient}]: it measures how the normals spread apart, and the parallel surfaces widen, as one moves from the wall into the fluid. And $2\kappa_E\,E_\mu E^\mu=n^\nu(E_\mu E^\mu)_{;\nu}$ [first of Eqs.~\eqref{eq:wall-identities}] is the derivative along the normal of the squared ideal velocity $E_\mu E^\mu$, equal to $E_iE^i$ on the wall: it measures how fast this squared velocity decreases as one moves into the fluid. The two effects combine into a single divergence: the integrand is $(H-\kappa_E)E_iE^i=-\tfrac12(E_\mu E^\mu n^\nu)_{;\nu}$, minus one half of the divergence, on the wall, of the unit normal field weighted by the squared ideal velocity. It vanishes on umbilic walls (planes and spheres), where all normal curvatures are equal. It is negative when the slip runs along the direction in which the wall is most convex towards the fluid (or least concave), as for the axial flow in a circular tube [$\kappa_E=0$, $H=-1/(2R)$] or for the flow across a fibre [$\kappa_E=1/R$, $H=1/(2R)$]. It is positive in the opposite situation, as for the flow along a fibre [$\kappa_E=0$, $H=1/(2R)$].

	The second one, $\int_{V_p}g^{\mu\nu}Q^{(0)}_{;\mu}Q^{(0)}_{;\nu}\,dV\ge0$, represents this back-reaction. It is (twice) the kinetic energy of the displacement flow induced in the whole pore volume by the transpiration of the boundary layer. It is driven by $E^k{}_{|k}$, that is, by the convergence or divergence of the ideal slip along the wall, as occurs at constrictions, at the entrance and exit of throats, and around stagnation regions. It vanishes for straight prismatic pores.

	Consequently, $\Sigma$ becomes negative when $\alpha_\infty d_2<-4/\Lambda^2$, that is, when the curvature term is sufficiently negative and is not compensated by the displacement energy. Both signs of $\Sigma$ can therefore occur; the negative values reported in Ref.~\cite{LBF} are an instance. For in-parallel disorder, $Q^{(0)}=0$ and the sign is decided by the curvature term alone: $\alpha_\infty d_2=-m_2$, while $4/\Lambda^2=4m_1^2$, so that $\Sigma<0$ when the narrowest pores, in which the concave wall reduces the deficit carried by the layer, dominate $m_2$. For the log-normal in-parallel distribution of Ref.~\cite{BLF}, $m_2/m_1^2=e^{\sigma_p^2}$, and $\Sigma<0$ for $\sigma_p>(\ln4)^{1/2}\simeq1.18$, the threshold found there. In-series disorder acts in the opposite direction, through the energy of a displacement flow described below: for a purely in-series chain of segments of equal length, $3/\Sigma=3\langle R^{-4}\rangle/\langle R^{-2}\rangle>0$, the averages being taken along the chain.

	\paragraph{The volume parameter $V$.} The combination $15/(4V)-12/(\Lambda\Sigma)+8/\Lambda^3=\alpha_\infty d_3$ is the sum of four surface terms:
	\begin{equation}
		|V_p|\,d_3=\int_{S_p}\Big[\underbrace{-\tfrac12H^2E_iE^i}_{\le0}\;\underbrace{-\,H(H-\kappa_E)E_iE^i}_{\text{curvature anisotropy}}\;\underbrace{-\,\tfrac12g_{ik}g^{j\ell}E^i{}_{|j}E^k{}_{|\ell}}_{\le0}\;+\underbrace{g_{ij}V^iV^j}_{\ge0}\Big]dS .
		\label{eq:d3-anatomy}
	\end{equation}
	The first term is a mean-curvature penalty. The second couples the mean curvature with the curvature anisotropy along the slip. The third measures the non-uniformity of the slip along the wall; it involves the full intrinsic gradient~\eqref{eq:surface-operators} of the slip on $S_p$, including shear of the slip pattern, and not only its surface divergence. It does not involve the normal component $E^3{}_{;j}=\Gamma^3_{jk}E^k$ of the three-dimensional derivative. The last term, with $V^i$ given by Eq.~\eqref{eq:Vi-def}, is the mismatch between two corrections to the slip at the edge of the boundary layer: the one produced by the curvature-induced transpiration, $\mathring\Gamma^i_{3j}E^j$, and the one produced by the displacement flow, $g^{ij}\partial_jQ^{(0)}$. On umbilic walls, $\mathring\Gamma^i_{3j}=0$ and the last term reduces to the tangential kinetic energy of the displacement flow on the wall.

	\paragraph{Sign of $V$ and constrictions.} Since $15/(4V)=\alpha_\infty d_3+(4/\Lambda)(3/\Sigma-2/\Lambda^2)$, a positive $V$ together with a negative $\Sigma$ requires $\alpha_\infty d_3>(4/\Lambda)(2/\Lambda^2-3/\Sigma)>8/\Lambda^3$: a large and positive third-order contribution, whereas $\alpha_\infty d_3=-1/(4R^3)$ for a circular pore and $d_3<0$ for any straight prismatic pore. In Eq.~\eqref{eq:d3-anatomy}, only two terms can contribute positively: the curvature-anisotropy term, where $H(H-\kappa_E)<0$, and the mismatch term, which is nonnegative and contains the displacement flow. For in-parallel disorder, $Q^{(0)}=0$ and $\alpha_\infty d_3=-m_3/4$, so that a negative $\Sigma$ entails a negative $V$: $m_2>4m_1^2$ implies $15/(4V)=8m_1^3-4m_1m_2-m_3/4<-8m_1^3-m_3/4<0$. This is the negative $V$ found in Ref.~\cite{BLF} under strong in-parallel disorder. As the disorder increases, $V$ becomes negative before $\Sigma$ does: the Cauchy--Schwarz inequality $m_2^2\le m_1m_3$ gives $15/(4V)\le8m_1^3-4m_1m_2-m_2^2/(4m_1)$, and this bound is negative as soon as $m_2>(4\sqrt6-8)\,m_1^2\simeq1.80\,m_1^2$, so that every bundle with $1.80\,m_1^2<m_2<4m_1^2$ has $V<0$ and $\Sigma>0$.

	\paragraph{Displacement flow in a chain of segments.} Unlike in-parallel disorder, in-series disorder brings a back-reaction even in a one-dimensional description. The displacement flow is created where the slip varies along the wall, $E^k{}_{|k}$ being the Neumann datum of $Q^{(0)}$ [Eq.~\eqref{eq:Qp0}]. In a chain of straight circular segments, this happens only at the junctions. Far from them, the ideal flow is uniform in each segment, with a slip $E$ such that $ES$ is the same in all the segments of a path, $S=\pi R^2$, and $E^k{}_{|k}=0$ on the straight walls. Near a junction between segments of radii $R_1$ and $R_2$, the slip passes from $E_1$ to $E_2$ over a transition zone, which comprises the step and extends over about one radius on either side, where the ideal flow adjusts through evanescent modes (decaying as $e^{-3.83|z|/R}$ for an axisymmetric junction). To apply the present smooth-wall analysis to this zone, the step must be rounded, at a scale large compared with $|\varepsilon|$; $E^k{}_{|k}$ is then a smooth function, localized in the transition zone. Beyond it, $Q^{(0)}$ is harmonic in a straight tube with zero Neumann datum on the wall and, up to evanescent modes of the same kind, has a uniform axial gradient $a$. Integrating $\nabla^2Q^{(0)}=0$ over the fluid between two cross-sections taken on either side of the transition zone, and writing the wall integral of $E^k{}_{|k}$ as the flux of the slip through the two edges of the corresponding part of the wall, one obtains
	\begin{equation}
		a_2S_2+2\pi R_2E_2=a_1S_1+2\pi R_1E_1,
		\label{eq:junction-balance}
	\end{equation}
	whatever the rounding profile. Since $2\pi R\,\varepsilon E$ is the deficit flux carried by the layer and $\varepsilon aS$ the deficit flux carried by the displacement flow in the core [Eq.~\eqref{eq:core-deficit}], Eq.~\eqref{eq:junction-balance} expresses the conservation of the total deficit flux: the core flow compensates what the layer gains or loses at the junction. Along a flow path, $aS+2\pi RE$ thus takes the same value in all the segments; this value is fixed by the vanishing of the net variation of $Q^{(0)}$ along the path. The model of Ref.~\cite{BLF} neglects the transition zones: the flow in each segment is taken to be that of an infinite tube up to the junctions, where it changes abruptly. For this idealized geometry, Eqs.~\eqref{eq:Sigma-def}--\eqref{eq:V-def} reproduce exactly the high-frequency coefficients of that model, provided the one-dimensional displacement flow just described is retained; with $Q^{(0)}=0$ in the segments, they would not. For a purely in-series chain of segments of equal length, these equations, with this displacement flow, give the value $3/\Sigma=3\langle R^{-4}\rangle/\langle R^{-2}\rangle$ quoted above. The idealization thus retains the far field of the displacement flow created at the junctions, but not the contributions of the transition zones themselves: the energy of the near field of $Q^{(0)}$, the end corrections of the ideal flow mentioned in Ref.~\cite{BLF}, and, in $d_2$ and $d_3$, the terms on the rounded wall, which is curved along the slip and on which the slip varies rapidly. These local contributions are small, of relative order $R/L$, for segments of length $L\gg R$ joined smoothly on the scale of their radii, but not for short constrictions or chambers, nor for abrupt steps. The convex edge of a step, across which the fluid flows, is a wedge of apex angle $\gamma=\pi/2$: for a sharp edge, the ideal field is singular, $|\mathbf E|\propto r^{-1/3}$, and the formal expansion contains a term in $\varepsilon^{4/3}$, which precedes the term in $\varepsilon^2$ (Appendix~\ref{app:wedge}, with $w=4/3$). For an edge rounded with a radius of curvature $\rho$, the smooth-wall expressions apply when $|\varepsilon|\ll\rho$, and the local contributions depend strongly on $\rho$. At the throats, windows and junctions of a three-dimensional pore space, similarly, the wall can be more curved along the slip than across it, which makes the curvature-anisotropy term positive, and the slip converges or diverges over distances comparable with the pore size, so that the displacement flow, which enters the mismatch term, becomes genuinely two- or three-dimensional.

	\paragraph{The foams of Ref.~\cite{LBF}.} The values of $\Lambda$, $\Sigma$ and $V$ fitted there for three polyurethane foams fabricated with graphite particles give, by Eq.~\eqref{eq:combinations}, $\alpha_\infty d_2\Lambda^2\simeq-4.7$, $-5.2$ and $-4.9$, and $\alpha_\infty d_3\Lambda^3\simeq69$, $35$ and $37$, against $-1$ and $-1/4$ for a circular pore of radius $\Lambda$. The second-order term has the sign and the magnitude expected from strong in-parallel disorder: for a bundle, $\alpha_\infty d_2\Lambda^2=-m_2/m_1^2$, and a bundle with some 11--14\% of the pore volume in tubes 21 to 28 times narrower than the others reproduces it together with the measured $M$ (the log-normal bundle of Ref.~\cite{BLF} does so for $\sigma_p\simeq1.25$, but with $M\simeq10^2$). The third-order term has the opposite sign to that of any bundle: since $m_1m_3\ge m_2^2$ by the Cauchy--Schwarz inequality, a bundle with the same second-order term has $\alpha_\infty d_3\Lambda^3\le-(m_2/m_1^2)^2/4\simeq-6$ ($-7.4$ to $-8.4$ for the two-population bundles). A positive third-order term therefore requires features that a bundle lacks: walls curved along the slip, as at throats, rims and junctions, or a back-reaction of the boundary layer at constrictions.

	The log-normal networks of Ref.~\cite{BLF} give $V<0$ under strong in-parallel disorder, a discrepancy with the measurements attributed there to transverse flow and three-dimensional interconnectivity; in these networks, the in-series disorder that can make $V$ positive while $\Sigma$ remains negative raises $M$ to at least several hundred and $\alpha_\infty$ to about 2--3, against $M\simeq9$--$19$ and $\alpha_\infty\simeq1.1$ measured. Marked in-series contrasts along part of the paths can match the measured values, but only formally: networks of the same idealized type, in which some of the paths contain constrictions or chambers, reproduce the four fitted values $\alpha_\infty$, $M$, $\alpha_\infty d_2\Lambda^2$ and $\alpha_\infty d_3\Lambda^3$ of each foam, together with the fitted $\Lambda$, only if some passages have radii of 1--3\,$\mu$m, against a few hundred micrometres for the wide segments. Such networks, with radius contrasts of order 100, are configurations possible within the model, not a representation of the actual geometry of the foams. At such contrasts the neglect of the transition zones is not justified, and these passages, narrower than the viscous skin depth over the whole measurement band ($|\varepsilon|\simeq4.3$--$5.8\,\mu$m), are not in the high-frequency regime: the exact response of these networks in the band generally differs markedly from the fitted expansion.

	At $100$\,kHz, where $|\varepsilon|\simeq4.9\,\mu$m, the third-order term of $1/\alpha$ is about half the first-order one, and larger than the second-order one: the expansion is only marginally ordered there, the volume parameter carries much of the fit of Ref.~\cite{LBF}, and the fitted $V$ is best regarded as an effective value. The membranes, moreover, are treated as rigid, as the en-bloc motion of the solid assumed in Sec.~\ref{sec2} requires, whereas thin films attached to a stiff frame can deform at the pore scale and modify the flow~\cite{VB1}. The local theory, finally, leaves out scattering: at $100$\,kHz the wavelength is only about ten times the typical size of the solid heterogeneities, $300\,\mu$m~\cite{LBF}, so that scattering, while probably small, cannot be excluded. Any motion of the membranes, and any scattering, would be absorbed into the fitted values. With these reservations, the measurements point to strong constrictions or enlargements along part of the flow paths, where the wall is strongly curved along the slip and the boundary layer acts back on the core flow, rather than to in-parallel disorder alone. Apertures in the membranes are natural candidates; their contribution must then be computed from the general expressions, including the local contributions of their transition zones and sharp rims (Appendix~\ref{app:wedge}), and the narrowest of them may not be in the high-frequency regime at $100$\,kHz.

	\paragraph{Straight prismatic pores.} For a straight pore of arbitrary smooth cross-section $\mathcal A$, of area $A$ and boundary curve $\partial\mathcal A$ of perimeter $P$, the ideal field is uniform, $\mathbf E=\mathbf e_z$, and $\alpha_\infty=1$. The wall is a cylinder: $K=0$, $\kappa_E=0$, $H=\kappa/2$ with $\kappa$ the signed curvature of $\partial\mathcal A$ (negative for a convex section with fluid inside), $E^i{}_{|j}=0$, $E^k{}_{|k}=0$ and $Q^{(0)}=0$. Equations~\eqref{eq:Lambda-def}, \eqref{eq:d2-compact} and~\eqref{eq:d3-compact} reduce to
	\begin{equation}
		\frac2\Lambda=\frac PA,
		\qquad
		d_2=\frac1{2A}\oint_{\partial\mathcal A}\kappa\,dl=-\frac{\pi}{A},
		\qquad
		d_3=-\frac1{8A}\oint_{\partial\mathcal A}\kappa^2\,dl .
		\label{eq:prismatic}
	\end{equation}
	The second coefficient is topological: for a simply connected section, the total curvature of the boundary is $-2\pi$ whatever its shape. The third is the bending (elastica) energy of the boundary curve. For a circular section, Eq.~\eqref{eq:prismatic} gives $\Lambda=R$, $\Sigma=R^2$, $V=R^3$; for an elliptic section, $\Sigma$ remains fixed by $P$ and $A$ alone, while $V$ depends on the distribution of curvature along the boundary. If the section develops corners, $\oint\kappa^2dl$ diverges while $\oint\kappa\,dl$ remains finite: Eq.~\eqref{eq:prismatic} then holds only for $|\varepsilon|$ small compared with the radii of curvature of the rounded corners. For sharp corners the axial flow remains regular ($\mathbf E=\mathbf e_z$), but each corner contributes at order $\varepsilon^2$ with a coefficient depending on its angle, as for the heat content of polygonal domains~\cite{VDBS}; the second coefficient is then no longer topological. For a square section of side $a$, for instance, the exact result $1/\alpha=1-4\varepsilon/a+16\varepsilon^2/(\pi a^2)$, up to exponentially small terms, replaces the value $\pi\varepsilon^2/a^2$ of the second-order term given by Eq.~\eqref{eq:prismatic}. The nonanalytic corrections $\varepsilon^{w}$, $1<w<2$, of Refs.~\cite{FGL,AA,CSGL} concern a different situation, the flow across an edge, where the ideal field itself is singular; their practical meaning is discussed in Appendix~\ref{app:wedge}.

	\section{Thermal Boundary-Layer Analysis in Gauss Coordinates}\label{sec8}

	The thermal deficit problem~\eqref{eq:thermal_deficit_gauss} is scalar and involves no pressure-like field. Its high-frequency analysis is a simplified version of that of Sec.~\ref{sec6}, and we present it briefly. Following Ref.~\cite{LBF} [Eq.~(10b) there], the coefficients are written
	\begin{equation}
		\alpha'(\omega)=1+\frac{2\varepsilon'}{\Lambda'}+\frac{3\varepsilon'^2}{\Sigma'}+\frac{15\,\varepsilon'^3}{4V'}+O(\varepsilon'^4),
		\label{eq:alpha-thermal-expansion}
	\end{equation}
	with the same normalization as in the visco-inertial case: $\Lambda'=R$, $\Sigma'=R^2$, $V'=R^3$ for a straight circular pore.

	The thermal deficit obeys exactly the same screened equation as the vortical part of the visco-inertial deficit, Eq.~\eqref{eq:vort-problem-mom}, but with a uniform datum and no pressure-like coupling: unlike the visco-inertial deficit, which is accompanied by the displacement flow $\varepsilon^2\nabla Q_\varepsilon$ in the whole pore volume, it is a pure boundary-layer field, exponentially small at distances from $S_p$ large compared with $|\varepsilon'|$. It could therefore be expanded in the stretched variable $\varsigma'=\zeta/\varepsilon'$ alone, the metric being expanded at the wall; the successive orders then contain powers of $\varsigma'$ multiplying $e^{-\varsigma'}$. These powers come only from the geometry, through the spreading $\mathcal J_S$ of the parallel surfaces, which varies on the scale of the radii of curvature. Treating $\zeta$ as a second, slow variable resums them into slowly varying amplitudes. We adopt this representation as the thermal counterpart of the non-secular closure chosen in Sec.~\ref{sec6-closure}; the two-scale substitution is used here only for this resummation, the abstract coefficients being functions of $(x^\mu;\varsigma')$, where $\varsigma'$, like $\varsigma$, is complex and ranges over the sector $\mathcal C$ of Eq.~\eqref{eq:sector}, since $\varepsilon'$ has the same phase as $\varepsilon$. With the scalar Laplacian~\eqref{eq:div-grad} in a near-wall patch, the overlined scalar Laplacian, built with the overlined derivatives~\eqref{eq:chain-rule} in which $\varepsilon$ and $\varsigma$ are replaced by $\varepsilon'$ and $\varsigma'$, has the same three-term structure as Eq.~\eqref{eq:three-term}. The non-secular closure, discussed in Sec.~\ref{sec6}, gives the WKB form
	\begin{align}
		&d_{\varepsilon'}=e^{-\zeta/\varepsilon'}\sum_{m\ge0}\varepsilon'^mC'_{(m)}(q,\zeta),
		\qquad
		2C'_{(m);3}+\Gamma^k_{3k}C'_{(m)}=-\Delta C'_{(m-1)}
		&&\text{in }\mathcal V_S,\nonumber\\
		&C'_{(0)}=1,\qquad C'_{(m)}=0\ \ (m\ge1)
		&&\text{on }S_p.
		\label{eq:thermal-WKB}
	\end{align}
	The leading amplitude is the geometric spreading factor, $(\mathcal J_S^{1/2}C'_{(0)})_{;3}=0$, and the next one follows by integration along the normal:
	\begin{equation}
		C'_{(0)}=\mathcal J_S^{-1/2},\qquad C'_{(1)}=\tfrac12\mathcal J_S^{-1/2}\int_0^\zeta\mathcal J_S^{1/2}\,\Delta C'_{(0)}\,d\zeta_1\qquad\text{in }\mathcal V_S.
		\label{eq:thermal-A01}
	\end{equation}
	The inverse thermal tortuosity follows from Eq.~\eqref{eq:thermal-tortuosity-deficit-average}:
	\begin{equation}
		\frac1{\alpha'(\omega)}=1-\frac1{|V_p|}\int_{S_p}dS\int_0^\infty d_{\varepsilon'}\,\mathcal J_S\,d\zeta+O(\varepsilon'^\infty).
	\end{equation}
	As in Sec.~\ref{sec6}, the normal derivatives of the scalars $\mathcal J_SC'_{(m)}$ are covariant derivatives: $d/d\zeta=-(\cdot)_{;3}$. With $(\mathcal J_S^{1/2})_{;3}=\tfrac12\Gamma^k_{3k}\mathcal J_S^{1/2}$ and $(\mathcal J_S^{1/2})_{;3;3}=-\tfrac12\mathring\Gamma^k_{3\ell}\mathring\Gamma^\ell_{3k}\mathcal J_S^{1/2}$, Eq.~\eqref{eq:div-grad} gives $\Delta C'_{(0)}=g^{jk}C'_{(0)|j|k}+\tfrac12\mathring\Gamma^k_{3\ell}\mathring\Gamma^\ell_{3k}C'_{(0)}$, equal to $\tfrac12[\mathring\Gamma^k_{3\ell}\mathring\Gamma^\ell_{3k}]_0$ on $S_p$, where $C'_{(0)}=1$. Carrying the recursion~\eqref{eq:thermal-WKB} one step further (the fourth-order term requires $C'_{(1)}$ to second order in $\zeta$ and $C'_{(2)}$ to first order), the layer moment is, at each point of $S_p$,
	\begin{multline}
		\int_0^\infty d_{\varepsilon'}\,\mathcal J_S\,d\zeta=\varepsilon'-\tfrac12\Gamma^k_{3k}\,\varepsilon'^2-\tfrac14\mathring\Gamma^k_{3\ell}\mathring\Gamma^\ell_{3k}\,\varepsilon'^3\\
		+\Big[-\tfrac14\Gamma^j_{3j}\,\mathring\Gamma^k_{3\ell}\mathring\Gamma^\ell_{3k}+c\,g^{jk}\big(\Gamma^\ell_{3\ell}\big)_{|j|k}\Big]\varepsilon'^4+O(\varepsilon'^5),
		\label{eq:thermal-layer-moment}
	\end{multline}
	all quantities being taken at $\zeta=0$, where the numerical constant $c$ need not be specified: $g^{jk}(\Gamma^\ell_{3\ell})_{|j|k}$ is a surface divergence, whose integral over the closed surface $S_p$ vanishes. Hence
	\begin{equation}
		\begin{gathered}
		\frac1{\alpha'(\omega)}=1-\big(d'_1\varepsilon'+d'_2\varepsilon'^2+d'_3\varepsilon'^3+d'_4\varepsilon'^4\big)+O(\varepsilon'^5),\\
		|V_p|\,d'_1=|S_p|,
		\quad
		|V_p|\,d'_2=-\tfrac12\int_{S_p}\Gamma^k_{3k}\,dS,\\
		|V_p|\,d'_3=-\tfrac14\int_{S_p}\mathring\Gamma^k_{3\ell}\mathring\Gamma^\ell_{3k}\,dS ,
		\quad
		|V_p|\,d'_4=-\tfrac14\int_{S_p}\Gamma^j_{3j}\,\mathring\Gamma^k_{3\ell}\mathring\Gamma^\ell_{3k}\,dS ,
		\end{gathered}
		\label{eq:thermal-d123}
	\end{equation}
	that is, with the mean and Gaussian curvatures~\eqref{eq:curvatures}, $|V_p|\,d'_2=\int_{S_p}H\,dS$, $|V_p|\,d'_3=\tfrac12\int_{S_p}(K-H^2)\,dS$ and $|V_p|\,d'_4=\int_{S_p}H\,(H^2-K)\,dS$.

	\paragraph{Checks.} For a straight circular pore of radius $R$ ($H=-1/(2R)$, $K=0$), Eq.~\eqref{eq:thermal-d123} gives $1/\alpha'=1-2\varepsilon'/R+\varepsilon'^2/R^2+\varepsilon'^3/(4R^3)+\varepsilon'^4/(4R^4)+O(\varepsilon'^5)$, in agreement with the exact result $1-2I_1(R/\varepsilon')/[(R/\varepsilon')I_0(R/\varepsilon')]$. For the exterior of a sphere of radius $a$, the exact deficit is $(a/r)e^{-(r-a)/\varepsilon'}$, whose layer moment is exactly $\varepsilon'+\varepsilon'^2/a$; Eq.~\eqref{eq:thermal-layer-moment} gives the same result, since $H=1/a$ and $K-H^2=0$, the fourth-order term vanishing as well. For the exterior of a circular cylinder of radius $R$, the exact deficit $K_0(r/\varepsilon')/K_0(R/\varepsilon')$ was integrated numerically; its layer moment agrees with $\varepsilon'+\varepsilon'^2/(2R)-\varepsilon'^3/(8R^2)+\varepsilon'^4/(8R^3)$ with a residual of order $\varepsilon'^5$.

	\section{Microgeometric Significance of the First Three Hierarchical Dimensional Thermal HF Parameters}\label{sec9}

	Since $\alpha'_\infty=1$, Eqs.~\eqref{eq:alpha-thermal-expansion} and~\eqref{eq:thermal-d123} give
	\begin{align}
		\frac2{\Lambda'}&=\frac{|S_p|}{|V_p|},\label{eq:Lambdap-def}\\
		\frac3{\Sigma'}&=\frac4{\Lambda'^2}+\frac1{|V_p|}\int_{S_p}H\,dS,\label{eq:Sigmap-def}\\
		\frac{15}{4V'}&=\frac{12}{\Lambda'\Sigma'}-\frac8{\Lambda'^3}-\frac1{8|V_p|}\int_{S_p}(\kappa_1-\kappa_2)^2\,dS ,\label{eq:Vp-def}
	\end{align}
	where we used $H^2-K=\tfrac12[\mathring\Gamma^k_{3\ell}\mathring\Gamma^\ell_{3k}]_0=\tfrac14(\kappa_1-\kappa_2)^2$. The first relation is the Champoux--Allard length. The thermal parameters thus have a purely geometrical meaning, independent of any field. The surface parameter $\Sigma'$ involves the integral of the mean curvature, which, together with $|V_p|$ and $|S_p|$, belongs to the Minkowski functionals (quermass integrals) of the pore space: $|S_p|$ and $\int_{S_p}H\,dS$ are the first coefficients of the Steiner expansion $|S_p|\zeta+\zeta^2\int_{S_p}H\,dS+\tfrac13\zeta^3\int_{S_p}K\,dS$ of the volume of the fluid layer of thickness $\zeta$ along the wall, the Jacobian $\mathcal J_S$ in Eq.~\eqref{eq:thermal-layer-moment} being precisely the Steiner density. The volume parameter $V'$ is not a Minkowski functional: it involves the Willmore-type energy $\int_{S_p}(H^2-K)\,dS$, which measures the departure of the wall from umbilicity; it differs from the Willmore energy $\int_{S_p}H^2dS$~\cite{WIL,WIL82} by the topological term $\int_{S_p}K\,dS$ discussed below. In terms of the unit normal field, $2(H^2-K)=n_{\alpha;\beta}n^{\alpha;\beta}-\tfrac12(n^\lambda{}_{;\lambda})^2$ is the squared trace-free part of its gradient (Appendix~\ref{app:greek}). By the Gauss--Bonnet theorem, $\int_{S_p}K\,dS=2\pi\chi(S_p)$ is topological ($\chi$ being the Euler characteristic of the interface within the periodic cell), and $d'_3=\{2\pi\chi(S_p)-\int_{S_p}H^2dS\}/(2|V_p|)$. Since $(\kappa_1-\kappa_2)^2\ge0$, Eq.~\eqref{eq:Vp-def} gives
	\begin{equation}
		\frac{15}{4V'}-\frac{12}{\Lambda'\Sigma'}+\frac8{\Lambda'^3}\le0,
	\end{equation}
	with equality if and only if the wall is umbilic everywhere (planes and spheres). Hence, by $15/(4V')=d'_3+(4/\Lambda')(3/\Sigma'-2/\Lambda'^2)$, the volume parameter $V'$ is negative whenever $3/\Sigma'-2/\Lambda'^2<\Lambda'|d'_3|/4$, in particular whenever $\Sigma'<0$.

	These results coincide with the short-time asymptotics of the heat content of the pore space. As noted in Sec.~\ref{sec5}, $\sum_nb'^2_ne^{-t/\Theta'_n}$ is the fraction of heat, initially uniform, that remains in $V_p$ at time $t$ when the walls are kept isothermal. Its expansion $\sum_n\beta_n\,(D't)^{n/2}/|V_p|$, with $D'=\kappa/(\rho_0c_P)$, has been obtained for a general smooth domain by van den Berg and Gilkey~\cite{VBG}, the mean-curvature term having been established first by van den Berg and Le Gall~\cite{VBLG}. Since $1/\alpha'$ is $-i\omega$ times the Laplace transform of this fraction, each term $t^{n/2}$ gives $\Gamma(1+n/2)\,\varepsilon'^n$, and $d'_n=-\Gamma(1+n/2)\,\beta_n/|V_p|$. With their coefficients, in which the second fundamental form refers to the inward normal, this reproduces Eqs.~\eqref{eq:Lambdap-def}--\eqref{eq:Vp-def} exactly; in particular, their coefficient of $t^{3/2}$ reduces, in principal curvatures, to $\int_{S_p}(\kappa_1-\kappa_2)^2dS/(6\sqrt\pi)$, which is our Willmore-type integral. The two derivations are independent: theirs rests on invariance theory and on functorial properties of the heat kernel, ours on the boundary-layer construction of Sec.~\ref{sec8}. Their next coefficient, $\beta_4$, agrees in the same way with the fourth-order term $|V_p|\,d'_4=\int_{S_p}H\,(H^2-K)\,dS$ of Eq.~\eqref{eq:thermal-d123}, again a local wall integral, in which the mean curvature weights the departure from umbilicity. The two methods differ in how they extend. In the invariance-theory approach, each coefficient is a combination of all the invariants of the appropriate homogeneity, with universal constants that must be determined anew at each order from special cases; the number of invariants grows rapidly, and the determination becomes laborious beyond the first orders. In the boundary-layer approach, the transport equations~\eqref{eq:thermal-WKB} give each amplitude recursively from the previous one, and Watson's lemma, applied to the layer moment, turns it into a wall integral: higher orders require more algebra but no new principle.

	For a straight pore of cross-section $\mathcal A$ ($K=0$, $H=\kappa/2$),
	\begin{equation}
		d'_2=\frac1{2A}\oint_{\partial\mathcal A}\kappa\,dl=-\frac\pi A,
		\qquad
		d'_3=-\frac1{8A}\oint_{\partial\mathcal A}\kappa^2\,dl,
		\label{eq:prismatic-thermal}
	\end{equation}
	the value of $d'_2$ holding for a simply connected section; for a section with holes, $\oint_{\partial\mathcal A}\kappa\,dl=-2\pi\chi(\mathcal A)$, with $\chi(\mathcal A)=1-$(number of holes), and $d'_2=-\pi\chi(\mathcal A)/A$, in agreement with the planar heat content of van den Berg and Le Gall~\cite{VBLG}. These values are identical to the visco-inertial values~\eqref{eq:prismatic}, as required by the exact equivalence of the two problems for prismatic pores. The two derivations are independent, so this agreement checks both the vector and the scalar analyses.

	The comparison between Eqs.~\eqref{eq:Sigma-def}--\eqref{eq:V-def} and~\eqref{eq:Sigmap-def}--\eqref{eq:Vp-def} shows how the visco-inertial parameters generalize the thermal ones. The thermal problem weights the wall uniformly; the visco-inertial problem weights it by $E_iE^i$. In $\Sigma'$, the wall term $H=-\tfrac12n^\nu{}_{;\nu}$ is minus one half of the divergence of the unit normal field, which measures the widening of the parallel surfaces; in $\Sigma$, its counterpart $-\tfrac12(E_\mu E^\mu n^\nu)_{;\nu}$ (Sec.~\ref{sec7}) is minus one half of the divergence of the same field weighted by the squared ideal velocity, whose variation along the normals brings in the curvature anisotropy along the slip ($\mathring\Gamma^i_{3j}$). The visco-inertial problem also adds the tangential non-uniformity of the slip ($E^i{}_{|j}$) and the back-reaction of the boundary layer through the displacement flow ($Q^{(0)}$); none of these has a thermal counterpart in the present local theory.

	\section{Spectral and Temporal Significance of the High-Frequency Parameters}\label{sec10}

	We now combine the relaxation-mode representation of Sec.~\ref{sec5} with the high-frequency coefficients of Secs.~\ref{sec6}--\ref{sec9}. The result is a dictionary between the high-frequency parameters and the analytic continuation of the spectral zeta function $\mathcal Z(s)$ to negative $s$. It shows which parameters fix the \emph{shape} of the short-relaxation-time end of the spectrum, where the high-frequency dispersion is generated, and which ones are \emph{global sum rules} over the whole spectrum, of the same nature as the low-frequency parameters. The same dictionary gives the short-time behaviour of the impulse response and of the pressure field. We present it for the visco-inertial response; the thermal case follows by replacing $(b_n,\sigma_n,\alpha_\infty,\Lambda,\Sigma,V)$ by $(b'_n,\sigma'_n,1,\Lambda',\Sigma',V')$.

	\subsection{Mellin dictionary}

	Consider $1/\alpha$ as a function of $x=\varepsilon^2$, given by Eq.~\eqref{eq:AT-Stieltjes}. Since $\int_0^\infty x^{s-1}(1+x/\sigma)^{-1}dx=\sigma^s\pi/\sin(\pi s)$ for $0<\operatorname{Re}s<1$~[Ref.~\cite{GR}, formula~3.222.2], its Mellin transform is
	\begin{equation}
		\int_0^\infty x^{s-1}\,\frac{1}{\alpha}\,dx=\frac{\pi}{\sin\pi s}\,\mathcal Z(s),
		\qquad 0<\operatorname{Re}s<1,
		\label{eq:Mellin}
	\end{equation}
	with $\mathcal Z$ the spectral zeta function~\eqref{eq:AT-zeta}. Conversely, $1/\alpha$ is recovered by the inverse transform
	\begin{equation}
		\frac1{\alpha}=\frac1{2\pi i}\int_{c-i\infty}^{c+i\infty}x^{-s}\,\frac{\pi}{\sin\pi s}\,\mathcal Z(s)\,ds,
		\qquad 0<c<1 .
		\label{eq:Mellin-inverse}
	\end{equation}
	By the standard Mellin-transform method for asymptotic expansions~\cite{BH,FGD}, the small-$x$ expansion of $1/\alpha$ is obtained by moving the line of integration to the left: each pole of the integrand that is crossed contributes its residue. A simple pole at $s=-\nu_0$, where $(\pi/\sin\pi s)\,\mathcal Z(s)$ has the residue $r$, thus contributes the term $r\,x^{\nu_0}$. The factor $\pi/\sin\pi s$ has simple poles at the integers, with residues $(-1)^j$ at $s=-j$. The function $\mathcal Z(s)$ has poles where the spectral function $G$ has power-law terms. If $G(\sigma)\sim\sum_kg_k\sigma^{\gamma_k}$ as $\sigma\to0$, then $dG\simeq\gamma_kg_k\sigma^{\gamma_k-1}d\sigma$ near $\sigma=0$, and the lower end of the integral~\eqref{eq:AT-zeta} behaves as $\gamma_kg_k/(s+\gamma_k)$: $\mathcal Z$ has simple poles at $s=-\gamma_k$, with residues $\gamma_kg_k$. We shall find that $\gamma_k=k/2$, with $k$ odd, for smooth walls. Comparing with the high-frequency expansion $1/\alpha=1/\alpha_\infty-\sum_kd_k\varepsilon^k$ of Sec.~\ref{sec6}, we conclude:
	\begin{enumerate}
		\item The pole of $\pi/\sin\pi s$ at $s=0$ gives $\mathcal Z(0)=1/\alpha_\infty$, as it must.
		\item Odd powers of $\varepsilon$ are half-integer powers of $x$; they cannot come from $\pi/\sin\pi s$ and must come from poles of $\mathcal Z$ at $s=-j-\tfrac12$, i.e.\ from terms $\sigma^{j+1/2}$ in $G$:
		\begin{equation}
			d_{2j+1}=(-1)^j\pi\operatorname*{Res}_{s=-j-\frac12}\mathcal Z(s),
			\qquad
			g_{2j+1}=\frac{(-1)^j\,d_{2j+1}}{\pi\,(j+\tfrac12)} .
			\label{eq:odd-dictionary}
		\end{equation}
		\item Even powers of $\varepsilon$ are integer powers of $x$. They come from the poles of $\pi/\sin\pi s$ at $s=-j$, where $\mathcal Z$ is regular; their coefficients are the values of the analytic continuation of $\mathcal Z$:
		\begin{equation}
			d_{2j}=(-1)^{j+1}\,\mathcal Z(-j) .
			\label{eq:even-dictionary}
		\end{equation}
		\item If $G$ contained an integer power $\sigma^j$, the pole of $\mathcal Z$ at $s=-j$ would coincide with that of $\pi/\sin\pi s$, and the resulting double pole would produce a term $\varepsilon^{2j}\ln\varepsilon$. The expansion of Sec.~\ref{sec6} contains no logarithm to the order considered. Hence $G$ has \emph{no} term in $\sigma$, and its short-$\sigma$ expansion proceeds by half-integer powers only.
	\end{enumerate}
	The value $\mathcal Z(-1)$ is the regularized value of the divergent sum $\sum_nb_n^2/\sigma_n$. Concretely,
	\begin{equation}
		\mathcal Z(-1)=\lim_{\sigma_0\to0}\Big[\sum_{\sigma_n>\sigma_0}\frac{b_n^2}{\sigma_n}-\frac{g_1}{\sigma_0^{1/2}}\Big],
		\label{eq:Z-finite-part}
	\end{equation}
	that is, the sum over the modes with relaxation time longer than $\sigma_0/\nu$, minus its divergent part, which is fixed by the leading tail coefficient $g_1$.

	With the parameters of Eqs.~\eqref{eq:Lambda-def}--\eqref{eq:V-def}, the dictionary reads
	\begin{align}
		G(\sigma)&\simeq\frac{4}{\pi\alpha_\infty\Lambda}\,\sigma^{1/2}
		-\frac{2}{3\pi\alpha_\infty}\Big(\frac{15}{4V}-\frac{12}{\Lambda\Sigma}+\frac8{\Lambda^3}\Big)\sigma^{3/2}+O(\sigma^{5/2}),
		\qquad\sigma\to0,
		\label{eq:G-tail}\\
		\mathcal Z(-1)&=\frac1{\alpha_\infty}\Big(\frac3\Sigma-\frac4{\Lambda^2}\Big).
		\label{eq:Z-minus-one}
	\end{align}
	The leading term of Eq.~\eqref{eq:G-tail} is the scaling law obtained by Avellaneda and Torquato for the cumulative distribution of relaxation times~\cite{AT}. Equation~\eqref{eq:G-tail} extends it to the next order. For the thermal problem, the same dictionary, with $\alpha_\infty$ replaced by $1$ and all quantities primed, gives Eqs.~\eqref{eq:G-tail}--\eqref{eq:Z-minus-one} for $G'$ and $\mathcal Z'$; with the geometric values~\eqref{eq:Sigmap-def}--\eqref{eq:Vp-def} of the combinations of $\Lambda'$, $\Sigma'$ and $V'$, they read
	\begin{align}
		G'(\sigma')&\simeq\frac{4}{\pi\Lambda'}\,\sigma'^{1/2}+\frac{1}{12\pi|V_p|}\int_{S_p}(\kappa_1-\kappa_2)^2dS\;\sigma'^{3/2}+O(\sigma'^{5/2}),\qquad\sigma'\to0,
		\label{eq:G-tail-thermal}\\
		\mathcal Z'(-1)&=\frac3{\Sigma'}-\frac4{\Lambda'^2}=\frac1{|V_p|}\int_{S_p}H\,dS .
		\label{eq:Z-minus-one-thermal}
	\end{align}
	For a finite periodic cell the spectrum is discrete, and $G$ is a staircase. It enters the responses, however, only through integrals such as~\eqref{eq:AT-Stieltjes} or the impulse response, whose kernels average it over a range of relaxation times around $|\varepsilon|^2$ or $\nu t$. Expansions such as~\eqref{eq:G-tail} are to be understood in this averaged sense, which is the one implicit in the Mellin method: they reproduce the asymptotic expansions of these transforms, but do not by themselves describe the pointwise deviation of the staircase from the smooth law.

	\begin{table}[t]
		\caption{The successive low- and high-frequency parameters as spectral data of the relaxation-time distribution, through the zeta function $\mathcal Z(s)=\sum_nb_n^2\sigma_n^s$. Positive integers: moments (convergent sums). Negative half-integers: poles, fixed by the short-relaxation-time tail of $G$. Negative integers: regularized values of divergent sums, global sum rules.}
		\label{tab:dictionary}
		\begin{ruledtabular}
			\begin{tabular}{lll}
				$s$ & spectral datum & visco-inertial parameter\\
				\hline
				$2$ & moment $\mathcal Z(2)$ & $\alpha_0\,(k_0/\phi)^2$\\
				$1$ & moment $\mathcal Z(1)$ & $k_0/\phi$\\
				$0$ & moment $\mathcal Z(0)$ & $1/\alpha_\infty$\\
				$-\tfrac12$ & residue & $2/(\pi\alpha_\infty\Lambda)$\\
				$-1$ & regularized value $\mathcal Z(-1)$ & $(3/\Sigma-4/\Lambda^2)/\alpha_\infty$\\
				$-\tfrac32$ & residue & $-\big(15/(4V)-12/(\Lambda\Sigma)+8/\Lambda^3\big)/(\pi\alpha_\infty)$\\
			\end{tabular}
		\end{ruledtabular}
	\end{table}

	\subsection{Short-time and whole-spectrum information}

	Table~\ref{tab:dictionary} separates the high-frequency parameters into two families.

	The odd-order coefficients---$\Lambda$, then the combination $15/(4V)-12/(\Lambda\Sigma)+8/\Lambda^3$---are residues. They depend only on the behaviour of $G$ as $\sigma\to0$, that is, on the modes of short relaxation time, confined near the walls, which generate the high-frequency dispersion. They fix the \emph{shape} of this end of the spectrum: $\Lambda$ its leading $\sigma^{1/2}$ law, and the combination containing $V$, equal to $\alpha_\infty d_3$ [Eq.~\eqref{eq:combinations}], its first correction [Eq.~\eqref{eq:G-tail}]. This correction enhances the tail when $\alpha_\infty d_3<0$, as for straight pores, and depletes it when $\alpha_\infty d_3>0$, as for the values fitted in Ref.~\cite{LBF} (Sec.~\ref{sec7}). In this respect, beyond $\Lambda$, the new information on the shape of the relaxation spectrum is concentrated in the single combination $15/(4V)-12/(\Lambda\Sigma)+8/\Lambda^3$.

	The even-order coefficients---first $3/\Sigma-4/\Lambda^2$---are regularized values of divergent sums. By Eq.~\eqref{eq:Z-finite-part}, they depend on the entire spectrum, long relaxation times included. They are global sum rules, the counterparts at negative $s$ of the moments $\mathcal Z(0)=1/\alpha_\infty$, $\mathcal Z(1)=k_0/\phi$ and $\mathcal Z(2)=\alpha_0k_0^2/\phi^2$. It is consistent with this global character that $3/\Sigma-4/\Lambda^2$ is the only coefficient involving a volume integral, the energy of the displacement flow in Eq.~\eqref{eq:Sigma-def}. The global character in the spectrum and the global character in space do not coincide, however: the odd-order coefficient $d_3$ also depends on the whole pore space, through the wall values of $Q^{(0)}$, which solves a Neumann problem in $V_p$.

	This distinction bears directly on the construction of full-frequency models. A model of $\alpha(\omega)$ of relaxational type amounts to a model of $G$. Such a model reproduces the high-frequency dispersion correctly only if its short-time tail has the geometric coefficients~\eqref{eq:G-tail}, and its global balance only if its moments and regularized moments take the geometric values of Table~\ref{tab:dictionary}. As an example, the Johnson--Koplik--Dashen function $\alpha=\alpha_\infty+(\phi\varepsilon^2/k_0)\sqrt{1+4\alpha_\infty^2k_0^2/(\phi^2\Lambda^2\varepsilon^2)}$~\cite{JKD} has the expansion
	\begin{equation}
		\frac{\alpha}{\alpha_\infty}=1+\frac{2\varepsilon}{\Lambda}+\frac{\Lambda\phi^2}{4\alpha_\infty^2k_0^2}\,\varepsilon^3+O(\varepsilon^5).
		\label{eq:JKD-expansion}
	\end{equation}
	It therefore imposes $3/\Sigma=0$, i.e.\ $\mathcal Z(-1)=-4/(\alpha_\infty\Lambda^2)$, and it ties the next tail coefficient of its spectrum to the static permeability, $15/(4V)=\Lambda\phi^2/(4\alpha_\infty^2k_0^2)$, instead of to the wall geometry of Eq.~\eqref{eq:V-def}. The departure of these model-implied values from the geometric ones measures how well the model represents the onset of the visco-inertial relaxation. This gives a spectral rationale for the observation of Ref.~\cite{LBF} that, in materials with numerous internal surfaces, the volume-related correction can be more significant than its asymptotic order suggests: the parameter $V$ controls the first correction to the shape of the part of the spectrum produced by the walls.

	\subsection{Short-time response and extended Curie--von Schweidler law}

	Since $\varepsilon^k=\nu^{k/2}(-i\omega)^{-k/2}$ and $(-i\omega)^{-\mu}$ is the Fourier transform of $H(t)\,t^{\mu-1}/\Gamma(\mu)$, Watson's lemma~\cite{BH} translates the high-frequency expansion into the short-time behaviour of the relaxation function~\eqref{eq:AT-relaxation-function}:
	\begin{equation}
		R(t)=\sum_nb_n^2e^{-t/\Theta_n}
		\begin{aligned}[t]
		&\simeq\frac1{\alpha_\infty}\Big[1-\frac4\Lambda\sqrt{\frac{\nu t}{\pi}}-\Big(\frac3\Sigma-\frac4{\Lambda^2}\Big)\nu t\\
		&\qquad-\frac{4}{3\sqrt\pi}\Big(\frac{15}{4V}-\frac{12}{\Lambda\Sigma}+\frac8{\Lambda^3}\Big)(\nu t)^{3/2}\Big]+O(t^2).
		\end{aligned}
		\label{eq:R-short-time}
	\end{equation}
	Its derivative is the initial deceleration of the mean flow after an impulsive start. It obeys an extended Curie--von Schweidler law~\cite{CUR,SCHW}, the power-law decay $t^{-n}$ of the current that follows a voltage step in dielectrics, here with $n=1/2$ and corrections in powers of $t^{1/2}$,
	\begin{equation}
		-\frac{dR}{dt}=\sum_n\frac{b_n^2}{\Theta_n}e^{-t/\Theta_n}
		\simeq\frac\nu{\alpha_\infty}\Big[\frac{2}{\Lambda}\frac{1}{\sqrt{\pi\nu t}}+\Big(\frac3\Sigma-\frac4{\Lambda^2}\Big)+\frac{2\sqrt{\nu t}}{\sqrt\pi}\Big(\frac{15}{4V}-\frac{12}{\Lambda\Sigma}+\frac8{\Lambda^3}\Big)\Big]+O(t).
		\label{eq:CvS}
	\end{equation}
	The divergent sum $\sum_nb_n^2/\Theta_n$ that one would obtain by setting $t=0$ is thereby resolved: its divergent part is the $t^{-1/2}$ law with amplitude $2\sqrt\nu/(\alpha_\infty\Lambda\sqrt\pi)$, and its finite part is the regularized value $\nu\mathcal Z(-1)$, carried by $\Sigma$.

\begin{figure}[t]
		\centering
		\includegraphics[width=0.48\textwidth]{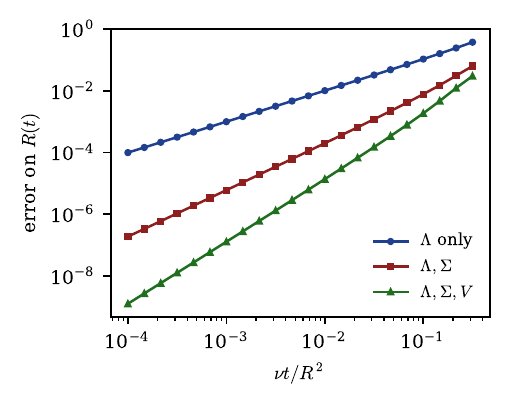}
		\caption{Straight circular pore: absolute error of the short-time expansion~\eqref{eq:R-short-time} of the relaxation function $R(t)=\sum_nb_n^2e^{-t/\Theta_n}$ (exact modal sum, $4\times10^5$ modes), truncated after the term in $\Lambda$, $\Sigma$ and $V$, respectively. The errors decrease as $t$, $t^{3/2}$ and $t^2$: each higher-order parameter improves the description of the onset of the visco-inertial relaxation.}
		\label{fig:tube}
	\end{figure}

	These formulas have been checked on the straight circular pore, for which $b_n^2=4/j_{0,n}^2$ and $\sigma_n=R^2/j_{0,n}^2$, $j_{0,n}$ being the zeros of $J_0$. With $4\times10^5$ modes, the error of Eq.~\eqref{eq:R-short-time} truncated after the $\Lambda$, $\Sigma$ and $V$ terms decreases as $t$, $t^{3/2}$ and $t^2$, respectively (Fig.~\ref{fig:tube}).

	Moreover, here $\mathcal Z(s)=4R^{2s}\sum_nj_{0,n}^{-2-2s}$ is a Bessel zeta function, which has poles only at half-integers. Its regularized values are $\mathcal Z(-1)=-1/R^2$ (since $\sum_nj_{0,n}^0=-1/4$ in the zeta-regularized sense) and $\mathcal Z(-2)=1/(4R^4)$ (evaluated numerically from McMahon's expansion of the zeros). By Eq.~\eqref{eq:even-dictionary} these give $d_2=-1/R^2$ and $d_4=-1/(4R^4)$, in agreement with the expansion of the exact result $1-1/\alpha=2I_1(R/\varepsilon)/[(R/\varepsilon)I_0(R/\varepsilon)]=2\varepsilon/R-\varepsilon^2/R^2-\varepsilon^3/(4R^3)-\varepsilon^4/(4R^4)+\cdots$ (Sec.~\ref{sec6-checks}).

	For the thermal problem, $R'(t)=\sum_nb'^2_ne^{-t/\Theta'_n}$ is the fraction of heat---or of walkers diffusing with diffusivity $D=\kappa/\rho_0c_P$ from an initially uniform distribution---not yet absorbed by the walls, and
	\begin{equation}
		R'(t)\simeq1-\frac{2}{\sqrt\pi}\frac{|S_p|}{|V_p|}\sqrt{Dt}-\frac{1}{|V_p|}\int_{S_p}H\,dS\;Dt+\frac{1}{6\sqrt\pi\,|V_p|}\int_{S_p}(\kappa_1-\kappa_2)^2dS\;(Dt)^{3/2}+O(t^2).
		\label{eq:R-thermal-short-time}
	\end{equation}
	Short-time expansions of this type, in which surface-to-volume ratio and curvature appear successively, are familiar in the theory of diffusion in confined geometries and of diffusion NMR~\cite{MSS}.

	\subsection{Structure of the pressure field}

	The same translation applies to the pressure. By Eqs.~\eqref{eq:AT-Q}, \eqref{eq:AT-impulse} and~\eqref{eq:expansion-principle}, $\varepsilon^2Q_\varepsilon=\sum_m\varepsilon^{m+1}Q^{(m)}$ is the Fourier transform of $\nu H(t)\sum_nb_nQ_ne^{-t/\Theta_n}$, so that
	\begin{multline}
		\nu\sum_nb_nQ_n\,e^{-t/\Theta_n}
		\simeq\sum_{m\ge0}\frac{\nu\,(\nu t)^{(m-1)/2}}{\Gamma\!\big(\frac{m+1}{2}\big)}\,Q^{(m)}\\
		=\sqrt{\frac\nu{\pi t}}\,Q^{(0)}+\nu\,Q^{(1)}+\frac{2\nu\sqrt{\nu t}}{\sqrt\pi}\,Q^{(2)}+\cdots\qquad\text{in }V_p .
		\label{eq:p-short-time}
	\end{multline}
	The visco-inertial pressure thus appears as a series in powers of $(\nu t)^{1/2}$ starting at $t^{-1/2}$, whose coefficients are the successive harmonic potentials of Sec.~\ref{sec6}. Since the pressure has no boundary-layer part (Sec.~\ref{sec6-pressure}), this expansion holds uniformly up to the wall. The visco-inertial pressure is therefore singular at $t=0^+$, in the volume as well as at the wall, but integrably so: it contains neither a $\delta(t)$ term nor any concentration on the wall. Similarly, at a fixed interior point, where the vortical part is exponentially small, Eq.~\eqref{eq:core-deficit} gives
	\begin{equation}
		\frac{\mathbf v(t)-\mathbf v_\infty(t)}{v_0}\simeq-2\sqrt{\frac{\nu t}{\pi}}\,\nabla Q^{(0)},
		\qquad
		\sum_n\frac{b_n}{\sigma_n}\bm\Psi_n\,e^{-t/\Theta_n}\simeq\frac{\nabla Q^{(0)}}{\sqrt{\pi\nu t}} .
		\label{eq:v-short-time}
	\end{equation}
	the second form following from the first by differentiation with respect to $t$, since $(\mathbf v-\mathbf v_\infty)/v_0=\sum_nb_n\bm\Psi_n(e^{-t/\Theta_n}-1)$ and $\Theta_n=\sigma_n/\nu$.

	The mechanism is the displacement effect of the nascent boundary layer. The displacement thickness of the layer grows as $(\nu t)^{1/2}$; wherever the wall slip converges or diverges ($E^k{}_{|k}\neq0$), the layer injects into the core a normal velocity growing like $(\nu t)^{1/2}$. In an incompressible fluid this injection is felt instantaneously throughout the pore volume, as the potential flow $\nabla Q^{(0)}$, whose acceleration, and hence pressure, behave as $t^{-1/2}$. The same mechanism produces the Basset history force, which, for a sphere started impulsively, behaves as $t^{-1/2}$ at short times~\cite{LL_FM}. The no-slip jump at the wall, by contrast, is created entirely by the vorticity sheet, that is, by the viscous term: the pressure gradient contributes to the velocity only at order $t^{1/2}$.

	This has a consequence for the modal sums. It has been argued that $\sum_nb_nQ_n=0$ and $\sum_nb_n\bm\Psi_n/\sigma_n=\mathbf 0$ in the pore volume~\cite{L-AT}, on the basis that only Bernoulli pressures should exist at $t=0^+$; these are quadratic in the velocity, and therefore vanish in the linearized description. In the Abel sense of Sec.~\ref{sec5-convergence}, Eqs.~\eqref{eq:AT-Q}, \eqref{eq:AT-deficit}, \eqref{eq:core-deficit} and~\eqref{eq:expansion-principle} give, at interior points,
	\begin{align*}
		\sum_n\frac{b_nQ_n}{1+\varepsilon^2/\sigma_n}&=Q_\varepsilon=\sum_m\varepsilon^{m-1}Q^{(m)},\\
		\sum_n\frac{b_n\bm\Psi_n/\sigma_n}{1+\varepsilon^2/\sigma_n}&=\frac{\mathbf w_\varepsilon}{\varepsilon^2}=\sum_m\varepsilon^{m-1}\nabla Q^{(m)}+O(\varepsilon^\infty),
	\end{align*}
	whose time-domain forms are Eqs.~\eqref{eq:p-short-time}--\eqref{eq:v-short-time}. The two sums therefore diverge, as $\varepsilon^{-1}$ or $t^{-1/2}$, wherever $Q^{(0)}$, respectively $\nabla Q^{(0)}$, does not vanish. The identities require $Q^{(0)}\equiv0$, that is, $E^k{}_{|k}\equiv0$ on the walls, and then also $Q^{(1)}\equiv0$, since the Abel values of the two sums are then $Q^{(1)}$ and $\nabla Q^{(1)}$. The identities hold for straight prismatic pores (for which the visco-inertial pressure vanishes identically), but not in general. The argument overlooks the displacement effect, whose bulk contribution had already been identified, at the level of $\Lambda$, in Ref.~\cite{CSGL}. What remains true is that the visco-inertial velocity vanishes in the volume as $t\to0^+$, and that the visco-inertial pressure has no impulsive part.

	The thermal problem offers an instructive contrast. Its deficit has no part outside the boundary layer (Sec.~\ref{sec8}), so that, at any interior point, the thermal analogues of these sums vanish as $t\to0^+$ faster than any power of $t$. The failure of the visco-inertial identities is thus a specific consequence of incompressibility, which converts a boundary-layer displacement into an instantaneous volume flow.

	\section{Conclusion}\label{sec11}

	Within the local theory, the question raised in the Introduction now has a definite answer. For an arbitrary smooth pore geometry, the high-frequency expansions of the dynamic tortuosity and of the dynamic compressibility follow from a recursive hierarchy of boundary-layer problems, which can be continued to any order, each coefficient being obtained from the solutions at the preceding orders. We have carried it out explicitly up to the third coefficient for the visco-inertial response, and up to the fourth for the thermal response. On the visco-inertial side, the surface and volume parameters $\Sigma$ and $V$ are given by Eqs.~\eqref{eq:Sigma-def}--\eqref{eq:V-def}: they involve only the wall trace of the ideal field, its tangential gradient, the principal curvatures of the walls, and a single harmonic potential, that of the displacement flow by which the boundary layer acts back on the pore space. The curvature term of $\Sigma$ is the integral over the wall of minus one half of the divergence of the unit normal field weighted by the squared ideal velocity; with a uniform weight, it would reduce to the integral mean curvature that enters $\Sigma'$. Applied to the parameters fitted in Ref.~\cite{LBF}, these expressions show that the second-order term $\alpha_\infty d_2$ has the sign and the magnitude expected from strong in-parallel disorder, which the series--parallel model of Ref.~\cite{BLF} identified as the origin of the large form factors and of the negative $\Sigma$. The third-order term $\alpha_\infty d_3$, by contrast, is large and positive, whereas any in-parallel arrangement makes it negative; it points to strong constrictions or enlargements along part of the flow paths, at which the boundary layer acts back on the core flow. This reading is qualitative: the expansion is only marginally ordered at the measurement frequencies, the fitted $V$ being an effective value; the local contributions of the junctions remain to be computed; and the fitted values would also absorb any motion of the membranes and any scattering, both neglected here.

	On the thermal side, where no such back-reaction exists, $\Sigma'$ and $V'$ reduce to the integral mean curvature and to a Willmore-type measure of the non-umbilicity of the walls [Eqs.~\eqref{eq:Sigmap-def}--\eqref{eq:Vp-def}], and the fourth coefficient $d'_4$ to the integral of $H(H^2-K)$ [Eq.~\eqref{eq:thermal-d123}]; up to this order, the thermal coefficients agree with the heat-content coefficients of van den Berg and Gilkey~\cite{VBG}. Straight prismatic pores give a common check of both analyses: the second coefficient is then topological, and the third is the bending energy of the boundary of the cross-section.

	The relaxation-mode representation of Avellaneda and Torquato places these results in a wider frame. It shows that the normalization $\pi=\Phi+\varepsilon^2Q_\varepsilon$ is the natural one, and that all the parameters, low- and high-frequency alike, are spectral data of a single function $\mathcal Z(s)$ of the relaxation-time distribution (Table~\ref{tab:dictionary}). They are not all of the same kind. The odd-order coefficients fix the shape of the short-relaxation-time end of the spectrum: at leading order through $\Lambda$, as found by Avellaneda and Torquato, and at the next order through the single combination $15/(4V)-12/(\Lambda\Sigma)+8/\Lambda^3$. The even-order ones are regularized negative moments of the whole spectrum, the counterparts of $\alpha_\infty$, $k_0$ and $\alpha_0$. This distinction tells how the new parameters should enter full-frequency relaxational models: as constraints on the shape of the spectral tail on the one hand, as global sum rules on the other. It also explains why classical models, which tie the higher coefficients to low-frequency data [Eq.~\eqref{eq:JKD-expansion}], can miss information that is essential in materials with many internal surfaces.

	In the time domain, the same analysis gives the onset of the visco-inertial relaxation after an impulsive excitation, an extended Curie--von Schweidler law, and the structure of the pressure field. The visco-inertial pressure has no boundary-layer part. At short times it behaves as $t^{-1/2}$ throughout the pore volume, because the displacement effect of the nascent boundary layer is transmitted instantaneously by incompressibility; yet it contains no impulsive term. Because of this $t^{-1/2}$ behaviour, the modal sums of pressures, and of velocities weighted by $1/\sigma_n$ (which represent the acceleration of the core flow), diverge in the pore volume in general, whereas the visco-inertial velocity vanishes there as $t\to0^+$. Their Abel limits are finite only when the ideal wall slip has zero surface divergence, and are then given by the next pressure potential $Q^{(1)}$ and its gradient.

	The smoothness of the walls, assumed throughout, is less restrictive than it may seem. Sharp edges across which the fluid flows, such as wedges of apex angle $\gamma$, make the ideal field singular, and the formal expansion then contains, besides the Johnson term in $\varepsilon$, a nonanalytic term in $\varepsilon^w$, with $w=2\pi/(2\pi-\gamma)$. For small angles, Appendix~\ref{app:wedge} shows that these two terms cannot be read separately: both diverge as $\gamma\to0$ (for the Johnson term, $\Lambda\to0$), with equal and opposite divergent parts, and their sum tends to the finite knife-edge law $\varepsilon\ln\varepsilon$, in agreement with the computations of Langlois~\cite{LANG}. They separate only at unphysical skin depths much smaller than $\ell\,e^{-2\pi/\gamma}$, with $\ell$ the geometrical length of Sec.~\ref{app:wedge-field}, such as the length of the faces or the pore aperture; for small angles, these skin depths lie far below the mean free path of the gas molecules. A vanishing $\Lambda$ is therefore not a breakdown of the high-frequency description: measurements yield apparent values of $\Lambda$, and likewise of $\Sigma$ and $V$, which depend only logarithmically on frequency and are those of the geometry with its edges rounded at the scale of the skin depth, to which the smooth-wall expressions apply. On the spectral side, the poles of $\mathcal Z$ at $s=-\tfrac12$ and $s=-w/2$ coalesce into a double pole, so that $G(\sigma)\propto\sigma^{1/2}\ln(1/\sigma)$ as $\sigma\to0$ [Eq.~\eqref{eq:wedge-G}]: the relaxation spectrum remains well defined when $\Lambda\to0$. This analysis concerns isolated edges; rough or fractal boundaries, in which singular features accumulate at all scales, would call for a different treatment.

	Several extensions remain within the local theory: testing on exact solutions the terms that involve tangential variations of the curvature; evaluating explicitly the next visco-inertial coefficient $d_4$, which, like its thermal counterpart $d'_4$, will be a global sum rule; treating the anisotropic case; and constructing full-frequency models able to incorporate measured values of $\Sigma$ and $V$. On this last point, the existing models are not sufficient as they stand. In the JCAPL model~\cite{LAF_PhD,LAF2,LAF0}, $\Sigma$ and $V$ are not free parameters: they are fixed by the form factors $M$ (Sec.~\ref{sec7}) and $P=M/[4(\alpha_0/\alpha_\infty-1)]$, $\alpha_0$ being the static viscous tortuosity (Sec.~\ref{sec5}), e.g.\ $3/\Sigma=-8(P-1)/(M\Lambda^2)$, so that independently measured values cannot be accommodated, as found in Ref.~\cite{LBF}. The relaxation-mode representation of Sec.~\ref{sec5} suggests a more open and better founded route: to model directly a positive relaxation-time distribution $G(\sigma)$, constrained by whichever spectral data are available or judged most relevant. On the low-frequency side, these are the moments $\mathcal Z(1)=k_0/\phi$ and, if desired, $\mathcal Z(2)=\alpha_0k_0^2/\phi^2$; on the high-frequency side, the tail data at $\sigma\to0$ and sum rules such as $\mathcal Z(-1)$, given by $\Sigma$ (Table~\ref{tab:dictionary}). How these data should be balanced---for instance, whether the high-frequency information should take precedence over $\alpha_0$---may depend on the material and on the frequency range of interest, and remains to be explored. Whatever the choice, causality and the relaxational character are then guaranteed by construction, and, since positivity constrains only $\Lambda$, negative values of $\Sigma$ and of $V$ are naturally allowed.

	Beyond these developments, the most significant perspective lies outside the local theory itself, which rests on the divergence-free simplification. In the nonlocal theory~\cite{LAF1,LAF2}, this simplification is removed, and the whole construction must be reconsidered from its starting point. The ideal field $\mathbf E$ itself is no longer the incompressible potential flow of a Neumann problem: the high-frequency limit, and with it $\alpha_\infty$, $\beta_\infty$ and the characteristic lengths, must be redefined and become dependent on the excitation~\cite{LAF_HF}. The separation between a boundary layer and an incompressible potential core, on which the present hierarchy is built, no longer holds as such, and the visco-inertial and thermal problems are no longer decoupled: the pressure, and with it the thermal source, varies within the pore space and is coupled to the flow, so that the thermal response acquires field-dependent contributions of its own and the asymmetry between the two problems disappears. The results obtained here should then be recovered as the local limit of this more general description, rather than extended term by term. The construction of full-frequency models will then also be considerably more demanding. In the local theory, the singularities of the response functions lie on the negative imaginary axis of the complex frequency plane, at $\omega=-i/\Theta_n$: the response is a superposition of decaying exponentials, monotonic in time, and a positive relaxation-time distribution suffices to describe it. In the nonlocal theory this restriction is lifted: the singularities may leave the imaginary axis, the modes acquire oscillatory, propagative components, and only causality and passivity remain as general constraints. The simple monotonic structure exploited here, and with it the direct spectral reading of the parameters, will then have to give way to a more general representation.

	\section*{Funding}
	This research received no specific grant from any funding agency, commercial or not-for-profit sectors.

	\section*{Declaration of interests}
	The authors report no conflict of interest.

	\section*{Data availability statement}
	The data shown in Fig.~\ref{fig:tube} can be reproduced from the closed-form expressions given in the text. No other data were created or analysed in this study.

	\section*{Use of AI tools}
	The authors used Claude (Anthropic) as an assistant for language editing, for restructuring and clarifying the exposition, for formatting the manuscript in LaTeX, and for checking parts of the calculations. All scientific content, results and interpretations are the authors' own; the authors reviewed and edited all AI-assisted text and take full responsibility for the content of the paper.

	\appendix
	\section{Cancellation of the singular terms at sharp wedges, and apparent parameters}\label{app:wedge}

\begin{figure}[t]
		\centering
		\includegraphics[width=0.6\textwidth]{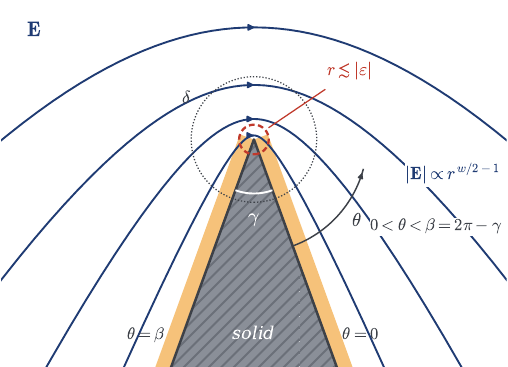}
		\caption{Flow across a solid wedge of apex angle $\gamma$ (schematic). The fluid occupies the angle $\beta=2\pi-\gamma$ around the apex, and the polar angle $\theta$ runs from one face ($\theta=0$) to the other ($\theta=\beta$). The ideal flow $\mathbf E$ (blue) turns around the apex, where it is singular, $|\mathbf E|\propto r^{w/2-1}$. Along the faces, the boundary layer (shaded) has the thickness $|\varepsilon|$. Within the inner region $r\lesssim|\varepsilon|$ (dashed circle), the layers of the two faces merge and the boundary-layer description fails. The intermediate radius $\delta$, with $|\varepsilon|\ll\delta\ll\ell$, separates the inner and outer parts of the analysis of Sec.~\ref{app:wedge-inner}.}
		\label{fig:wedge}
	\end{figure}

	This Appendix condenses unpublished calculations by one of the authors, and follows their order. We consider pore geometries invariant along one direction, with the flow across this direction, and with wedge-shaped solid asperities of apex angle $\gamma$, as in Refs.~\cite{FGL,CSGL,LANG} (Fig.~\ref{fig:wedge}). All integrals are taken per unit length along the invariant direction, so that $S_p$ reduces to the boundary curves of the cross-section and $|V_p|$ to its area. We consider a single apex; the contributions of several apices add.

	The argument proceeds in five steps. We first describe the singularity of the ideal field at the apex (Sec.~\ref{app:wedge-field}), and how it enters the Johnson integral (Sec.~\ref{app:wedge-johnson}). We then ask which part of that integral is meaningful (Sec.~\ref{app:wedge-content}). The answer suggests the form of the high-frequency expansion: nonanalytic terms that pair with the regular ones, and coefficients whose divergences, as $\gamma\to0$, can compensate only if their dependence on $\gamma$ contains logarithms (Sec.~\ref{app:wedge-expansion}). Only then do we turn to the flow near the apex, which shows that this compensation does take place and gives the constants (Sec.~\ref{app:wedge-inner}). The consequences follow: the real and imaginary parts and the comparison with the computations of Langlois (Sec.~\ref{app:wedge-ReIm}), the frequencies at which the terms separate (Sec.~\ref{app:wedge-separation}), the apparent parameters (Sec.~\ref{app:wedge-apparent}), the next group of terms (Sec.~\ref{app:wedge-next}), and the spectral reading (Sec.~\ref{app:wedge-spectral}).

	\subsection{The wedge and the singularity of the ideal field}\label{app:wedge-field}

	Let $(r,\theta)$ be polar coordinates centred on the apex, the fluid occupying the angle $0<\theta<\beta$, with $\beta=2\pi-\gamma$. The ideal field is the gradient of the potential $u=\mathbf e\cdot\mathbf x-\varPhi$, a harmonic function with zero normal derivative on the two faces $\theta=0$ and $\theta=\beta$. Near the apex, separation of variables gives
	\begin{equation}
		u=u_0+\sum_{m\ge1}A_m\,r^{m\pi/\beta}\cos\frac{m\pi\theta}{\beta},
		\label{eq:wedge-psi}
	\end{equation}
	each term satisfying the Neumann condition on both faces, since $\partial_\theta\cos(m\pi\theta/\beta)$ vanishes at $\theta=0$ and at $\theta=\beta$. The exponents $m\pi/\beta$ are the multiples of $\pi/\beta=w/2$, where
	\begin{equation}
		w=\frac{2\pi}{2\pi-\gamma},\qquad w-1=\frac{\gamma}{2\pi-\gamma}=\frac{\gamma}{2\pi}+\frac{\gamma^2}{4\pi^2}+\cdots,\qquad \frac1{w-1}=\frac{2\pi}\gamma-1 .
		\label{eq:wedge-w}
	\end{equation}
	For $0<\gamma<\pi$, one has $1<w<2$: $w\to1$ for a knife edge ($\gamma\to0$), and $w\to2$ for a flat wall ($\gamma\to\pi$). The coefficients $A_m$ are not given by the local analysis; they depend on the whole pore geometry, through the global potential problem.

	The leading term, $m=1$, gives the field
	\begin{equation}
		E_r=A_1\frac w2\,r^{w/2-1}\cos\frac{w\theta}2,\qquad
		E_\theta=-A_1\frac w2\,r^{w/2-1}\sin\frac{w\theta}2,\qquad
		|\mathbf E|^2=A_1^2\Big(\frac w2\Big)^2r^{w-2},
		\label{eq:wedge-E-local}
	\end{equation}
	whose modulus does not depend on $\theta$. Since $w/2-1<0$, the ideal field is singular at the apex: the ideal fluid turns around it with an infinite velocity. On the face $\theta=0$, where $\cos(w\theta/2)=1$, the slip $E_r$ is directed away from the apex; on the face $\theta=\beta$, where $\cos(w\beta/2)=\cos\pi=-1$, it is directed towards it. The flow comes along one face, turns around the apex, and leaves along the other. For a knife edge, $|\mathbf E|\propto r^{-1/2}$, with the same exponent as the electric field near the edge of a thin conducting sheet~\cite{LALIel}; for a right-angled edge ($\gamma=\pi/2$, $w=4/3$), $|\mathbf E|\propto r^{-1/3}$. On both faces,
	\begin{equation}
		|\mathbf E|^2=E_0^2\Big(\frac r\ell\Big)^{w-2}\Big[1+O\Big(\big(r/\ell\big)^{w/2}\Big)\Big],\qquad E_0^2=A_1^2\Big(\frac w2\Big)^2\ell^{w-2},
		\label{eq:wedge-E2}
	\end{equation}
	where $\ell$ is a geometrical length over which this local form holds, such as the length of the faces or the pore aperture. On each face, the correction comes from the cross term between $m=1$ and $m=2$. It has opposite signs on the two faces, so that on the two faces taken together the first correction is $O\big((r/\ell)^{w}\big)$; it comes from the square of the $m=2$ field and from the cross term between $m=1$ and $m=3$. Only the combination $E_0^2\ell^{2-w}=A_1^2(w/2)^2$, the intensity of the apex singularity, is intrinsic; a change of $\ell$ will be absorbed into the constant $c$ of Eq.~\eqref{eq:wedge-knife}.

	\subsection{The Johnson integral near the apex}\label{app:wedge-johnson}

	The first coefficient of Eq.~\eqref{eq:inverse-tortuosity-d} is the Johnson integral, $d_1=|V_p|^{-1}\int_{S_p}|\mathbf E|^2dS$ [Eq.~\eqref{eq:d123}]. Near the apex, $|\mathbf E|^2\propto r^{w-2}$ with $w-2>-1$: the integrand is singular but integrable, and $\Lambda$ is finite for any $\gamma>0$. To isolate the contribution of the apex, we split the integral into the two faces, $0<r<\ell$, and the rest of the wall. On each face, the leading form~\eqref{eq:wedge-E2} gives
	\begin{equation}
		\int_0^\ell\Big(\frac r\ell\Big)^{w-2}dr=\ell^{2-w}\Big[\frac{r^{w-1}}{w-1}\Big]_0^\ell=\frac{\ell}{w-1},
	\end{equation}
	and, with a factor 2 for the two faces,
	\begin{equation}
		d_1=\frac1{|V_p|}\int_{S_p}|\mathbf E|^2dS=\frac{a}{w-1}+b,\qquad a=\frac{2E_0^2\ell}{|V_p|},
		\label{eq:wedge-d1}
	\end{equation}
	where $b$ collects the regular part,
	\begin{equation}
		b=\frac1{|V_p|}\Big\{\int_{S_p\setminus\text{faces}}|\mathbf E|^2dS+\int_{\text{faces}}\Big[|\mathbf E|^2-E_0^2\Big(\frac r\ell\Big)^{w-2}\Big]dS\Big\}.
		\label{eq:wedge-b}
	\end{equation}
	By Eq.~\eqref{eq:wedge-E2}, the integrand of the second integral is $O(r^{3w/2-2})$ on each face, which is integrable for $w>2/3$, so that $b$ remains bounded as $\gamma\to0$. With the exact relation $1/(w-1)=2\pi/\gamma-1$, Eq.~\eqref{eq:wedge-d1} reads
	\begin{equation}
		d_1=\frac{2\pi a}{\gamma}+b-a :
	\end{equation}
	$d_1$ diverges as $1/\gamma$, and $\Lambda=2/(\alpha_\infty d_1)\simeq\gamma/(\pi\alpha_\infty a)$ tends to zero as the wedge sharpens into a knife edge.

	\subsection{What $\Lambda$ contains, and what it does not}\label{app:wedge-content}

	Where, along the faces, does the large value of $d_1$ come from? By Eq.~\eqref{eq:wedge-E2}, the part of the apex integral that comes from the distances $r<\rho$ is the fraction
	\begin{equation}
		\int_0^\rho\Big(\frac r\ell\Big)^{w-2}dr\Big/\int_0^\ell\Big(\frac r\ell\Big)^{w-2}dr=\Big(\frac\rho\ell\Big)^{w-1}.
		\label{eq:wedge-fraction}
	\end{equation}
	For a knife edge ($w=1$), each decade of distance to the apex would contribute equally, and for small angles it nearly does. Half of the apex contribution comes from $r<\ell\,2^{-1/(w-1)}=\ell\,2^{1-2\pi/\gamma}$, that is, from $r<\ell/8$ for $\gamma=\pi/2$, from $r<\ell/128$ for $\gamma=\pi/4$, and from $r<2\times10^{-19}\ell$ for $\gamma=0.1$. As the wedge sharpens, $\Lambda$ is controlled by smaller and smaller distances from the apex.

	The Johnson integral, however, rests on a local result. Along a face, at a distance $r$ from the apex, the vortical part of the deficit is the Stokes layer $\mathbf w_{\mathrm{vort},\varepsilon}\simeq\mathbf E\,e^{-\zeta/\varepsilon}$, whose contribution to the projection~\eqref{eq:tortuosity-vortical-projection} is, per unit length of wall,
	\begin{equation}
		\int_0^\infty\mathbf E\cdot\mathbf w_{\mathrm{vort},\varepsilon}\,d\zeta=\varepsilon\,|\mathbf E|^2\big[1+O(|\varepsilon|/r)\big] .
		\label{eq:wedge-local-Johnson}
	\end{equation}
	This is the origin of the term $d_1\varepsilon$, and it holds only where the slip varies slowly over a distance $|\varepsilon|$. Near the apex, the slip varies over the distance $r$ itself: the relative correction $O(|\varepsilon|/r)$ is no longer small at distances $r\lesssim|\varepsilon|$, where the layers of the two faces merge (Fig.~\ref{fig:wedge}). There, Eq.~\eqref{eq:wedge-local-Johnson} loses its meaning.

	The Johnson integral therefore contains a part that should not be there. Here and until Sec.~\ref{app:wedge-ReIm}, $\varepsilon$ is taken real and positive, as on the positive imaginary frequency axis (Sec.~\ref{sec6}), and the results are then continued analytically, $\varepsilon^w$ being taken on the principal branch. Unlike Sec.~\ref{sec6}, where only the normal distance is stretched and the stretched variable may be complex [Eq.~\eqref{eq:sector}], this appendix uses $\varepsilon$ as a length in the plane of the wedge: in the region $r<\varepsilon$ below, and in the scaling $\mathbf x=\varepsilon\mathbf X$ of Sec.~\ref{app:wedge-inner}, which poses the inner problem in a real domain, the infinite wedge, only for real $\varepsilon$. By Eq.~\eqref{eq:wedge-fraction}, the region $r<\varepsilon$ contributes to the term $d_1\varepsilon$ the amount
	\begin{equation}
		\frac{a\,\varepsilon}{w-1}\Big(\frac{\varepsilon}{\ell}\Big)^{w-1}=\frac{a\,\ell^{1-w}}{w-1}\,\varepsilon^w=\frac{\varepsilon}{|V_p|}\,2E_0^2\int_0^{\varepsilon}\Big(\frac r\ell\Big)^{w-2}dr ,
		\label{eq:wedge-spurious}
	\end{equation}
	which is of order $\varepsilon^w$ and diverges as $1/(w-1)$ when $\gamma\to0$. This spurious part is not small: for $\varepsilon/\ell=10^{-3}$, it amounts to 10\%, 37\% and 89\% of the apex contribution for $\gamma=\pi/2$, $\pi/4$ and $0.1$, respectively. Conversely, the Johnson integral does not contain the true contribution of that region. Within that region, the true velocity vanishes on the walls and remains bounded, so that the deficit is at most of the order of the ideal field itself; the true contribution is then at most of the order of $\int|\mathbf E|^2dV$ over the region, that is, of $\int_0^{\varepsilon}r^{w-2}\,r\,dr=\varepsilon^w/w$. It is of order $\varepsilon^w$ as well, but its coefficient remains bounded as $\gamma\to0$. The difference between the two lies in the weighting: Eq.~\eqref{eq:wedge-local-Johnson} weights each element $dr$ of the faces by the thickness $\varepsilon$ of the layer, even where $r<\varepsilon$, whereas the true weight of that region is its own area, $r\,dr\,d\theta$.

	Removing the spurious part and adding the true one are thus both corrections of order $\varepsilon^w$. The high-frequency expansion of a wedge geometry must therefore contain a nonanalytic term in $\varepsilon^w$, whose coefficient contains, with the opposite sign, the divergent factor $a\,\ell^{1-w}/(w-1)$ of Eq.~\eqref{eq:wedge-spurious}.

	\subsection{The expansion and its nonanalytic terms}\label{app:wedge-expansion}

	Generalizing this observation, the high-frequency expansion of a geometry with wedges is sought in the form
	\begin{equation}
		\frac1{\alpha(\omega)}=\frac1{\alpha_\infty}-\big(d_1\varepsilon+d_w\varepsilon^w\big)-\big(d_2\varepsilon^2+d_{1+w}\varepsilon^{1+w}+d_{2w}\varepsilon^{2w}\big)-\cdots .
		\label{eq:wedge-expansion}
	\end{equation}
	The coefficients $d_1$ and $d_2$ of the integer powers are given by the smooth-wall expressions~\eqref{eq:d123}, applied to the wedge geometry, whose ideal field has the apex singularity of Sec.~\ref{app:wedge-field}. The first one, the Johnson integral~\eqref{eq:wedge-d1}, converges at the apex. The second one does not: its surface terms vanish on the straight faces, but its volume term $\int_{V_p}g^{\mu\nu}Q^{(0)}_{;\mu}Q^{(0)}_{;\nu}\,dV$ [Eq.~\eqref{eq:d2-final}] diverges at the apex. The coefficient $d_2$ is the smooth-wall expression with this volume term replaced by its finite part: the limit, as $\rho\to0$, of the integral over $r>\rho$ once its term in $\rho^{w-2}$, which diverges, is removed. The term so removed is cancelled by the contribution of the apex region (Sec.~\ref{app:wedge-next}).

	The nonanalytic terms come from the apex region. The first, in $\varepsilon^w$, is the one just met: its exponent follows from the area $\varepsilon^2$ of the region and the field $|\mathbf E|^2\propto\varepsilon^{w-2}$ at its edge. The term in $\varepsilon^{2w}$ arises in the same way from the next terms of the local field, which contribute to $|\mathbf E|^2$ the terms in $r^{2w-2}$ of Eq.~\eqref{eq:wedge-E2}. The term in $\varepsilon^{1+w}$ arises from the displacement flow, which changes at relative order $\varepsilon$ the intensity of the apex singularity seen by the apex region (Sec.~\ref{app:wedge-next}). In general, the exponents are of the form $jw+k$, with $j$ and $k$ nonnegative integers: the multiples of $w$ follow from the exponents $mw/2$ of the local field~\eqref{eq:wedge-psi}, and the integers from the smooth-wall expansion. The exponent $3w/2$ of the cross term between $m=1$ and $m=2$ does not appear, because this term is odd under the reflection that exchanges the two faces, whereas the wedge is symmetric.

	All the coefficients are real: on the positive imaginary $\omega$-axis, $\varepsilon$ and $\alpha$ are real (Sec.~\ref{sec6}), and so are all the coefficients of the asymptotic expansion of a real function in powers of a real variable.

	Each bracket of Eq.~\eqref{eq:wedge-expansion} groups a regular term with the nonanalytic terms it merges with when the wedge sharpens: as $w\to1$, $\varepsilon^w\to\varepsilon$, and $\varepsilon^{1+w}$ and $\varepsilon^{2w}$ tend to $\varepsilon^2$. The terms are written in their formal order, $\varepsilon$, $\varepsilon^w$, $\varepsilon^2$, $\varepsilon^{1+w}$, $\varepsilon^{2w}$, \dots, that of decreasing magnitude as $\varepsilon\to0$ at a fixed angle; but this order becomes effective only at extremely high frequencies (Sec.~\ref{app:wedge-separation}). At the frequencies of interest and for small angles, $(\varepsilon/\ell)^{w-1}$ remains close to 1: the terms of each group are of the same magnitude, and they must be taken together.

	\paragraph{The expected compensation.} As $\gamma\to0$, $d_1\simeq2\pi a/\gamma$ diverges, and, by Sec.~\ref{app:wedge-content}, $d_w$ must contain $-a\,\ell^{1-w}/(w-1)\simeq-2\pi a/\gamma$. If this is the whole divergent part of $d_w$, the first pair reads
	\begin{equation}
		d_1\varepsilon+d_w\varepsilon^w=\frac{a}{w-1}\,\varepsilon\Big[1-\Big(\frac\varepsilon\ell\Big)^{w-1}\Big]+O(\varepsilon),
		\label{eq:wedge-pair}
	\end{equation}
	where the terms $O(\varepsilon)$ remain bounded as $\gamma\to0$. Each of the two terms diverges, but their sum does not: since $(\varepsilon/\ell)^{w-1}=e^{(w-1)\ln(\varepsilon/\ell)}=1+(w-1)\ln(\varepsilon/\ell)+\cdots$, it tends to $a\,\varepsilon\ln(\ell/\varepsilon)$, a finite knife-edge law in $\varepsilon\ln\varepsilon$. The divergence of $\Lambda^{-1}$ would then be an artefact of reading the Johnson term alone.

	\paragraph{How the coefficients must depend on the angle.} The original calculations approached this compensation in the opposite direction: rather than deriving the sum from the coefficients, they asked how the coefficients must depend on $\gamma$ for the sum to remain finite. In practice, everything then happens as if the usual smooth-surface characteristic parameters took effective values (Sec.~\ref{app:wedge-apparent}). Following this route, we write the first pair in the form
	\begin{equation}
		\frac{\alpha_\infty}{\alpha(\omega)}=1-\frac{2\varepsilon}{\Lambda}+\Big(\frac{2\varepsilon}{\Lambda_w}\Big)^w-\cdots,\qquad \frac2\Lambda=\alpha_\infty d_1,\qquad \Big(\frac2{\Lambda_w}\Big)^w=-\alpha_\infty d_w ,
		\label{eq:wedge-lengths}
	\end{equation}
	where $\Lambda$ is the viscous characteristic length and $\Lambda_w$ a second length, positive since $d_w<0$ for small angles. When $w\to1$, the wedge term becomes $2\varepsilon/\Lambda_w$: the two terms then have the same form, and differ only by their lengths. Since $2/\Lambda$ diverges as $1/\gamma$ (Sec.~\ref{app:wedge-johnson}), the simplest assumption is
	\begin{equation}
		\frac2\Lambda=\frac p\gamma+t+\cdots,\qquad \frac2{\Lambda_w}=\frac{p_w}\gamma+t_w+\cdots .
		\label{eq:wedge-simple}
	\end{equation}
	To compare the two terms, we write $(2\varepsilon/\Lambda_w)^w=(2\varepsilon/\Lambda_w)\,(2\varepsilon/\Lambda_w)^{w-1}$ and expand the power,
	\begin{equation}
		\Big(\frac{2\varepsilon}{\Lambda_w}\Big)^{w-1}=\exp\Big[(w-1)\ln\frac{2\varepsilon}{\Lambda_w}\Big]=1+(w-1)\ln\frac{2\varepsilon}{\Lambda_w}+\frac12(w-1)^2\ln^2\frac{2\varepsilon}{\Lambda_w}+\cdots .
		\label{eq:wedge-power}
	\end{equation}
	By Eq.~\eqref{eq:wedge-w}, $2(w-1)/\Lambda_w=p_w/2\pi+O(\gamma)$ and $\ln(2/\Lambda_w)=\ln(p_w/\gamma)+O(\gamma)$. Hence
	\begin{equation}
		-\frac{2\varepsilon}\Lambda+\Big(\frac{2\varepsilon}{\Lambda_w}\Big)^w=\Big[\frac{p_w-p}{\gamma}+t_w-t\Big]\varepsilon+\frac{p_w}{2\pi}\,\varepsilon\ln(p_w\varepsilon)-\frac{p_w}{2\pi}\,\varepsilon\ln\gamma+\cdots,
		\label{eq:wedge-trial}
	\end{equation}
	where the dots stand for terms that vanish with $\gamma$, such as the terms in $\gamma\ln^2\gamma$ given by the third term of Eq.~\eqref{eq:wedge-power}. The divergence in $1/\gamma$ disappears if $p_w=p$, but another one remains: the term $-(p/2\pi)\,\varepsilon\ln\gamma$, which diverges logarithmically. The simple forms~\eqref{eq:wedge-simple} are therefore insufficient: the inverse lengths must contain logarithms of $\gamma$,
	\begin{equation}
		\frac2\Lambda=\frac p\gamma+q\ln\gamma+t+\cdots,\qquad \frac2{\Lambda_w}=\frac{p_w}\gamma+q_w\ln\gamma+t_w+\cdots .
		\label{eq:wedge-log-expansions}
	\end{equation}
	These add $(q_w-q)\,\varepsilon\ln\gamma$ to Eq.~\eqref{eq:wedge-trial}, the products $2(w-1)/\Lambda_w$ and $(w-1)\ln(2/\Lambda_w)$ acquiring only corrections that vanish with $\gamma$. The sum is therefore finite as $\gamma\to0$ if and only if
	\begin{equation}
		p_w=p,\qquad q_w-q=\frac{p}{2\pi},
		\label{eq:wedge-conditions}
	\end{equation}
	and its limit is then the knife-edge law
	\begin{equation}
		\lim_{\gamma\to0}\Big[-\frac{2\varepsilon}\Lambda+\Big(\frac{2\varepsilon}{\Lambda_w}\Big)^w\Big]=\frac{p}{2\pi}\,\varepsilon\ln(p\varepsilon)+(t_w-t)\,\varepsilon ,
		\label{eq:wedge-knife-limit}
	\end{equation}
	the further terms of the expansions in $\gamma$, in $\gamma\ln^2\gamma$, $\gamma\ln\gamma$, $\gamma$, \dots, only adding terms that vanish with $\gamma$. The pair thus keeps the smooth-surface form $-2\varepsilon/\Lambda$, with $2/\Lambda$ replaced by the effective value $(p/2\pi)\ln[1/(p\varepsilon)]+t-t_w$, which depends logarithmically on the frequency (Sec.~\ref{app:wedge-apparent}).

	Truncating the expansion~\eqref{eq:wedge-power} after its first terms assumes $|(w-1)\ln(2|\varepsilon|/\Lambda_w)|<1$. At high frequencies, $\ln(2|\varepsilon|/\Lambda_w)$ is large and negative, and the condition reads $|\varepsilon|>(\Lambda_w/2)\,e^{-1/(w-1)}$. Beyond this limit the two terms separate, but only at unrealistic frequencies unless the angle is large (Sec.~\ref{app:wedge-separation}).

	This argument shows what must happen: $2/\Lambda$ and $2/\Lambda_w$ must have the same divergence in $1/\gamma$, and logarithms of $\gamma$ whose coefficients differ by $p/2\pi$. It does not show that they do. This requires the flow near the apex.

	\subsection{What happens at the apex}\label{app:wedge-inner}

	We now determine the true contribution of the apex region. We introduce an intermediate distance $\delta$, with $|\varepsilon|\ll\delta\ll\ell$ (Fig.~\ref{fig:wedge}), and split the projection~\eqref{eq:tortuosity-vortical-projection} into the contributions of $r>\delta$ and $r<\delta$. The potential part of the deficit, the displacement flow, must be left out of this splitting: its projection vanishes over the whole pore space, but not over each region, to which it contributes opposite fluxes through the circle $r=\delta$, of the same order as the terms retained below.

	\emph{Outer part} ($r>\delta$). Eq.~\eqref{eq:wedge-local-Johnson} holds, and the Johnson integral is truncated at $r=\delta$. By Eq.~\eqref{eq:wedge-fraction},
	\begin{equation}
		\frac{\varepsilon}{|V_p|}\int_{S_p,\,r>\delta}|\mathbf E|^2dS=\frac{a\,\varepsilon}{w-1}\Big[1-\Big(\frac\delta\ell\Big)^{w-1}\Big]+b\,\varepsilon .
		\label{eq:wedge-outer}
	\end{equation}

	\emph{Inner part} ($r<\delta$). Near the apex, the only length available is $|\varepsilon|$: the wedge itself has none. The flow there obeys an inner problem obtained by the scaling $\mathbf x=\varepsilon\mathbf X$. Writing $R=|\mathbf X|=r/\varepsilon$ and
	\begin{equation}
		\mathbf E=E_0\Big(\frac\varepsilon\ell\Big)^{w/2-1}\hat{\mathbf E}(\mathbf X),\qquad
		\mathbf w_\varepsilon=E_0\Big(\frac\varepsilon\ell\Big)^{w/2-1}\hat{\mathbf W}(\mathbf X),\qquad
		\varepsilon Q_\varepsilon=E_0\Big(\frac\varepsilon\ell\Big)^{w/2-1}\hat P(\mathbf X),
		\label{eq:wedge-inner-scaling}
	\end{equation}
	where $|\hat{\mathbf E}|=R^{w/2-1}$ on the faces, by Eq.~\eqref{eq:wedge-E2}. Since $\varepsilon\nabla=\nabla_X$, the terms $\varepsilon^2\nabla Q_\varepsilon$ and $\varepsilon^2\nabla^2\mathbf w_\varepsilon$ of Eq.~\eqref{eq:deficit-momentum} become, up to the common factor $E_0(\varepsilon/\ell)^{w/2-1}$, $\nabla_X\hat P$ and $\nabla_X^2\hat{\mathbf W}$, and the deficit problem~\eqref{eq:deficit-momentum}--\eqref{eq:deficit-wall} becomes
	\begin{equation}
		\nabla_X^2\hat{\mathbf W}+\nabla_X\hat P-\hat{\mathbf W}=\mathbf 0,\qquad
		\nabla_X\cdot\hat{\mathbf W}=0\qquad\text{in }0<\theta<\beta,\qquad
		\hat{\mathbf W}=\hat{\mathbf E}\quad\text{on }\theta=0,\beta .
		\label{eq:wedge-inner-problem}
	\end{equation}
	This problem is posed in the infinite wedge and depends only on $w$; it contains the full viscous dynamics of the apex region, including its pressure and its displacement flow. In these variables, the potential part $\varepsilon^2\nabla Q_\varepsilon$ of the deficit becomes $\nabla_X\hat P$, and its vortical part $\hat{\mathbf W}-\nabla_X\hat P$. With $dV=\varepsilon^2d\mathbf X$, the inner contribution is
	\begin{equation}
		\frac1{|V_p|}\int_{r<\delta}\mathbf E\cdot\mathbf w_{\mathrm{vort},\varepsilon}\,dV=\frac{E_0^2\ell^{2-w}}{|V_p|}\,\varepsilon^w F_w\Big(\frac\delta\varepsilon\Big),
		\qquad F_w(\Delta)=\int_{R<\Delta}\hat{\mathbf E}\cdot\big(\hat{\mathbf W}-\nabla_X\hat P\big)\,d\mathbf X .
		\label{eq:wedge-inner}
	\end{equation}
	The factor $\varepsilon^w$ appears here directly: it is the product of the area $\varepsilon^2$ and of the squared field $(\varepsilon/\ell)^{w-2}$ of Eq.~\eqref{eq:wedge-inner-scaling}. For $R\gg1$, the vortical part reduces to Stokes layers along the two faces, with slip $|\hat{\mathbf E}|=R^{w/2-1}$, and by Eq.~\eqref{eq:wedge-local-Johnson} each of them contributes $R^{w-2}$ per unit length. Hence, integrating up to $R=\Delta$,
	\begin{equation}
		F_w(\Delta)=\frac{2\Delta^{w-1}}{w-1}+f_w+o(1),\qquad\Delta\to\infty,
		\label{eq:wedge-matching}
	\end{equation}
	where the constant $f_w$ is determined by the full inner solution.

	\emph{Sum.} Adding Eqs.~\eqref{eq:wedge-outer} and~\eqref{eq:wedge-inner}, the contribution of the first term of Eq.~\eqref{eq:wedge-matching}, equal to $a\,\varepsilon\,(\delta/\ell)^{w-1}/(w-1)$, cancels the term in $\delta$ of Eq.~\eqref{eq:wedge-outer}. The intermediate distance disappears, as it must, and
	\begin{equation}
		\frac1{\alpha(\omega)}=\frac1{\alpha_\infty}-d_1\varepsilon-d_w\varepsilon^w+\cdots,\qquad d_w=\frac{E_0^2\ell^{2-w}}{|V_p|}\,f_w .
		\label{eq:wedge-dw-fw}
	\end{equation}
	This confirms the form of the first pair in Eq.~\eqref{eq:wedge-expansion}.

	\emph{Behaviour of $d_w$ as $\gamma\to0$.} At fixed $\Delta$, $F_w(\Delta)$ is the contribution of a bounded region of the inner problem. Near $R=0$, $\hat{\mathbf E}\cdot\hat{\mathbf W}=O(R^{w-2})$, and $\int R^{w-2}\,R\,dR$ converges for every $w\ge1$, the knife edge $w=1$ included: as anticipated in Sec.~\ref{app:wedge-content}, the true contribution of the apex region is bounded. The potential part enters only through a flux: since $\hat{\mathbf E}$ is divergence-free and tangent to the faces, $\int_{R<\Delta}\hat{\mathbf E}\cdot\nabla_X\hat P\,d\mathbf X$ reduces to the integral of $\hat P\hat E_R$ over the circle $R=\Delta$, the contribution of a small circle around the apex vanishing with its radius for $w>1$. Assuming, as is natural, that the inner solution depends continuously on the opening angle, $F_w(\Delta)$ tends to a finite limit $F_1(\Delta)$ as $w\to1$. Now, by Eq.~\eqref{eq:wedge-matching},
	\begin{equation*}
		f_w=\lim_{\Delta\to\infty}\Big[F_w(\Delta)-\frac{2(\Delta^{w-1}-1)}{w-1}\Big]-\frac2{w-1},
	\end{equation*}
	and, as $w\to1$, $2(\Delta^{w-1}-1)/(w-1)\to2\ln\Delta$, so that the limit as $\Delta\to\infty$ tends to a finite constant. Hence
	\begin{equation}
		f_w=-\frac{2}{w-1}+\hat f+o(1),\qquad \hat f=\lim_{\Delta\to\infty}\big[F_1(\Delta)-2\ln\Delta\big].
		\label{eq:wedge-fw}
	\end{equation}
	Substituting into Eq.~\eqref{eq:wedge-dw-fw}, with $2E_0^2\ell^{2-w}/|V_p|=a\,\ell^{1-w}$, we obtain
	\begin{equation}
		d_w=-\frac{a\,\ell^{1-w}}{w-1}+b_w ,\qquad b_w=\frac{a\,\ell^{1-w}}2\Big[f_w+\frac2{w-1}\Big]\;\to\;\frac{a\hat f}2 .
		\label{eq:wedge-dw}
	\end{equation}
	The singular part of $d_w\varepsilon^w$, $-a\,\ell^{1-w}\varepsilon^w/(w-1)$, is exactly minus the spurious part~\eqref{eq:wedge-spurious} of the Johnson integral, the part coming from the region $r<\varepsilon$; the bounded remainder $b_w\varepsilon^w$ is the true contribution of that region. This is what the argument of Sec.~\ref{app:wedge-expansion} assumed.

	\emph{Cancellation and knife-edge limit.} Collecting the divergent parts of Eqs.~\eqref{eq:wedge-d1} and~\eqref{eq:wedge-dw}, and writing $\varepsilon^w=\varepsilon\,\ell^{w-1}(\varepsilon/\ell)^{w-1}$,
	\begin{equation}
		\frac1{\alpha(\omega)}=\frac1{\alpha_\infty}-\frac{a}{w-1}\,\varepsilon\Big[1-\Big(\frac{\varepsilon}{\ell}\Big)^{w-1}\Big]-b\,\varepsilon-b_w\varepsilon^w+\cdots .
		\label{eq:wedge-sum}
	\end{equation}
	This is Eq.~\eqref{eq:wedge-pair}, now derived. Since $(\varepsilon/\ell)^{w-1}=\exp[-(w-1)\ln(\ell/\varepsilon)]=1-(w-1)\ln(\ell/\varepsilon)+\tfrac12(w-1)^2\ln^2(\ell/\varepsilon)+\cdots$,
	\begin{equation}
		\frac{a}{w-1}\,\varepsilon\Big[1-\Big(\frac\varepsilon\ell\Big)^{w-1}\Big]=a\,\varepsilon\ln\frac\ell\varepsilon\Big[1-\frac{w-1}2\ln\frac\ell\varepsilon+\cdots\Big],
		\label{eq:wedge-u-expansion}
	\end{equation}
	while $b_w\varepsilon^w\to b_w\varepsilon$. Eq.~\eqref{eq:wedge-sum} therefore tends to the knife-edge law
	\begin{equation}
		\frac1{\alpha(\omega)}=\frac1{\alpha_\infty}-a\,\varepsilon\ln\frac{\ell}{\varepsilon}-c\,\varepsilon+o(\varepsilon),\qquad c=b+b_w ,
		\label{eq:wedge-knife}
	\end{equation}
	with the principal logarithm and real $a$, $c$, $\ell$; the slow dependence of $a$, $b$, $b_w$ and $\ell$ on $\gamma$ only adds terms that vanish with $\gamma$.

	\emph{The expansions in $\gamma$ recovered.} Eqs.~\eqref{eq:wedge-d1} and~\eqref{eq:wedge-dw} give the two lengths of Eq.~\eqref{eq:wedge-lengths}. With $1/(w-1)=2\pi/\gamma-1$,
	\begin{equation}
		\frac2\Lambda=\alpha_\infty\Big[\frac a{w-1}+b\Big]=\frac p\gamma+t,\qquad p=2\pi\alpha_\infty a,\qquad t=\alpha_\infty(b-a),
		\label{eq:wedge-Lambda-micro}
	\end{equation}
	with $a$ and $b$ taken at $\gamma=0$. If $a$, $b$ and $\ell$ have Taylor expansions in $\gamma$, $2/\Lambda$ contains no logarithm ($q=0$). For the wedge term,
	\begin{equation*}
		\Big(\frac2{\Lambda_w}\Big)^w=\alpha_\infty\Big[\frac{a\,\ell^{1-w}}{w-1}-b_w\Big]=\frac p\gamma\,\ell^{1-w}\Big[1-\frac\gamma{2\pi}\Big(1+\frac{b_w\ell^{w-1}}a\Big)\Big].
	\end{equation*}
	The logarithm appears when the $w$th root is taken. Since $(w-1)/w=\gamma/2\pi$ exactly, the factor $(p/\gamma)^{1/w}\ell^{(1-w)/w}$ equals $(p/\gamma)\exp[-(\gamma/2\pi)\ln(p\ell/\gamma)]$, which contains the term $(p/2\pi)\ln\gamma$; hence
	\begin{equation}
		\frac2{\Lambda_w}=\frac p\gamma+\frac p{2\pi}\ln\gamma-\frac p{2\pi}\Big[\ln(p\ell)+1+\frac{b_w\ell^{w-1}}a\Big]+O(\gamma\ln^2\gamma).
		\label{eq:wedge-Lambdaw-micro}
	\end{equation}
	Thus $p_w=p$ and $q_w-q=p/2\pi$: the two conditions~\eqref{eq:wedge-conditions} are satisfied. The logarithm that Sec.~\ref{app:wedge-expansion} found necessary is the trace, in $\Lambda_w$, of the exponent $w$ of the wedge term. Finally, $t_w-t=-(p/2\pi)[\ln(p\ell)+c/a]$, and Eq.~\eqref{eq:wedge-knife-limit} is Eq.~\eqref{eq:wedge-knife} multiplied by $\alpha_\infty$. A first-order term in the expansion of $a$ in powers of $\gamma$ adds the same constant to $t$ and to $t_w$, and leaves $t_w-t$ unchanged.

	\subsection{Real and imaginary parts, and comparison with the computations of Langlois}\label{app:wedge-ReIm}

	With $\varepsilon=|\varepsilon|\,e^{i\pi/4}$, $|\varepsilon|=(\nu/\omega)^{1/2}$, and $k=\phi\varepsilon^2/\alpha$, we have $\varepsilon^2=i|\varepsilon|^2$, $\varepsilon^3=|\varepsilon|^3(i-1)/\sqrt2$ and $\ln(\ell/\varepsilon)=L-i\pi/4$, with $L=\ln(\ell/|\varepsilon|)$. Eq.~\eqref{eq:wedge-knife} gives
	\begin{equation}
		k=\frac{i\phi|\varepsilon|^2}{\alpha_\infty}+\frac{\phi|\varepsilon|^3}{\sqrt2}\,(1-i)\Big[a L+c-\frac{i\pi a}{4}\Big],
		\label{eq:wedge-k-complex}
	\end{equation}
	and, since $(1-i)(X-iY)=(X-Y)-i(X+Y)$,
	\begin{align}
		\Re k&\simeq\frac{\phi|\varepsilon|^3}{\sqrt2}\Big[a\ln\frac{\ell}{|\varepsilon|}+c-\frac{\pi a}{4}\Big],\nonumber\\
		\Im k&\simeq\frac{\phi|\varepsilon|^2}{\alpha_\infty}-\frac{\phi|\varepsilon|^3}{\sqrt2}\Big[a\ln\frac{\ell}{|\varepsilon|}+c+\frac{\pi a}{4}\Big].
		\label{eq:wedge-ReIm}
	\end{align}
	The real part has the form $\omega^{-3/2}(\ln\omega+\mathrm{const})$ obtained numerically by Langlois~\cite{LANG}. In the parametrization introduced in Ref.~\cite{CSGL} and used by Langlois,
	\begin{equation*}
		\Re\frac{k}{k_0}\to\frac{\sqrt M}2\,\tilde\omega^{-3/2}\Big[1+C_1\tilde\omega^{(1-w)/2}\Big],\qquad\tilde\omega=\frac\omega{\omega_c},\qquad\omega_c=\frac{\nu\phi}{k_0\alpha_\infty},\qquad M=\frac{8k_0\alpha_\infty}{\phi\Lambda^2},
	\end{equation*}
	the term $1$ of the bracket, with its prefactor, is the Johnson term $\phi d_1|\varepsilon|^3/\sqrt2$. Taking the real part of $-\phi\varepsilon^2(d_1\varepsilon+d_w\varepsilon^w)$, with $\Re\varepsilon^{2+w}=|\varepsilon|^{2+w}\cos[(2+w)\pi/4]$ and $|\varepsilon|=\ell_c\tilde\omega^{-1/2}$, $\ell_c=(\nu/\omega_c)^{1/2}$, identifies
	\begin{equation}
		C_1=-\sqrt2\cos\frac{(2+w)\pi}{4}\;\frac{d_w}{d_1}\;\ell_c^{\,w-1}.
		\label{eq:wedge-C1}
	\end{equation}
	By Eqs.~\eqref{eq:wedge-d1} and~\eqref{eq:wedge-dw}, $d_w/d_1=-\ell^{1-w}[1-(w-1)c/a+O((w-1)^2)]$, and $-\sqrt2\cos[(2+w)\pi/4]=1+(w-1)\pi/4+\cdots$, so that, in the notation of Ref.~\cite{LANG},
	\begin{equation}
		C_1=-1+C_{C_1}\gamma+o(\gamma),\qquad C_{C_1}=\frac1{2\pi}\Big[\frac ca+\ln\frac{\ell}{\ell_c}-\frac\pi4\Big].
		\label{eq:wedge-CC1}
	\end{equation}
	The cancellation thus appears as the limit $C_1\to-1$ of the coefficient of the wedge correction. Writing $w=1+C_w\gamma+o(\gamma)$, as Langlois does, the expansion of $[1+C_1\tilde\omega^{(1-w)/2}]$ gives $\gamma[C_{C_1}+(C_w/2)\ln\tilde\omega]$. By Eq.~\eqref{eq:wedge-w}, $C_w=1/2\pi\approx0.16$, close to his numerical value $0.17$. The constant $C_{C_1}$, by contrast, is not universal: it depends on the geometry through $c/a$ and $\ell/\ell_c$. The logarithmic terms of the two parts in Eq.~\eqref{eq:wedge-ReIm} have the same coefficient, while their constants differ by $\pi a/2$, without any adjustable parameter.

		\subsection{When do the two terms separate?}\label{app:wedge-separation}

	For a small but finite angle, by Eq.~\eqref{eq:wedge-u-expansion}, Eq.~\eqref{eq:wedge-knife} holds up to corrections of relative order $(w-1)\ln(\ell/|\varepsilon|)\simeq(\gamma/2\pi)\ln(\ell/|\varepsilon|)$. The ordering of the formal expansion, a term $\varepsilon$ followed by a smaller term $\varepsilon^w$, requires $|\varepsilon/\ell|^{w-1}=e^{-(w-1)\ln(\ell/|\varepsilon|)}\ll1$, and becomes effective only when $(w-1)\ln(\ell/|\varepsilon|)$ is large. Since $1/(w-1)=2\pi/\gamma-1$ and $|\varepsilon|^2=\nu/\omega$, the crossover $(w-1)\ln(\ell/|\varepsilon|)=1$ lies at $|\varepsilon|=\ell\,e^{-(2\pi/\gamma-1)}$, so that the ordering requires $\omega\gg(\nu/\ell^2)\,e^{4\pi/\gamma}$, up to a factor $e^{-2}$, or equivalently $|\varepsilon|\ll\ell\,e^{-2\pi/\gamma}$. This is, up to the prefactor, the limit met in Sec.~\ref{app:wedge-expansion} for the expansion~\eqref{eq:wedge-power}. Even for $\gamma=\pi/4$, for which $e^{4\pi/\gamma}=e^{16}\approx10^7$, and for $\ell$ of the order of $100\,\mu$m in air, the ordering requires skin depths much smaller than $0.1\,\mu$m, that is, frequencies far into the gigahertz range, where the skin depth becomes comparable with the mean free path of the molecules and the continuum description itself fails. For smaller angles the requirement is out of reach altogether, as the estimate of Langlois for $\gamma=0.1$ also shows~\cite{LANG}. At any physically relevant frequency, the two terms must therefore be taken together. The wedge thus gives an extreme instance of the point of principle made in the Introduction: being of higher order in the formal expansion does not make a term small at the frequencies that matter.

\subsection{Apparent parameters}\label{app:wedge-apparent}

	For a smooth boundary, Eq.~\eqref{eq:wedge-ReIm} would read
	\begin{equation*}
		\Re k\simeq\frac{\phi|\varepsilon|^3}{\sqrt2}\,\frac2{\alpha_\infty\Lambda},\qquad
		\Im k\simeq\frac{\phi|\varepsilon|^2}{\alpha_\infty}-\frac{\phi|\varepsilon|^3}{\sqrt2}\,\frac2{\alpha_\infty\Lambda}.
	\end{equation*} Comparing, the measured response defines an apparent characteristic length,
	\begin{equation}
		\frac2{\Lambda_{\mathrm{app}}(\omega)}=\alpha_\infty\Big[a\ln\frac{\ell}{|\varepsilon|}+c\Big],
	\end{equation}
	the real and imaginary parts giving this value up to the shifts $\mp\pi a/4$, which reflect the phase of $\ln\varepsilon$ and are small compared with the logarithm. The bracket $a\ln(\ell/|\varepsilon|)+c$ is the Johnson integral truncated at a distance of order $|\varepsilon|$ from the apex: by Eq.~\eqref{eq:wedge-outer}, the truncation of $d_1$ at $r=\rho$ gives, as $\gamma\to0$, $a\ln(\ell/\rho)+b$, which coincides with $a\ln(\ell/|\varepsilon|)+c$ for $\rho=|\varepsilon|\,e^{-b_w/a}\to|\varepsilon|\,e^{-\hat f/2}$. The apparent length $\Lambda_{\mathrm{app}}$ is finite, depends only logarithmically on frequency, and is of the order of the geometrical lengths: everything happens as if the apex were rounded at the scale of the viscous skin depth, consistently with the truncated-domain approximation proposed in Ref.~\cite{LANG}. The same holds for the higher-order parameters, whose smooth-boundary expressions diverge at an apex (Sec.~\ref{app:wedge-next}): the measured values are those of the geometry rounded at that scale.

		\subsection{The next group of terms, and the parameter $\Sigma$ near an apex}\label{app:wedge-next}

	The second bracket of Eq.~\eqref{eq:wedge-expansion} can be read in the same way, although its inner analysis has not been carried out. Two features of the smooth-wall expression of the second coefficient, Eq.~\eqref{eq:Sigma-def}, appear near a sharp apex.

	First, this expression diverges at the apex, whatever the angle. On the straight faces the curvature term vanishes; but the slip varies along the faces as $r^{w/2-1}$, so that the Neumann datum $E^k{}_{|k}$ of the first pressure potential [Eq.~\eqref{eq:Qp0}] behaves as $r^{w/2-2}$, with opposite signs on the two faces. Near the apex, $Q^{(0)}$ then behaves as $r^{w/2-1}$, and its energy density $|\nabla Q^{(0)}|^2\propto r^{w-4}$ makes the integral $\int|\nabla Q^{(0)}|^2dV$ diverge as $\int r^{w-3}dr$. This divergence, a negative power of the distance to the apex, again comes from the region $r\lesssim|\varepsilon|$: cut there, it contributes a term of order $\varepsilon^2|\varepsilon|^{w-2}$, that is, of the order $\varepsilon^w$ of the first nonanalytic term, which absorbs it. In the matching of Sec.~\ref{app:wedge-inner}, it is cancelled by a term in $\Delta^{w-2}$ contained in the remainder $o(1)$ of Eq.~\eqref{eq:wedge-matching}. The coefficient $d_2$ of $\varepsilon^2$ is then the finite part of the divergent integral. For an apex rounded with a radius of curvature $\rho\gg|\varepsilon|$, the smooth-wall expression is finite, but its apex contribution grows as $\rho^{w-2}$ when $\rho$ decreases.

	Second, a divergence as $\gamma\to0$ appears at this order too. The next terms of the local field, $m=2$ and $m=3$, have slips varying as $r^{w-1}$ and $r^{3w/2-1}$; they produce pressure potentials in $r^{w-1}$ and $r^{3w/2-1}$. The squared gradient of the potential in $r^{w-1}$ is in $r^{2w-4}$; so is the cross term between the gradient of the potential in $r^{3w/2-1}$ and that of the leading potential, in $r^{w/2-1}$. Both give terms in $r^{2w-3}$ in the second-order integrand. Their integral, $\int_0^\ell r^{2w-3}dr=\ell^{2w-2}/(2w-2)$, converges, but diverges as $1/(2w-2)$ when $w\to1$, and its part coming from $r<|\varepsilon|$ is of order $\varepsilon^2|\varepsilon|^{2w-2}=|\varepsilon|^{2w}$. Besides, near the apex $Q^{(0)}$ contains, in addition to its part in $r^{w/2-1}$, a term in $r^{w/2}\cos(w\theta/2)$, of the form of the leading ideal potential, with a coefficient set by the whole pore space: the displacement flow changes, at relative order $\varepsilon$, the intensity of the apex singularity. Its slip, multiplied by the ideal one, gives in $d_2$ a Johnson-type integral, $\int_0^\ell r^{w-2}dr=\ell^{w-1}/(w-1)$, whose part coming from $r<|\varepsilon|$ is of order $\varepsilon^{1+w}$. Like $d_1$, the coefficient $d_2$ thus diverges as $1/\gamma$. The mechanism of the first pair is therefore expected to operate again: the divergent parts of $d_2$, $d_{1+w}$ and $d_{2w}$ cancel, $\varepsilon^{1+w}$ and $\varepsilon^{2w}$ merge with $\varepsilon^2$, and the sum tends to a term in $\varepsilon^2\ln\varepsilon$, with a finite coefficient.

\subsection{Spectral reading}\label{app:wedge-spectral}

	By Eq.~\eqref{eq:AT-zeta}, a tail $G(\sigma)\simeq g_w\sigma^{w/2}$ makes $\mathcal Z(s)$ behave as $(w/2)g_w/(s+w/2)$: a simple pole at $s=-w/2$, with residue $(w/2)g_w$ (Sec.~\ref{sec10}). In the inverse transform~\eqref{eq:Mellin-inverse}, with $x=\varepsilon^2$, this pole contributes the residue of $x^{-s}(\pi/\sin\pi s)\mathcal Z(s)$ at $s=-w/2$. Since $x^{w/2}=\varepsilon^w$ and $\sin(-\pi w/2)=-\sin(\pi w/2)$, it is the term $-[\pi(w/2)g_w/\sin(\pi w/2)]\,\varepsilon^w$ of $1/\alpha$, that is, the term $-d_w\varepsilon^w$ of Eq.~\eqref{eq:wedge-expansion}. Hence
	\begin{equation}
		\operatorname*{Res}_{s=-w/2}\mathcal Z=\frac{d_w}{\pi}\sin\frac{\pi w}{2},\qquad
		\operatorname*{Res}_{s=-1/2}\mathcal Z=\frac{d_1}{\pi},
		\label{eq:wedge-residues}
	\end{equation}
	the second relation being the case $j=0$ of Eq.~\eqref{eq:odd-dictionary}. As $\gamma\to0$, the two poles coalesce, with residues $\pm a/[\pi(w-1)]$ that diverge and cancel. Since $w/2=\tfrac12+\tfrac{w-1}2$,
	\begin{equation}
		\frac{a}{\pi(w-1)}\Big[\frac1{s+\frac12}-\frac1{s+\frac12+\frac{w-1}2}\Big]\;\to\;\frac{a}{2\pi}\,\frac1{(s+\frac12)^2},
		\label{eq:wedge-double-pole}
	\end{equation}
	a double pole. Since $\int_0\sigma^s\,d\big[\sigma^{1/2}\ln(1/\sigma)\big]$ has the double pole $\tfrac12(s+\tfrac12)^{-2}$, this corresponds to
	\begin{equation}
		G(\sigma)\simeq\frac{a}{\pi}\,\sigma^{1/2}\ln\frac1\sigma+O(\sigma^{1/2}),\qquad\sigma\to0.
		\label{eq:wedge-G}
	\end{equation}
	This tail is positive and increasing near $\sigma=0$, as a nondecreasing $G$ [Eq.~\eqref{eq:AT-G}] requires: the relaxation spectrum remains well defined when $\Lambda\to0$. As a check, the residue at $s=-\tfrac12$ of $x^{-s}(\pi/\sin\pi s)\,(a/2\pi)(s+\frac12)^{-2}$, in Eq.~\eqref{eq:Mellin-inverse}, is $a\,\varepsilon\ln\varepsilon$, the logarithmic term of Eq.~\eqref{eq:wedge-knife}.

	\section{The high-frequency coefficients and parameters in terms of the unit normal field}\label{app:greek}

	The compact forms~\eqref{eq:d2-compact}--\eqref{eq:d3-compact} and the thermal coefficients~\eqref{eq:thermal-d123} are written with the Latin indices of the wall. Since the ideal slip is tangent to the wall, they can also be written with Greek indices only, in terms of the unit normal $n^\mu$, extended to $\mathcal V_S$ along the normals, of its covariant derivative, and of the tangential projector. This Appendix gives these forms and the identities on which they rest; their meaning is discussed in Secs.~\ref{sec7} and~\ref{sec9}.

	\subsection{The normal field and the tangential projector}\label{app:greek-normal}

	In the near-wall patches, $n^\mu=\delta^\mu_3$ and $n_\mu=\delta_\mu^3=\partial_\mu x^3$: the unit normal is the gradient of $x^3=-\zeta$. Since $\Gamma^\lambda_{33}=0$ and $\Gamma^3_{3\nu}=0$, its covariant derivative,
	\begin{equation}
		n^\mu{}_{;\nu}=\Gamma^\mu_{\nu3}=\Gamma^\mu_{3\nu},
		\label{eq:B-normal-gradient}
	\end{equation}
	has the $\Gamma^i_{3j}$ as its only nonzero components. Lowered, $n_{\mu;\nu}=x^3{}_{;\mu;\nu}$ is symmetric; $n_\mu n^\mu{}_{;\nu}=0$, $n$ having unit length, and $n^\mu{}_{;\nu}n^\nu=0$, the normals being straight lines. Its trace is $n^\lambda{}_{;\lambda}=\Gamma^k_{3k}$, equal to $-2H$ on the wall [Eq.~\eqref{eq:curvatures}]. With the tangential projector $P^\mu{}_\nu=\delta^\mu_\nu-n^\mu n_\nu$, whose only nonzero components are $P^i{}_j=\delta^i_j$, the trace-free part~\eqref{eq:Gamma-trace-free} becomes the tensor
	\begin{equation}
		\mathring\Gamma^\mu_{3\nu}=n^\mu{}_{;\nu}-\tfrac12\,n^\lambda{}_{;\lambda}\,P^\mu{}_\nu,
		\qquad
		\mathring\Gamma^\alpha_{3\beta}\mathring\Gamma^\beta_{3\alpha}=n_{\alpha;\beta}n^{\alpha;\beta}-\tfrac12\big(n^\lambda{}_{;\lambda}\big)^2,
		\label{eq:B-ring}
	\end{equation}
	whose only nonzero components are the $\mathring\Gamma^i_{3j}$. On the wall, $\mathring\Gamma^\alpha_{3\beta}\mathring\Gamma^\beta_{3\alpha}=2(H^2-K)$, and the second of Eqs.~\eqref{eq:cayley-hamilton} reads $\mathring\Gamma^\mu_{3\lambda}\mathring\Gamma^\lambda_{3\nu}=\tfrac12\big(\mathring\Gamma^\alpha_{3\beta}\mathring\Gamma^\beta_{3\alpha}\big)P^\mu{}_\nu$.

	On $S_p$, the slip has no normal component, $E^3=E_3=0$. Its contractions with these tensors can therefore be written with Greek indices,
	\begin{equation}
		E_iE^i=E_\mu E^\mu,\qquad
		E_i\Gamma^i_{3j}E^j=E_\mu n^\mu{}_{;\nu}E^\nu,\qquad
		E_i\mathring\Gamma^i_{3j}E^j=E_\mu\mathring\Gamma^\mu_{3\nu}E^\nu,\qquad
		E^iE_{i;3}=\tfrac12n^\nu\big(E_\lambda E^\lambda\big)_{;\nu},
		\label{eq:B-contractions}
	\end{equation}
	and, the $\Gamma^i_{jk}$ being the intrinsic symbols on the wall, its intrinsic gradient is the tangential part of its three-dimensional covariant derivative:
	\begin{equation}
		g_{ik}g^{j\ell}E^i{}_{|j}E^k{}_{|\ell}=P^{\mu\alpha}P^{\nu\beta}E_{\mu;\nu}E_{\alpha;\beta},
		\qquad
		E^k{}_{|k}=P^{\mu\nu}E_{\mu;\nu}
		\qquad\text{on }S_p .
		\label{eq:B-surface-gradient}
	\end{equation}
	A product of two derivatives, by contrast, is not a contraction with the slip: $E_{\mu;3}E^\mu{}_{;3}=E_{i;3}E^i{}_{;3}+(E_{3;3})^2$, and $E_{3;3}$ does not vanish on $S_p$.

	\subsection{The wall identities}\label{app:greek-wall}

	The ideal field is a gradient, $E_{\mu;\nu}=E_{\nu;\mu}$; it is divergence-free, $E^\mu{}_{;\mu}=0$; and $E_\mu n^\mu=0$ along $S_p$. Differentiating the last relation along the wall, in a tangent direction $t^\nu$, gives $t^\nu E_{\alpha;\nu}n^\alpha+t^\nu n_{\alpha;\nu}E^\alpha=0$, that is, by the symmetry of $E_{\mu;\nu}$ and of $n_{\mu;\nu}$, $t^\nu\big(E_{\nu;\alpha}n^\alpha+n_{\nu;\alpha}E^\alpha\big)=0$ for every tangent vector $t^\nu$. Since the second term is already tangent, this relation gives the tangential part of $E_{\nu;\alpha}n^\alpha$, and the divergence, with $g^{\mu\nu}=P^{\mu\nu}+n^\mu n^\nu$, gives its normal part $n^\mu n^\nu E_{\mu;\nu}$:
	\begin{equation}
		P_\mu{}^\nu E_{\nu;\alpha}n^\alpha=-n_{\mu;\alpha}E^\alpha,
		\qquad
		n^\mu n^\nu E_{\mu;\nu}=-P^{\mu\nu}E_{\mu;\nu}
		\qquad\text{on }S_p ,
		\label{eq:B-wall}
	\end{equation}
	which are the first two of Eqs.~\eqref{eq:wall-identities}. Contracted with $E^\mu$, the first gives, with the normal curvature $\kappa_E$ of the wall along the slip (Sec.~\ref{sec7}),
	\begin{equation}
		n^\nu\big(E_\mu E^\mu\big)_{;\nu}=2\kappa_E\,E_\mu E^\mu\qquad\text{on }S_p .
		\label{eq:B-kappaE}
	\end{equation}

	\subsection{Visco-inertial coefficients}\label{app:greek-visco}

	By Eqs.~\eqref{eq:B-normal-gradient}, \eqref{eq:B-contractions} and~\eqref{eq:B-kappaE}, the surface integrand of Eq.~\eqref{eq:d2-final}, equal to that of Eq.~\eqref{eq:d2-compact} at every point of the wall, takes the divergence form
	\begin{equation}
		E_\mu\mathring\Gamma^\mu_{3\nu}E^\nu=-\tfrac12\big(E_\mu E^\mu n^\nu\big)_{;\nu}=(H-\kappa_E)\,E_\mu E^\mu .
		\label{eq:B-d2-div}
	\end{equation}
	The compact forms become
	\begin{align}
		|V_p|\,d_2&=\int_{S_p}E_\mu\mathring\Gamma^\mu_{3\nu}E^\nu\,dS+\int_{V_p}g^{\mu\nu}Q^{(0)}_{;\mu}Q^{(0)}_{;\nu}\,dV,\label{eq:B-d2}\\
		|V_p|\,d_3&=\int_{S_p}\Big[-\tfrac18\big(n^\lambda{}_{;\lambda}\big)^2E_\mu E^\mu+\tfrac12n^\lambda{}_{;\lambda}\,E_\mu\mathring\Gamma^\mu_{3\nu}E^\nu-\tfrac12P^{\mu\alpha}P^{\nu\beta}E_{\mu;\nu}E_{\alpha;\beta}+g_{\mu\nu}V^\mu V^\nu\Big]dS,\label{eq:B-d3}
	\end{align}
	with the tangent vector $V^\mu=P^{\mu\nu}Q^{(0)}_{;\nu}-\mathring\Gamma^\mu_{3\nu}E^\nu$, whose components are the $V^i$ of Eq.~\eqref{eq:Vi-def}; $P^{\mu\nu}Q^{(0)}_{;\nu}$ is the surface gradient of the wall values of $Q^{(0)}$. The projectors can be removed. Splitting the symmetric tensor $E_{\mu;\nu}$ into its tangential and normal parts, and using Eqs.~\eqref{eq:B-wall} and the Neumann condition $n^\nu Q^{(0)}_{;\nu}=E^k{}_{|k}$,
	\begin{equation}
		P^{\mu\alpha}P^{\nu\beta}E_{\mu;\nu}E_{\alpha;\beta}=E_{\mu;\nu}E^{\mu;\nu}-2E_\mu n^\mu{}_{;\lambda}n^\lambda{}_{;\nu}E^\nu-\big(n^\mu Q^{(0)}_{;\mu}\big)^2 ,
	\end{equation}
	and, expanding $g_{\mu\nu}V^\mu V^\nu$ and $n^\mu{}_{;\lambda}n^\lambda{}_{;\nu}$ with Eq.~\eqref{eq:cayley-hamilton},
	\begin{multline}
		|V_p|\,d_3=\int_{S_p}\Big[\Big(\tfrac18\big(n^\lambda{}_{;\lambda}\big)^2+\mathring\Gamma^\alpha_{3\beta}\mathring\Gamma^\beta_{3\alpha}\Big)E_\mu E^\mu+\tfrac32n^\lambda{}_{;\lambda}\,E_\mu\mathring\Gamma^\mu_{3\nu}E^\nu-\tfrac12E_{\mu;\nu}E^{\mu;\nu}\\
		+g^{\mu\nu}Q^{(0)}_{;\mu}Q^{(0)}_{;\nu}-\tfrac12\big(n^\mu Q^{(0)}_{;\mu}\big)^2-2Q^{(0)}_{;\mu}\mathring\Gamma^\mu_{3\nu}E^\nu\Big]dS ,
		\label{eq:B-d3-full}
	\end{multline}
	in which only the full gradients of the ideal field and of the first potential appear on the wall.

	\subsection{Thermal coefficients}\label{app:greek-thermal}

	The thermal coefficients~\eqref{eq:thermal-d123} involve the unit normal alone, whose gradient has no normal component, so that no projector is needed:
	\begin{equation}
		\begin{gathered}
		|V_p|\,d'_1=\int_{S_p}dS,\qquad
		|V_p|\,d'_2=-\tfrac12\int_{S_p}n^\lambda{}_{;\lambda}\,dS,\\
		|V_p|\,d'_3=-\tfrac14\int_{S_p}\Big[n_{\alpha;\beta}n^{\alpha;\beta}-\tfrac12\big(n^\lambda{}_{;\lambda}\big)^2\Big]dS,\\
		|V_p|\,d'_4=-\tfrac14\int_{S_p}n^\kappa{}_{;\kappa}\Big[n_{\alpha;\beta}n^{\alpha;\beta}-\tfrac12\big(n^\lambda{}_{;\lambda}\big)^2\Big]dS .
		\end{gathered}
		\label{eq:B-thermal}
	\end{equation}

	\subsection{The parameters}\label{app:greek-parameters}

	With Eqs.~\eqref{eq:B-d2-div}--\eqref{eq:B-d3} and~\eqref{eq:B-thermal}, the parameters~\eqref{eq:Lambda-def}--\eqref{eq:V-def} and~\eqref{eq:Lambdap-def}--\eqref{eq:Vp-def} read
	\begin{align}
		\frac2\Lambda&=\frac{\alpha_\infty}{|V_p|}\int_{S_p}E_\mu E^\mu\,dS,\label{eq:B-Lambda}\\
		\frac3\Sigma&=\frac4{\Lambda^2}+\frac{\alpha_\infty}{|V_p|}\Big\{-\frac12\int_{S_p}\big(E_\mu E^\mu n^\nu\big)_{;\nu}\,dS+\int_{V_p}g^{\mu\nu}Q^{(0)}_{;\mu}Q^{(0)}_{;\nu}\,dV\Big\},\label{eq:B-Sigma}\\
		\frac{15}{4V}&=\frac{12}{\Lambda\Sigma}-\frac8{\Lambda^3}+\frac{\alpha_\infty}{|V_p|}\int_{S_p}\Big[-\tfrac18\big(n^\lambda{}_{;\lambda}\big)^2E_\mu E^\mu+\tfrac12n^\lambda{}_{;\lambda}\,E_\mu\mathring\Gamma^\mu_{3\nu}E^\nu\nonumber\\
		&\qquad-\tfrac12P^{\mu\alpha}P^{\nu\beta}E_{\mu;\nu}E_{\alpha;\beta}+g_{\mu\nu}V^\mu V^\nu\Big]dS,\label{eq:B-V}
	\end{align}
	and
	\begin{gather}
		\frac2{\Lambda'}=\frac{|S_p|}{|V_p|},\qquad
		\frac3{\Sigma'}=\frac4{\Lambda'^2}-\frac1{2|V_p|}\int_{S_p}n^\lambda{}_{;\lambda}\,dS,\nonumber\\
		\frac{15}{4V'}=\frac{12}{\Lambda'\Sigma'}-\frac8{\Lambda'^3}-\frac1{4|V_p|}\int_{S_p}\Big[n_{\alpha;\beta}n^{\alpha;\beta}-\tfrac12\big(n^\lambda{}_{;\lambda}\big)^2\Big]dS .
		\label{eq:B-thermal-parameters}
	\end{gather}

	\bibliography{biblio_Param-v-th_HF}

\end{document}